\documentclass[]{aa}

\usepackage{graphicx}
\usepackage{txfonts}
\usepackage[colorlinks=true,citecolor=blue,linkcolor=black]{hyperref}
\usepackage{siunitx}
\usepackage{float}
\usepackage{bm}
 \usepackage{soul}
\setstcolor{red}
\usepackage{hyperref}
\usepackage{booktabs}
\usepackage{adjustbox} 
\usepackage{comment}
\usepackage{subcaption}
\usepackage{threeparttablex} 
\usepackage{natbib}
\usepackage{lscape} 
\usepackage{multirow}
\usepackage[dvipsnames]{xcolor}
\newcommand{\oiii}{\textup{[O\,\textsc{iii}]}}
\newcommand{\nii}{\textup{[N\,\textsc{ii}]}}
\newcommand{\sii}{\textup{[S\,\textsc{ii}]}}
\newcommand{\oi}{\textup{[O\,\textsc{i}]}}
\newcommand{\siii}{\textup{[S\,\textsc{iii}]}}
\newcommand{\oii}{\textup{[O\,\textsc{ii}]}}
\newcommand{\civ}{\textup{C\,\textsc{iv}}}
\newcommand{\hii}{\textup{H}\,\textsc{ii}}
\newcommand{\ha}{\textup{H}\ensuremath{\alpha}}
\newcommand{\hb}{\textup{H}\ensuremath{\beta}}
\newcommand{\hg}{\textup{H}\ensuremath{\gamma}}

\defcitealias{curti2017}{C17}
\defcitealias{guseva2011}{G11}
\defcitealias{nakajima2022}{N22}
\defcitealias{PG16}{PG16}

\begin{document}

   \title{Metallicity calibrations based on auroral lines from PHANGS--MUSE data\thanks{The catalogue of dust-corrected line flux measurements and other quantities of interest for the analysis conducted in this work is only available at the CDS via anonymous ftp to \url{cdsarc.u-strasbg.fr} (130.79.128.5) or via \url{http://cdsweb.u-strasbg.fr/cgi-bin/qcat?J/A+A/.}}}

   \author{Matilde Brazzini
          \inst{1, 2, 3}
          \and
          Francesco Belfiore\inst{1}
          \and
          Michele Ginolfi\inst{1, 4}
          \and
          Brent Groves\inst{5}
          \and
          Kathryn Kreckel\inst{6}
          \and
          Ryan J. Rickards Vaught\inst{7}
          \and 
          Dalya Baron\inst{8}
          \and 
          Frank Bigiel\inst{9}
          \and
          Guillermo A. Blanc\inst{8,10}
          \and 
          Daniel~A.~Dale\inst{11}
          \and
          Kathryn~Grasha\inst{12}
          \and
          Eric Habjan \inst{13}
          \and
          Ralf S. Klessen\inst{14,15}
          \and
          J. Eduardo M\'endez-Delgado\inst{16}
          \and
          Karin Sandstrom\inst{17}
          \and
          Thomas G. Williams\inst{18}
          }

    \institute{ INAF -- Arcetri Astrophysical Observatory, Largo E. Fermi 5, I-50125 Florence, Italy \\
              \email{matilde.brazzini@inaf.it}
    \and
    Dipartimento di Fisica, Università di Trieste, Sezione di Astronomia, Via G.B. Tiepolo 11, I-34143 Trieste, Italy 
    \and
    INAF -- Osservatorio Astronomico di Trieste, Via G.B. Tiepolo 11, I-34143 Trieste, Italy
    \and
    Dipartimento di Fisica e Astronomia, Università di Firenze, Via G. Sansone 1, I-50019, Sesto F.no (Firenze), Italy
    \and 
    International Centre for Radio Astronomy Research, University of Western Australia, 7 Fairway, Crawley, 6009 WA, Australia
    \and 
    Astronomisches Rechen-Institut, Zentrum für Astronomie der Universität Heidelberg, Mönchhofstraße 12-14, 69120 Heidelberg, Germany
    \and
    Space Telescope Science Institute, 3700 San Martin Drive, Baltimore, MD 21218, USA
    \and
    The Observatories of the Carnegie Institution for Science, 813 Santa Barbara St., Pasadena, CA, 91101, USA
    \and
    Argelander-Institut für Astronomie, Universität Bonn, Auf dem H\"ugel 71, D-53121 Bonn, Germany
    \and
    Departamento de Astronom\'ia, Universidad de Chile, Camino del Observatorio 1515, Las Condes, Santiago, Chile
    \and
    Department of Physics \& Astronomy, University of Wyoming, 1000 E. University Ave., Laramie, WY, 82071, USA
    \and 
    Research School of Astronomy and Astrophysics, Australian National University, Canberra, ACT 2611, Australia
    \and
    University of Connecticut, Department of Physics, 196A Auditorium Road, Unit 3046, Storrs, CT 06269, USA
    \and
    Universit\"at Heidelberg, Zentrum f\"ur Astronomie, Institut f\"ur Theoretische Astrophysik, Albert-Ueberle-Str. 2, 69120, Heidelberg, Germany
    \and
    Universit\"at Heidelberg, Interdisziplin\"ares Zentrum f\"ur Wissenschaftliches Rechnen, Im Neuenheimer Feld 205, D-69120 Heidelberg, Germany
    \and
    Astronomisches Rechen-Institut, Zentrum f\"ur Astronomie der Universit\"at Heidelberg, M\"onchhofstraße 12-14, 69120 Heidelberg, Germany
    \and
    Department of Astronomy \& Astrophysics, University of California, San Diego, 9500 Gilman Drive, La Jolla, CA 92093
    \and
    Sub-department of Astrophysics, Department of Physics, University of Oxford, Keble Road, Oxford OX1 3RH, UK
}

   \date{}

  \abstract{
  We present a chemical analysis of selected \hii\ regions from the PHANGS-MUSE nebular catalogue. Our intent is to empirically re-calibrate strong-line diagnostics of gas-phase metallicity, applicable across a wide range of metallicities within nearby star-forming galaxies. To ensure reliable measurements of auroral line fluxes, we carried out a new spectral fitting procedure whereby only restricted wavelength regions around the emission lines of interest are taken into account: this assures a better fit for the stellar continuum. No prior cuts to nebulae luminosity were applied to limit biases in auroral line detections. Ionic abundances of O$^+$, O$^{2+}$, N$^+$, S$^+$, and S$^{2+}$ were estimated by applying the direct method. We integrated the selected PHANGS-MUSE sample with other existing auroral line catalogues, appropriately re-analysed to obtain a homogeneous dataset. This was used to derive strong-line diagnostic calibrations that span from 12+log(O/H) = 7.5 to 8.8. We investigate their dependence on the ionisation parameter and conclude that it is likely the primary cause of the significant scatter observed in these diagnostics. We apply our newly calibrated strong-line diagnostics to the total sample of \hii\ regions from the PHANGS-MUSE nebular catalogue, and we exploit these indirect metallicity estimates to study the radial metallicity gradient within each of the 19 galaxies of the sample. We compare our results with the literature and find good agreement, validating our procedure and findings. With this paper, we release the full catalogue of auroral and nebular line fluxes for the selected \hii\ regions from the PHANGS-MUSE nebular catalogue. This is the first catalogue of direct chemical abundance measurements carried out with PHANGS-MUSE data. 
  }


\maketitle

\section{Introduction} \label{section-introduction}

    The evolution of heavy element abundances (metallicity) in the interstellar medium (ISM) of galaxies across different environments and cosmic epochs provides unique information on stellar yields, feedback processes, and the evolutionary histories of galaxies \citep{MaiolinoMannucci2019}. Optical spectroscopy has traditionally been used as a powerful tool to study ISM abundances, as the optical wavelength range contains line ratios probing key properties of the ISM, including density, temperature, and properties of the illuminating radiation field \citep{kewley2019}. 

    The ratios of specific forbidden, collisionally excited line fluxes (referred to as auroral to nebular ratios) emitted from the same ion are particularly sensitive to the electron temperatures, $T_e$, within the emitting region. In the optical wavelength range, the most widely used temperature-sensitive line ratios are \nii$\lambda 5756$/\nii$\lambda 6584$, \sii$\lambda 4969$/\sii$\lambda 6717,6731$, \siii$\lambda 6312$/\siii$\lambda 9069$, \oii$\lambda \lambda 7320,7330$/\oii$\lambda \lambda 3726,3729$, and \oiii$\lambda 4363$/\oiii$\lambda 5007$. 
    In the low-density regime, the flux ratios of collisionally excited forbidden lines to a hydrogen recombination line (usually \hb) depend solely upon temperature and abundance. 
    Hence, these ratios can be used to infer ionic abundances if a $T_e$ estimate is provided, for instance, through one or more of the previously listed temperature-dependent line ratios. 
    This procedure for estimating chemical abundances has long been considered the gold standard and is referred to as the direct method. 

    Because ionised nebulae are characterised by a complex internal structure, different zones within the same nebula are traced by different ions. Hence, multiple ions have to be considered to calculate total chemical abundances. For example, in the popular three-zone model approximation to \hii\ regions, the inner, high-ionisation zone is mostly traced by O$^{2+}$, the intermediate-ionisation zone is bright in S$^{2+}$, and the outer low-ionisation zone in O$^+$, S$^+$ and N$^+$. Hence, a comprehensive chemical study of typical \hii\ regions benefits from access to several temperature tracers. Moreover, temperature and density inhomogeneities may more substantially affect specific ionisation zones, and therefore bias temperature measurements from specific ions \citep{mendez-delgado+2023_density, mendez-delgado+2023, vaught+2023}.

      The major limitation of the direct method is the intrinsic faintness of the auroral lines, which are typically $\sim$100 times fainter than the corresponding nebular lines. In fact,  auroral to nebular line ratios decrease exponentially with decreasing temperature. Since  optical collisionally excited lines are major coolants for ionised gas, the electron temperature is anti-correlated with metallicity, making the detection of auroral lines extremely challenging in solar-metallicity environments. As a consequence, direct abundance studies are available only for a relatively small number of bright sources, generally at sub-solar metallicity (see, for example, \citealt{bresolin2004, bresolin2005, zurita2012, gusev2012, berg2015, toribiosancipriano2016}), and have proved extremely challenging in high-redshift objects in the pre-JWST era (e.g. \citealt{christensen2012, sanders2020, gburek2022}).  
    
    In the absence of auroral line detections, chemical abundances studies rely on different methods to infer metallicities. A possible alternative procedure consists of studying stacked, rather than individual, spectra of galaxies \citep{andrewsmartini2013, curti2017, bian2018} with the intent of obtaining a higher signal-to-noise ratio (S/N) and facilitating the detection of weak emission features. 
    
    Alternatively, specific line ratios between strong collisionally excited emission lines show a dependence on metallicity. These strong line ratios have been calibrated and used extensively as metallicity indicators \citep{pagel1979}.  Such strong-line calibrations can be obtained either empirically, for samples in which oxygen abundance has been previously derived with the direct method \citep{pilyugin+thuan2005, PG16, curti2017, nakajima2022}, or theoretically, where the oxygen abundance is estimated via photoionisation models \citep{kewleydopita2002, kobulnicky+kewley2004, dopita2016}. 
    The main caveat in using strong-line diagnostics is that, apart from a dependence on metallicity, they show further dependencies on other physical quantities associated with emitting gaseous sources, such as the ionisation parameter (that is, the ratio of the number of ionising photons to electron densities), the abundance pattern (in particular, the N/O ratio, \citealp{perez-montero2014}), the hardness of the ionising spectrum (e.g. \citealp{stasinska2010}. As a consequence, they are hard to calibrate, and different strong-line ratios provide different metallicity estimates, with discrepancies of up to $\sim 0.6$ dex in most extreme cases \citep{kewleyallison2008}. Using multiple diagnostics can help to minimise the uncertainties arising when a single line ratio is used, but simple photoionisation models nonetheless struggle to reproduce all the line ratios simultaneously \citep{mingozzi2020, marconi2024}.
    
    Another important source of uncertainty concerns the calibration dataset. Theoretical calibrations produce metallicity estimates that are systematically higher than the empirical ones for a fixed compilation of line ratios \citep{blanc2015}. Although still unclear, the origin of these discrepancies could be attributed to oversimplified assumptions made in photoionisation models, for example, about the geometry of the nebula and the lack of small-scale temperature and density fluctuations. This caveat is even more significant at high redshift, where the majority of chemical abundance measurements have been obtained using strong-line diagnostics calibrated on data from local galaxies. Since the ionisation conditions of the ISM affect strong-line metallicity calibrations, these calibrations may evolve from $z \sim 0$ to $z \sim 2$ and even more to $z \gtrsim 6$ \citep{sanders2020, sanders2021, Sanders2023, nakajima+2023}.
    
    In summary, both the direct and strong-line methods pose some challenges: auroral lines are difficult to detect at high S/N, while strong-line diagnostics are difficult to calibrate with high precision due to additional dependencies that are hard to predict and model. 
    Nevertheless, developing robust and precise methods to measure chemical abundances in low-redshift sources is essential for accurately interpreting high-redshift data, including those  obtained with JWST (e.g. \citealt{rogers2023, Laseter2024}). 
    
    In this work, we aim to provide a comprehensive, unbiased catalogue of low-redshift direct abundance measurements extending into the solar-metallicity regime. We draw our data from the nebular catalogue of \cite{groves2023}, which contains more than $31000$ spectra of \hii\ regions in nearby galaxies. These spectra were obtained with the Multi Unit Spectrograph Explorer (MUSE, \citealp{Bacon2010}) at the Very Large Telescope (VLT) in the context of the Physics at High-Angular resolution in Nearby GalaxieS (PHANGS) programme.
    
    In Sect. \ref{section-phangs-muse-survey}, we describe the PHANGS--MUSE Survey and our analysis of the nebular catalogue optimised for auroral line detections. To complement the PHANGS--MUSE dataset in the low-metallicity regime, in Sect. \ref{section-other-data} we present four additional literature catalogues of auroral lines. In Sect. \ref{section:methods}, we describe the methodology used to derive electron temperatures and ionic abundances with the direct method. Strong-line diagnostics are calibrated in Sect. \ref{section-results}, where we also discuss their dependence on the ionisation parameter. 
    A discussion of metallicity gradients within the 19 galaxies of the PHANGS--MUSE sample is presented in Sect. \ref{section-metallicity-gradients}. Finally, Sect. \ref{section-conclusions} summarises our main results.

\section{Data } 

\subsection{The PHANGS-MUSE survey}
\label{section-phangs-muse-survey}
    
    
   The PHANGS survey was designed specifically to resolve galaxies into the individual elements of the star formation process: molecular clouds, \hii\ regions, and stellar clusters. Driven by this aim, the full PHANGS sample was originally determined by selecting southern-sky accessible and low-inclination sources, so that they could be observed almost face-on with both MUSE (PHANGS--MUSE Survey, \citealp{emsellem2022}) and ALMA (PHANGS--ALMA Survey, \citealp{leroy2021a}). The same sample has more recently been targeted with the \textit{Hubble} Space Telescope (PHANGS--HST, \citealp{lee2022}), the \textit{James Webb} Space Telescope (PHANGS--JWST, \citealp{lee2023, williams2024}), and the Ultraviolet Imaging Telescope (UVIT) on the \textit{AstroSat} satellite (the recent PHANGS-AstroSat atlas, \citealp{hassani2024}). 
   
   The 19 galaxies in the PHANGS--MUSE \citep{emsellem2022} sample were selected to be nearby, massive, and star-forming ($5.2 \leq D /\text{Mpc}  \leq 19.0$, $9.40 \leq \log M_\star / M_\odot \leq 10.99$ and $ -0.56 \leq \log \text{SFR} \text{[}  M_\odot \text{yr}^{-1} \text{]} \leq 1.23$, respectively). In particular, as all of the galaxies are within $\sim 20$ Mpc, structures down to $\sim 100 $ pc can be isolated within the galactic discs at a median resolution of $\sim 0.7"$. 

 The PHANGS--MUSE observations were  carried out in wide field mode, with a field of view (FoV) of $\sim 1 $ arcmin$^2$, either in seeing-limited mode or with ground layer adaptive optics. Individual galaxies were covered by a variable number of telescope pointings, depending on their angular size and ranging from $3$ (NGC~7496) to $15$ (NGC~1433). For each pointing, the exposure time was set to $\sim 43$ minutes. The procedures of pointing alignment, flux calibration, sky subtraction, the generation of final mosaics, and line maps are described in detail in \cite{emsellem2022}. 
    
    Maps of \ha\ intensity were used by \cite{groves2023} to define \hii\ regions and other ionised nebulae leveraging the \mbox{HIIphot} algorithm \citep{thilker2000}. 
    The final catalogue provided in \cite{groves2023} and used as a starting point in this work is composed of 31497 nebulae across the 19 PHANGS-MUSE galaxies. The catalogue comprises various types of nebulae, including \hii\ regions, planetary nebulae, and supernova remnants. In this catalogue, we do not attempt to subtract the contribution of the diffuse ionised gas (DIG) along the line of sight. However, since in this work we focus only on the most luminous \hii\ regions with a high S/N detection of auroral lines (see Sect. \ref{section-sample-selection}), the DIG contribution does not affect our results for the metallicity calibrations. 
    This hypothesis finds support in the results reported in \cite{congiu2023}, in which they employed a different methodology from \cite{groves2023} to construct an analogous nebular catalogue of PHANGS-MUSE gaseous regions, taking the DIG into consideration. Here, they show that the most brilliant \hii\ regions are the least affected by DIG contamination, validating our approach. In fact, the impact of DIG subtraction on the strong-line metallicities is found to be small for all except the faintest \hii\ regions (Tova et al., in prep.).

\subsubsection{Spectral fitting} 
\label{section-spectral-fitting}

     \begin{figure*}[t]
   \centering
   \includegraphics[width=\linewidth]{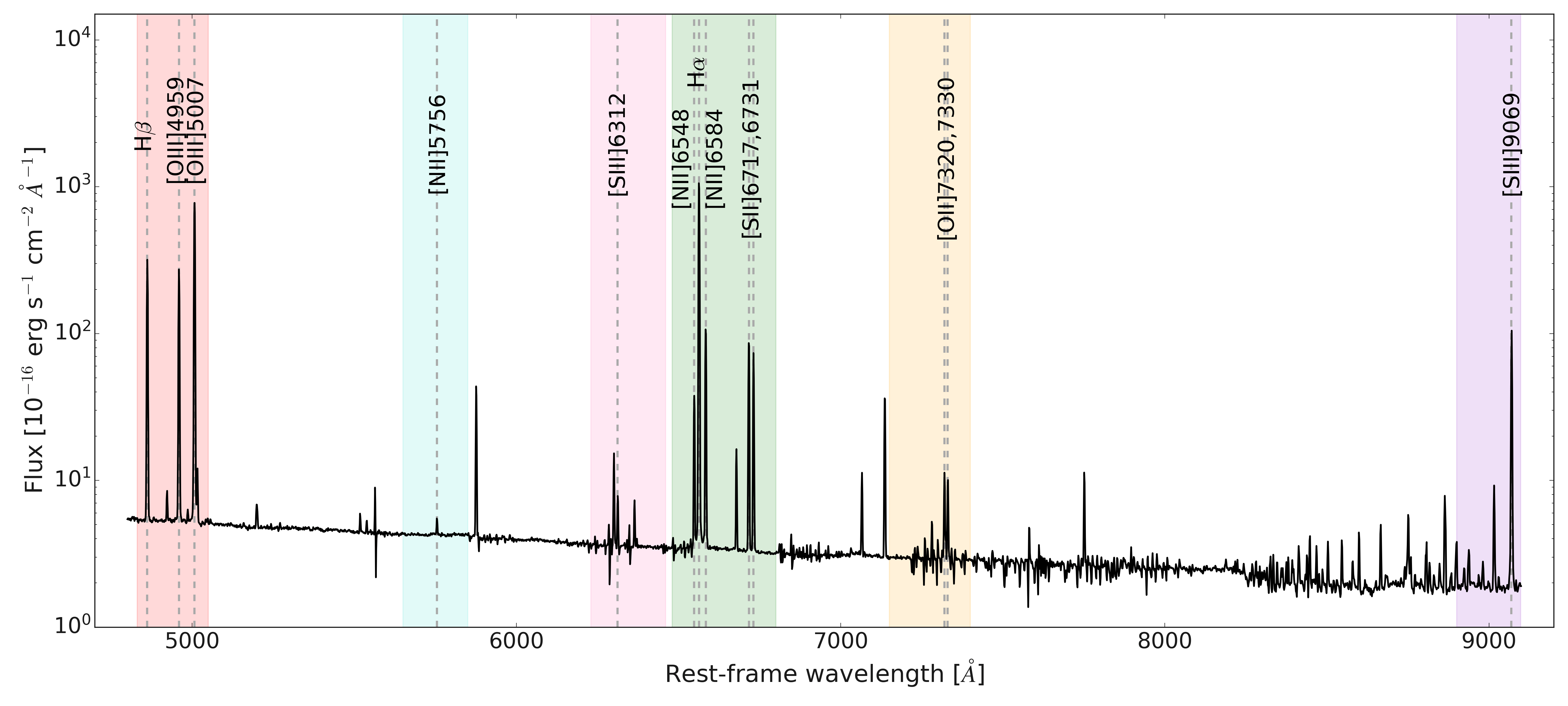}
   \caption{Spectrum of an \hii\ region (Galaxy: NGC5068, region ID: 268) with the six spectral regions from Table \ref{table-spectral-regions} highlighted in different colours. Auroral lines are the faintest lines of the spectrum, with typical intensities $\sim 100$ times less intense than nebular lines and $\sim 1000$ times less intense than hydrogen recombination lines. }
   \label{figure-HIIreg-spectrum}
    \end{figure*}

\begin{figure*}
  \centering
  \begin{subfigure}{0.33\linewidth}
    \centering
    \includegraphics[width=\linewidth]{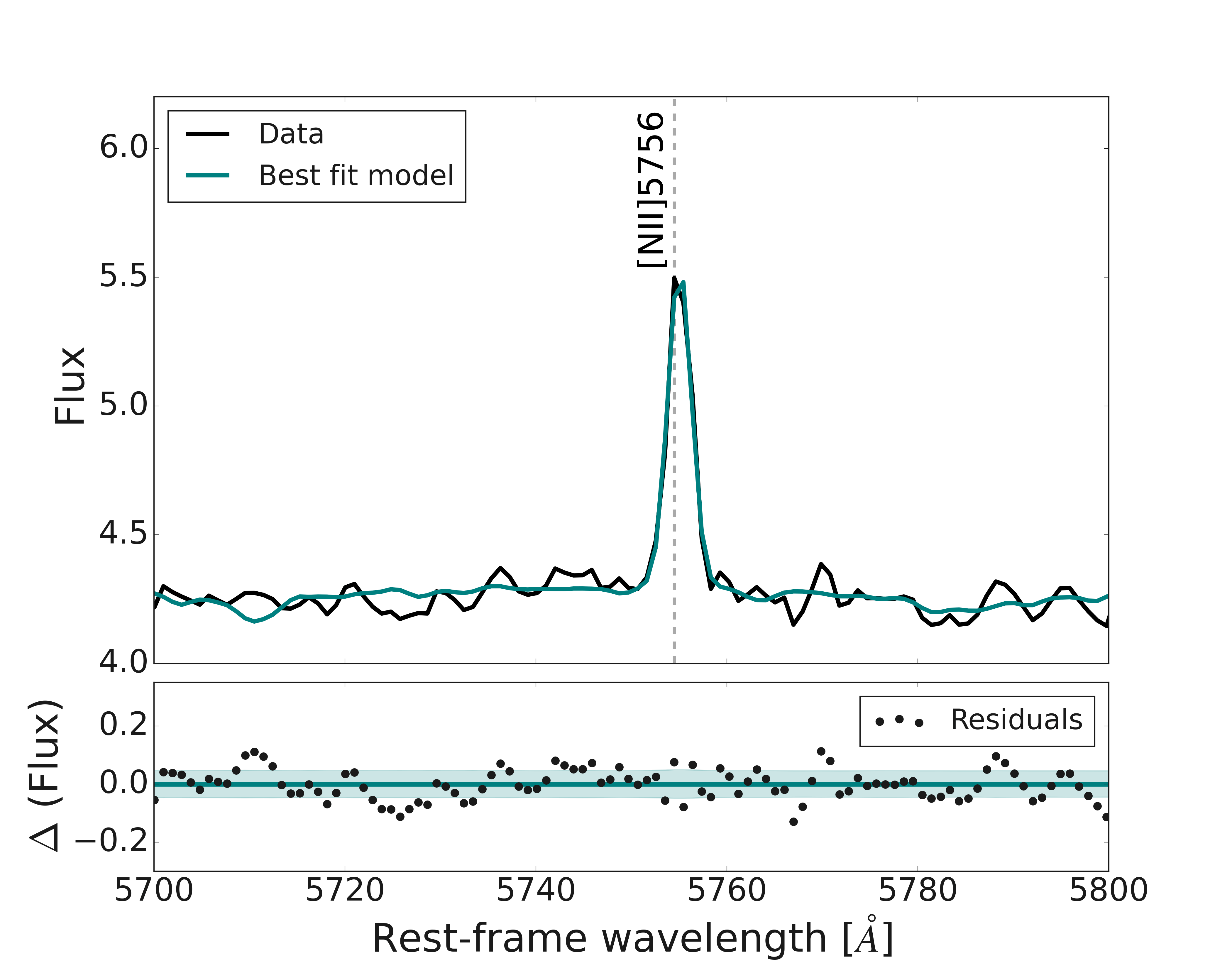}
    \label{figure-NII-galNGC5068-regID268}
  \end{subfigure}
  \begin{subfigure}{0.33\linewidth}
    \includegraphics[width=\linewidth]{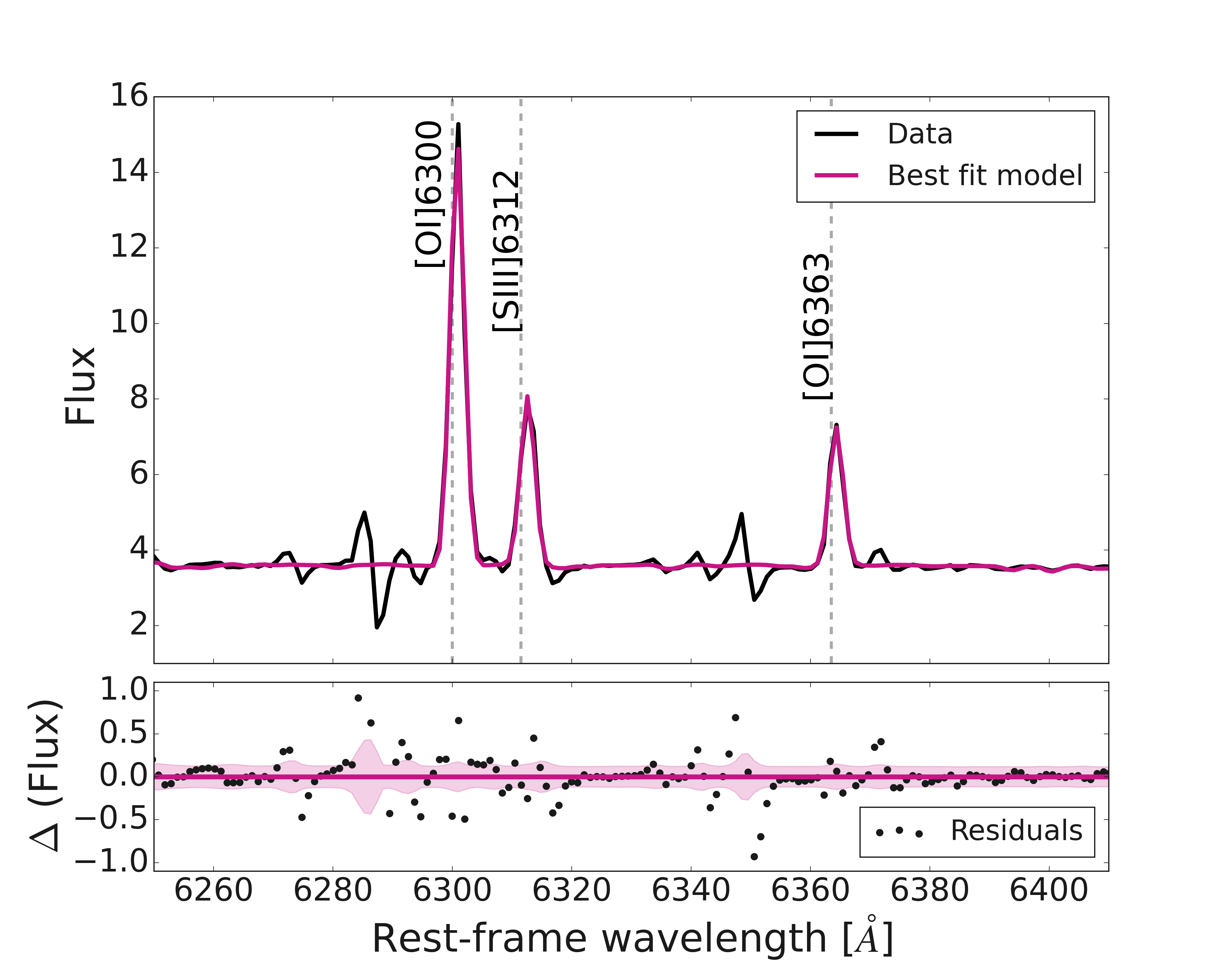}
    \label{figure-SIII-galNGC5068-regID268}
  \end{subfigure}
  \begin{subfigure}{0.33\linewidth}
    \includegraphics[width=\linewidth]{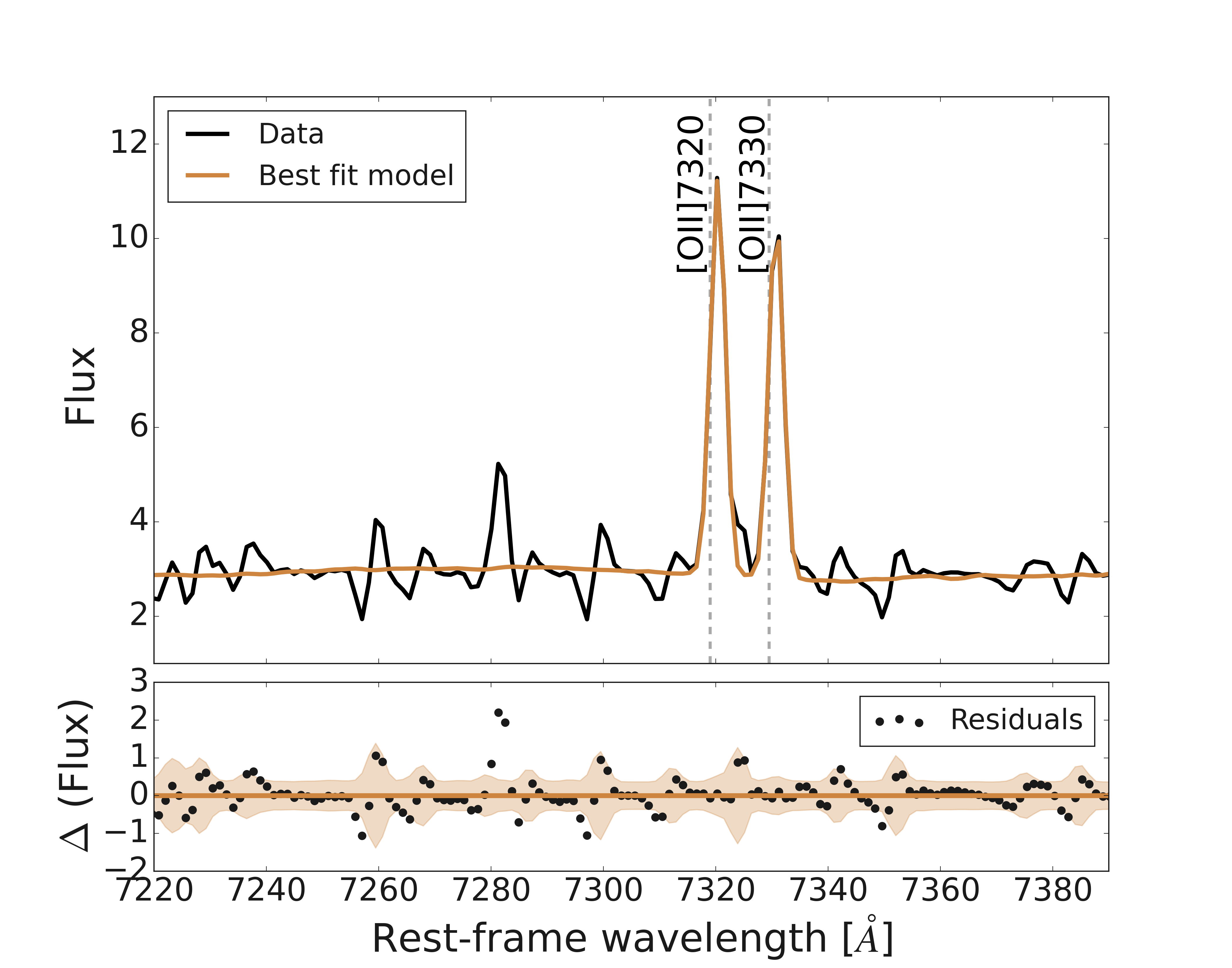}
    \label{figure-OII-galNGC5068-regID268}
  \end{subfigure}

    \begin{subfigure}{0.33\linewidth}
    \centering
    \includegraphics[width=\linewidth]{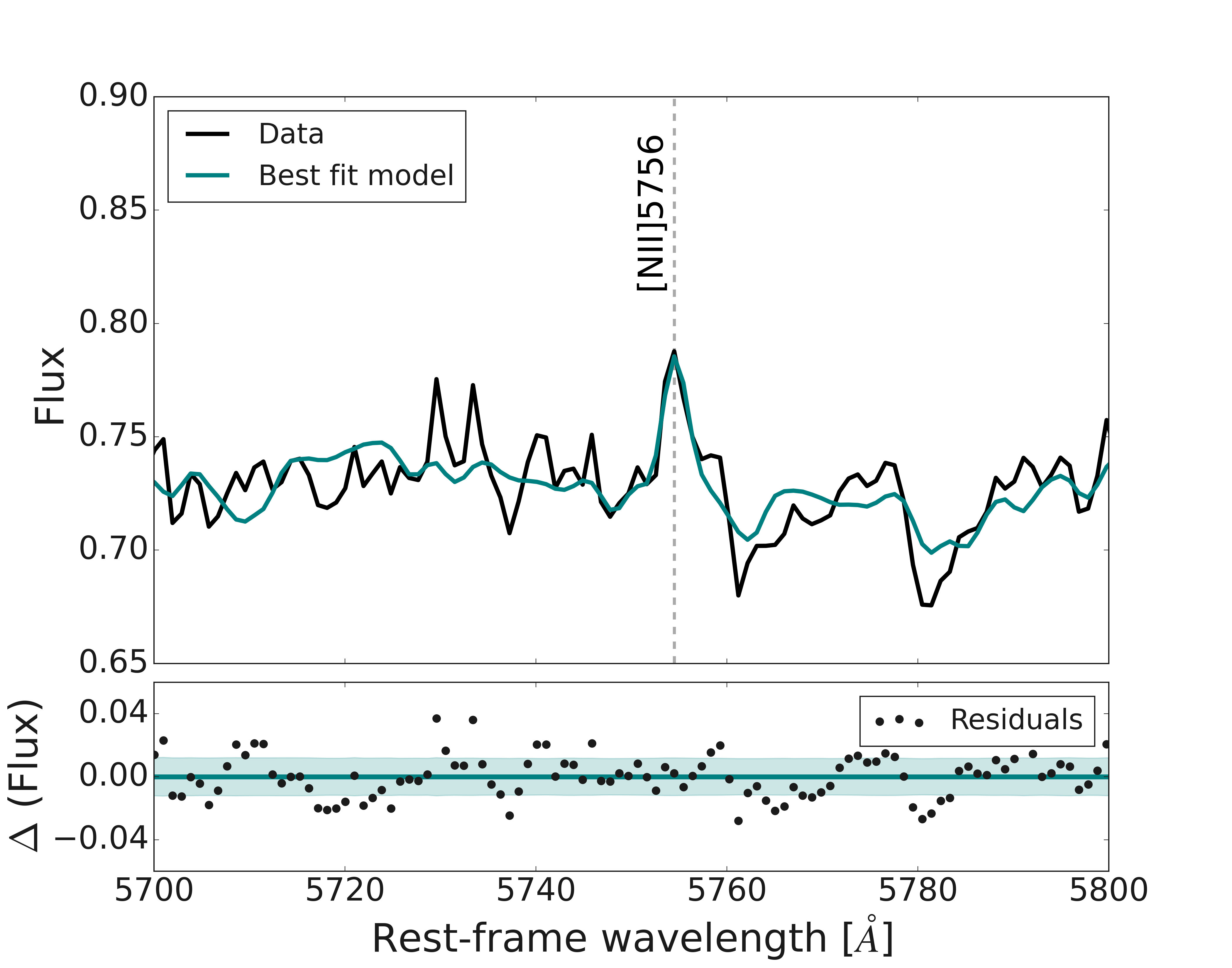}
    \label{figure-NII-galNGC5068-regID817}
  \end{subfigure}
  \begin{subfigure}{0.33\linewidth}
    \includegraphics[width=\linewidth]{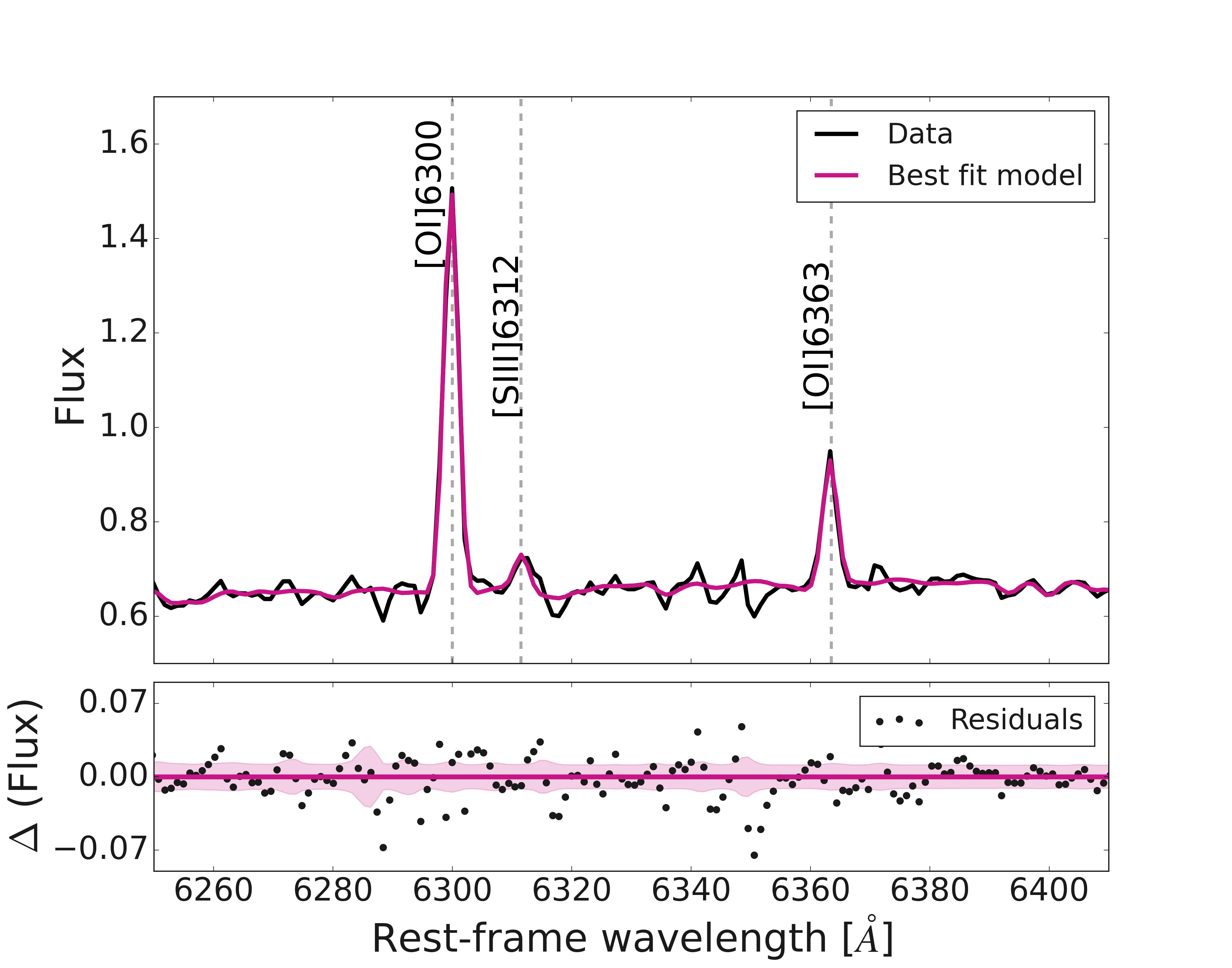}
    \label{figure-SIII-galNGC5068-regID817}
  \end{subfigure}
  \begin{subfigure}{0.33\linewidth}
    \includegraphics[width=\linewidth]{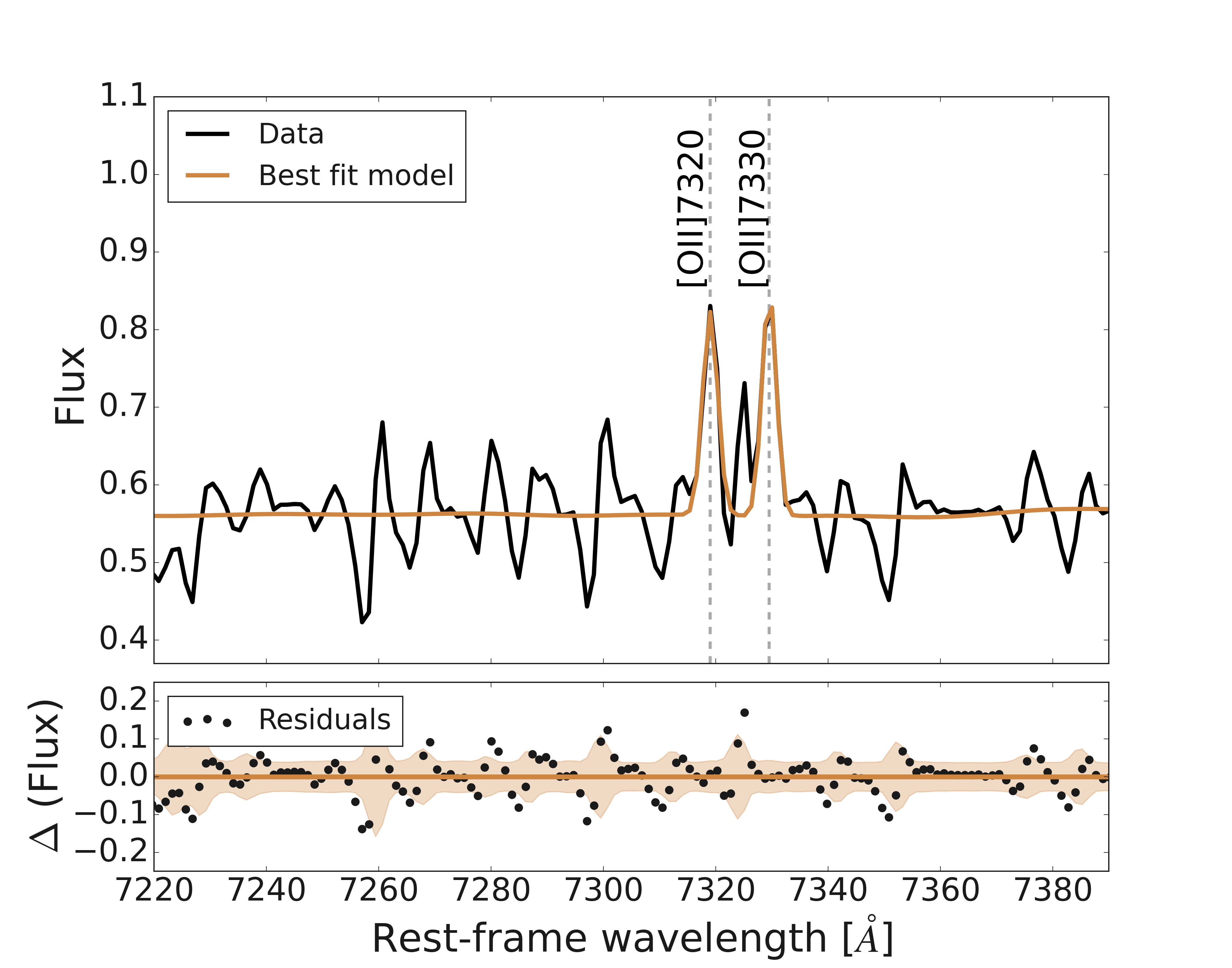}
    \label{figure-OII-galNGC5068-regID817}
  \end{subfigure}
    
   \caption{Example fits (both stellar continuum and emission lines) of spectral regions around auroral lines for two \hii\ regions in NGC5068 (IDs: 268 in the \textit{upper panels}, the same as Fig. \ref{figure-HIIreg-spectrum}, and 817 in the \textit{lower panels}). 
   The [NII]$\lambda 5756$, [SIII]$\lambda 6312$ and [OII]$\lambda \lambda 7320,7330$ auroral lines are detected with corresponding S/N of 29, 39, 22, and 20 for region 268 and 6, 7, 7, and 7 for region 817. The flux in the \textit{y} axis is expressed in the same units as Fig. \ref{figure-HIIreg-spectrum}. In the lower panels, we additionally report as a shaded region the spectral error band, already re-scaled by the residuals' standard deviation.}
  \label{figure-fitting-results-auroral-lines}
\end{figure*}

  \begin{table} 
    \caption{Rest-frame wavelength ranges identified for detailed spectral fits, along with the most important emission lines within each region. }
    \centering
    \adjustbox{max width=0.5 \textwidth}{
    \begin{tabular}{cc}
        \toprule
       Spectral Interval (\AA) & Emission Lines\\
        \midrule
        \midrule
        $4830-5050$ & \hb, \oiii$4960$, \oiii$\lambda 5007$\\
         \midrule
         $5650-5850$ & \nii$\lambda 5756$\\
         \midrule
         $6230-6460$ &  \oi$6300$, \siii$\lambda 6312$, \oi$6363$ \\
         \midrule
         $6480-6800$ & \ha, \nii$\lambda\lambda 6548, 84$, \\
          &   \sii$\lambda\lambda 6717, 31$, \\
          \midrule
         $7150-7400$ &  \oii$\lambda 7320$, \oii$\lambda 7330$\\
         \midrule
          $8900- 9098$  & \siii$\lambda 9069$\\
        \bottomrule
    \end{tabular}}
    \vspace{6pt}
  \par \smallskip \raggedright 

    \label{table-spectral-regions}
\end{table}
    We fitted the integrated spectra of all the nebular regions in the \cite{groves2023} catalogue using the penalised PiXel-Fitting (\textsc{pPXF}) python package \citep{CappellariEmsellem2004, cappellari2017}, which fits the stellar continuum with a combination of simple stellar population (SSP) templates (in our case E-MILES SSP models, \citealp{vazdekis2016}) and gas emission lines with Gaussian profiles. In this fit, we did not consider potential continuum nebular emission, but we added eighth-order multiplicative polynomials to account for any remaining inaccuracies in the continuum subtraction. We used the pPXF wrapper provided by the PHANGS data analysis pipeline.\footnote{\url{https://gitlab.com/francbelf/ifu-pipeline}} 
    
   In this work, we have fitted individual spectral regions containing specific lines of interest (Table \ref{table-spectral-regions}) rather than the full spectrum, as was done in \cite{groves2023}. In our experiments (see Appendix \ref{appendixA}), we found that using restricted spectral regions always resulted in a more accurate fit of the stellar continuum in regions of interest rather than performing the continuum fit over the full spectrum, a result also confirmed in \cite{vaught+2023}. Another advantage of this approach is that in our restricted wavelength intervals the potential contribution of nebular continuum does not influence the continuum fitting as much as if fitting the total spectrum.
   The spectrum of a representative \hii\ region is shown in Fig. \ref{figure-HIIreg-spectrum}, while examples of auroral line fits are presented in Fig. \ref{figure-fitting-results-auroral-lines}. 

With respect to the fitting procedure adopted in \cite{groves2023}, the choice of fitting restricted wavelength intervals rather than the entire spectrum prevented us from tying the line kinematics over the total spectrum. We therefore imposed kinematic constraints within the individual spectral regions by fixing the velocity and velocity dispersion of weak lines to strong lines, if available; for example, the \siii$\lambda 6312$ auroral line to the \oi$\lambda 6300$ nebular line, or nitrogen and sulphur nebular lines to \ha. Details of the lines fitted in each wavelength intervals are presented in Table \ref{table-spectral-regions}.

The most compelling check for evaluating the goodness of our procedure concerns the consistency of the different kinematic parameters calculated independently for each spectral region. 
Hence, we verified that both the velocity and the velocity dispersion evaluated independently for each spectral region are consistent, at least for the brightest emission lines (\ha, \hb, \oiii$\lambda$5007).

\subsubsection{Auroral line detections} 
\label{section-sample-selection}

The task of selecting reliable auroral line detections is particularly delicate due to their relative faintness and the fact that they can be easily confused with noise spikes.
To identify reliable detections, we first required a threshold of 5 for each line in the amplitude-over-noise ratio (ANR), which we considered as a proxy of the S/N. We evaluated the amplitude as the line peak, while the noise was estimated as the standard deviation of the residuals between the data and the best fit around the line of interest.  Requiring a simultaneous 5$\sigma$ detection of \nii$\lambda$5756, \siii$\lambda$6312 and \oii $\lambda\lambda$7320, 7330 leads to a sample of $116$ regions. For this sub-sample, the $\ha$ and $\hb$ lines have S/N$>$400, reaching values of several thousand for \ha; the \oiii$\lambda 5007$ and \nii$\lambda 6584$ nebular lines are always detected with S/N$>$130 and the \siii$\lambda 9069$ nebular line, which is slightly weaker than the other two, generally shows S/N$>$40. For these bright regions, potential contributions from DIG can be safely neglected.


Secondly, we checked that the intrinsic width of the best-fit Gaussian for each auroral line does not significantly differ from the one for \ha. In particular, we required that $| \sigma - \sigma_{\text{\ha}} | \leq 0.4 \sigma_{\text{\ha}}$. This condition allowed us to exclude any residual noise spikes and reduced our sample of regions with detections in all three auroral lines to 95, that is, 0.3\% of the parent sample. 


\begin{table*}
\caption{Total number of nebulae, along with the number of \hii\ regions meeting the selection criterion individually for each auroral line and simultaneously for all lines, for each galaxy in the PHANGS-MUSE sample.}
    \centering
    \adjustbox{max width=\textheight}{ 
    \begin{tabular}{cccccc}
        \toprule
        Galaxy  & Number of  & Number of detections & Number of detections & Number of detections   & All auroral lines\\
         & regions & of \nii $\lambda$ 5756 &  of  \siii$\lambda$6312 &  of \oii$\lambda, \lambda $7320,7330  & \\
        \midrule
    IC5332 & 816 & 12 (1.47\%) & 10 (1.23\%) & 8 (0.98\%) & 5 (0.61\%) \\
    NGC0628 & 2869 & 34 (1.19\%) & 18 (0.63\%) & 19 (0.66\%) & 9 (0.31\%) \\
    NGC1087 & 1011 & 46 (4.55\%) & 9 (0.89\%) & 49 (4.85\%) & 7 (0.69\%) \\
    NGC1300 & 1478 & 20 (1.35\%) & 0 & 7 (0.47\%) & 0\\
    NGC1365 & 1455 & 48 (3.30\%) & 6 (0.41\%) & 21 (1.44\%) & 2 (0.14\%) \\
    NGC1385 & 1029 & 93 (9.03\%) & 7 (0.68\%) & 39 (3.79\%) &6 (0.58\%) \\
    NGC1433 & 1736 & 20 (1.15\%) & 1 (0.06\%) & 1 (0.06\%) & 1 (0.06\%) \\
    NGC1512 & 632 & 8 (1.27\%) & 0 & 0 (0.00\%) & 0 \\
    NGC1566 & 2404 & 122 (5.08\%) & 10 (0.42\%) & 28 (1.17\%) & 7 (0.29\%) \\
    NGC1672 & 1581 & 72 (4.56\%) & 7 (0.44\%) & 24 (1.52\%) &6 (0.38\%) \\
    NGC2835 & 1121 & 42 (3.75\%) & 14 (1.25\%) & 67 (5.98\%) & 9 (0.80\%)\\
    NGC3351 & 1284 & 5 (0.39\%) & 0& 1 (0.08\%) & 0\\
    NGC3627 & 1635 & 11 (0.67\%) & 0  & 3 (0.18\%) & 0 \\
    NGC4254 & 2960 & 139 (4.69\%) & 3 (0.10\%) & 38 (1.28\%) & 0\\
    NGC4303 & 3067 & 168 (5.48\%) & 3 (0.10\%) & 18 (0.59\%) & 3 (0.10\%) \\
    NGC4321 & 1847 & 32 (1.73\%) & 0  & 0  & 0  \\
    NGC4535 & 1938 & 20 (1.03\%) & 3 (0.15\%) & 2 (0.10\%) & 1 (0.05\%) \\
    NGC5068 & 1857 & 50 (2.69\%) & 79 (4.26\%) & 83 (4.47\%) & 36 (1.94\%)\\
    NGC7496 & 777 & 27 (3.48\%) & 3 (0.39\%) & 19 (2.44\%) & 3 (0.39\%) \\

        \midrule
         & 31497 & 969 (3.08\%) & 173 (0.55\%) & 427 (1.36\%) & 95 (0.30\%)\\
        \bottomrule
    \end{tabular}}
    
    \label{table-good-regions}
\end{table*}

Finally, since the catalogue from \cite{groves2023} includes gaseous regions of various nature, we used the Baldwin-Phillips-Terlevich (BPT) diagrams \citep{baldwin1981} to select \hii\ regions. We verified that the 95 spectra that match our previous selection criteria are classified as \hii\ regions in the \nii/\ha\ versus \oiii/\hb\ BPT diagram according to the demarcation line of \cite{kauffmann2003}, and are therefore all bona fide \hii\ regions.

In Table \ref{table-good-regions}, we summarise the statistics of auroral line detections in \hii\ regions within our sample of galaxies.
Reliable auroral lines detections (with S/N>5 and with reliable kinematics parameters) are obtained only for a few percent ($\sim0.1 $\% to $\sim10$\%) of the nebulae in each galaxy. The \nii$\lambda 5756$ line is the most commonly detected line, with a total of 969 detections over 31497 regions ($\sim3\%$ of the sample). Conversely, \siii$\lambda 6312$ is the most difficult line to detect, and our analysis provides only for 173 good line detections ($0.55\%$ of the total sample). 
Despite the fact that the mean spectral noise does not vary significantly along the available spectral coverage, such differences in auroral line detections arise because of the different continuum contamination due to telluric lines around auroral lines, which significantly affect the estimation of residuals' standard deviation, and hence the line ANR. This effect is particularly evident for the \oii\ auroral lines (right panels in Fig. \ref{figure-fitting-results-auroral-lines}), around which the standard deviation of residuals is typically a factor of two larger than around the \nii\ auroral line (see also Fig. \ref{figure-rms-comparison-hist}). 
We observe that in Fig. \ref{figure-fitting-results-auroral-lines} the spectral error is represented as a shaded region overplotted to the residuals between the flux and our model. 
In particular, the error is corrected by the normalisation factor, the standard deviation of residuals over median of the error vector, to take into account for the understimate in the spectral noise \citep{emsellem2022}. 

\subsection{Ancillary datasets} 
\label{section-other-data}

We complemented the PHANGS--MUSE auroral lines dataset with high-quality, homogeneous catalogues of auroral line detections from the literature in order to compare with measurements of $T_e$ and chemical abundances over a wider metallicity range.
Since the PHANGS--MUSE galaxies cover the metallicity range around solar \citep{groves2023}, we specifically complemented them with the catalogue from \cite{guseva2011} (\citetalias{guseva2011} from now on) targeting low-metallicity objects, including also blue compact dwarf (BCD) galaxies. 
We also included the catalogues described in \cite{nakajima2022} and \cite{isobe2022} (denoted as \citetalias{nakajima2022} in the following), which target extremely metal-poor galaxies (EMPGs) from the EMPRESS (Extremely Metal-Poor Representatives Explored by the Subaru Survey) project.  
For comparison purposes, we considered the data from the CHAOS (CHemical Abundances Of Spirals) project. CHAOS encompasses observations of nearby, star-forming spiral galaxies carried out with the Multi-Object Double Spectrograph (MODS, \citealp{pogge2010}) at the Large Binocular Telescope (LBT). In this work, we included the catalogues of \hii\ regions from the six galaxies presented in \cite{berg2020} and \cite{rogers2021, rogers2022}.
Finally, we also compared our data with the line ratios obtained by \cite{curti2017} (denoted as \citetalias{curti2017} in the following) by stacking Sloan Digital Sky Survey (SDSS) Data Release 7 \citep{sdss-dr7}, with
galaxies stacked according to their values of reddening-corrected \oii$\lambda \lambda 3726,3729$/\hb~ and \oiii$\lambda 5007$/\hb~ line ratios. 
In general, the line fluxes in \citetalias{curti2017} will include contributions from multiple \hii\ regions, DIG, and other ionisation sources. 

To obtain a homogeneous set of physical parameters, especially temperatures and ionic abundances, we re-analysed all the selected literature catalogues (although without re-fitting their spectra), trying to adhere to the same selection and data analysis procedures applied to the PHANGS--MUSE sample to the greatest extent possible. 
This implies using only the same emission lines available for the PHANGS--MUSE data, when possible. For this reason, we tried to avoid using the \oiii$\lambda 4363$ auroral line present in all the other catalogues. 
The details of this analysis are presented in Appendix \ref{appendixB}. 
We also applied to all the catalogues the S/N threshold of $5$ for line detections.

Additionally, we selected some of the most recent high-redshift publications reporting auroral line detection with NIRSpec (Near Infrared Spectrograph) on JWST, in particular from \cite{Sanders2023} and \cite{laseter2023}. Both summarise and further expand on the auroral line detections reported in the JWST Early Release Observations (ERO) programme \citep{pontoppidan2022, arellano-cordova2022, schaerer2022, curti2023a}.
In particular, \cite{Sanders2023} report detections of the \oiii$\lambda 4363$ auroral line for a sample of $16$ galaxies at z = $1.4 - 8.7$ from the Cosmic Evolution Early Release Science (CEERS) survey programme \citep{finkelstein2023-1, finkelstein2023-2}, while \cite{laseter2023} presents \oiii$\lambda 4363$ detections for $10$ galaxies at $z = 1.7 - 9.4$ from the JWST Advanced Deep Extragalactic Survey (JADES) programme \citep{eisenstein2023}. 
The procedure followed for determining oxygen abundances from these high-redshift data differs from that of the low-redshift catalogues because of the limited number of available emission lines, and is detailed in Appendix \ref{appendixB}. 

\section{Methods} \label{section:methods}

\begin{table*}
\caption{Temperature-sensitive lines available for each catalogue.  }
    \centering
    \adjustbox{max width=\textwidth}{
    \begin{tabular}{ccccc}
        \toprule
        Temperature  & $T_e$\oiii&  $T_e$\nii& $T_e$\siii & $T_e$\oii \\
        Auroral line  & \oiii$\lambda 4363$ &  \nii$\lambda 5756$ & \siii$\lambda 6312$ & \oii$\lambda \lambda 7320,7330$ \\
        Nebular line & (\oiii$\lambda 5007$) & (\nii$\lambda 6584$) & (\siii$\lambda 9069$) & (\oii$\lambda \lambda 3726,3729$)   \\
        \midrule
        PHANGS-MUSE & X  & \checkmark & \checkmark & X \\
        CHAOS & \checkmark & \checkmark & \checkmark  & \checkmark\\
        \citetalias{curti2017} & \checkmark & \checkmark & X & \checkmark \\
        \citetalias{guseva2011} & \checkmark & \checkmark & \checkmark & \checkmark\\
        \citetalias{nakajima2022} & \checkmark & X & X & X\\
        
        \bottomrule
    \end{tabular}}
    \label{table-literature-catalogues-lines}
    \vspace{6pt} 
  \par \smallskip \raggedright \textit{Notes.}  We use 'X' when either  the auroral or nebular lines are absent, since both of them are necessary for measuring $T_e$. 
\end{table*}

\subsection{Electron temperatures}
\label{section-electron-temperatures}

 \begin{figure}
   \centering
   \includegraphics[width=1.1\linewidth]{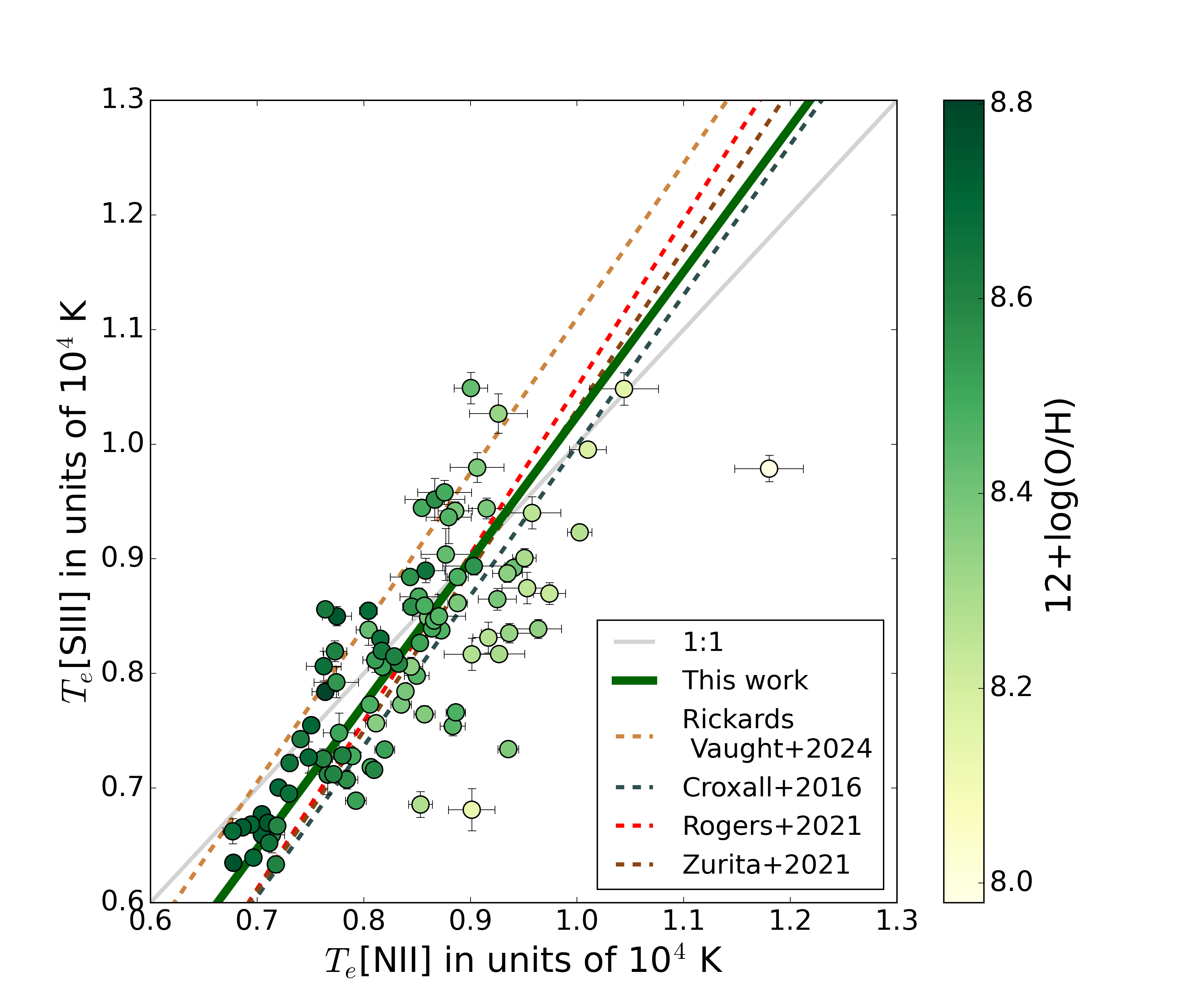}
      \caption{Resulting MCMC best fit (in dark green) of the $T_e$\nii\-$T_e$\siii\ relation calibrated on PHANGS-MUSE data, which are reported and colour-coded according to their metallicity value (see Sect. \ref{section-abundance-determination}). 
      Other calibrations from previous literature works are reported for comparison; the five curves show comparable behaviours. 
              }
         \label{figure-TSIII-TNII}
   \end{figure}

The optimal approach to determine $T_e$ consists of simultaneously constraining both electron temperature and electron density using specific temperature- and density-dependent line ratios (but see \citealp{kreckel2022} for the introduction of an alternative approach). The temperature-sensitive line ratios available in the datasets considered in this work are summarised in Table \ref{table-literature-catalogues-lines}. For density measurements, we consistently used the \sii$\lambda 6731$/\sii$\lambda 6717$ ratio.

For the PHANGS--MUSE data, and other catalogues that have not already been corrected for dust extinction, we have adopted the extinction law from \cite{odonnell1994} with $R_V = 3.1$ and an intrinsic Balmer ratio of \ha/\hb $=2.86$ to perform a reddening correction. 
For PHANGS--MUSE, we obtain $A_V $ values spanning the range from $0$ to $1.3$ mag, with a median of $\sim 0.7$ mag. 

With regard to density tracers, the \sii\ line ratio is sensitive to density in the range of $n_e \sim 10^2-10^4$ cm$^{-3}$. However, the majority of our selected \hii\ regions are characterised by $n_e<100$ cm$^{-3}$, that is,\ near or well below the low-density limit. For line ratios below the low-density limit, which was set at $n_e = 40$ cm$^{-3}$ following \cite{kewley2019-1}, we imposed a fixed $n_e$ value of 20 cm$^{-3}$. In this regime, varying the electron density does not significantly affect the final ionic abundance results.

In PHANGS--MUSE, we inferred $T_e$ from both the \nii$\lambda 5756$/\nii$\lambda 6584$ ($T_e$\nii) and the \siii$\lambda 6312$/\siii$\lambda 9069$ ($T_e$\siii) auroral-to-nebular line ratios. We calculated $n_e$, $T_e$\siii\ and $T_e$\nii\ for the 95 selected PHANGS--MUSE \hii\ regions using the \textsc{Pyneb} python package (\citealp{luridiana2012}, \citealp{luridiana2015}), version 1.1.15. Specifically, we used the \textsc{getTemDen} function to infer the temperatures when $n_e$ was fixed at 20 cm$^{-3}$ (54\% of selected \hii\ regions), and \textsc{getCrossTemDen} to get both $n_e$ and $T_e$ for regions outside the low-density limit. In our analysis, we do not find density values exceeding 200 cm$^{-3}$.

We assumed $T_e$\nii\ to be representative of the whole low-ionisation zone, and used it to estimate the N$^+$, S$^+$, and O$^+$ abundances, since MUSE does not allow a direct measurement of these electron temperatures. This assumption is supported by predictions from photoionisation models, although various studies \citep{rogers2021, mendez-delgado+2023, vaught+2023} show some discrepancies between $T_e$ estimates derived from the \nii, \oii\ and \sii\ temperature-dependent line ratios. Finally, we used $T_e$\siii\ to determine the S$^{2+}$ abundance within the intermediate-ionisation zone. 

We exploited the two temperature measurements in the PHANGS--MUSE data to re-calibrate the $T_e$\siii\--$T_e$\nii\ relation. Using a Monte Carlo Markov Chain (MCMC) approach, we obtained the following best-fit relation: 
\begin{equation}
    T_e \siii = 1.22 (\pm 0.01)~ T_e  \nii - 0.20(\pm 0.01) \label{eq-te-te-rel},
\end{equation}
with temperatures expressed in units of $10^4$K.
This relation (Fig. \ref{figure-TSIII-TNII}) is consistent within 1$\sigma$ with previous calibrations based on analyses of \hii\ regions in nearby galaxies, encompassing  also the CHAOS sample (\citealp{croxall2016, rogers2021}) and a restricted sub-sample of seven galaxies within the PHANGS-MUSE dataset \citep{vaught+2023}. 
The consistency with the latter calibration is significant as it represents a completely independent analysis of auroral lines in PHANGS-MUSE HII regions. 
Nevertheless, the correspondence between the two calibrations is not perfect, as there is a clear offset.
The major difference between our work and the one of \cite{vaught+2023} lies in the identification of \hii\ region borders and subsequent extraction of line fluxes, with their apertures being typically larger than ours because of the larger PSF of the Keck Cosmic Web Imager (KCWI) instrument relative to MUSE (see their Sect. 2 for more details). 
The different morphology of \hii\ regions will have a greater impact on low-ionisation ions found towards the outer regions of ionised nebulae, whereas ions in the intermediate- and high-ionisation zones are expected to be less influenced by differences in \hii\ region boundaries. 
We tested this hypothesis by cross-matching in co-ordinates the \hii\ regions from \cite{vaught+2023} and the ones defined as in \cite{groves2023} and used in this work. We obtained 32 matches, for which we compared the $T_e$\siii\ and $T_e$\nii\ estimates. 
We found that the two $T_e$\siii\ estimates are in good agreement, while the $T_e$\nii\ values derived in \cite{vaught+2023} show a median bias of -570 K with respect to ours.
This result explains the offset between the two $T_e$\siii--$T_e$\nii\ calibrations observed in Fig. \ref{figure-TSIII-TNII}, and allows us to conclude that the main difference in the two works is due to the different setting of \hii\ region borders. 

We estimated the intrinsic dispersion $\sigma _{int}$ about the best-fit $T_e$\siii\--$T_e$\nii\ relation following the same approach outlined in \cite{rogers2021}. 
We fixed the slope and intercept of the relation to the best-fit parameters of the linear fit; then we sampled the parameter space using an MCMC to determine the $\sigma _{int}$ value that maximises the likelihood function. 
We find $\sigma _{int} = 724 \pm 55$ K, which is higher than the corresponding values found in \cite{rogers2021} ($\sigma _{int} = 173$ K), but lower than the one found in \cite{vaught+2023} ($\sigma _{int} = 997$ K). 
In general, $\sigma _{int}$ is significantly higher than the typical errors associated with our PHANGS--MUSE $T_e$\siii\ error measurements, which have a median value of $\sim 70 $ K.

We carried out a similar procedure (Appendix \ref{appendixB}) to evaluate $n_e$ and $T_e$ for the literature catalogues. We used these to re-calibrate the $T_e$\oiii\--$T_e$\siii\ relation using temperature measurements from the CHAOS and \citetalias{guseva2011} catalogues, the only datasets in our compilation where $T_e$\oiii\ and $T_e$\siii\ are directly measured (Fig. \ref{figure-TOIII-TSIII}). We find a $T_e$\oiii\--$T_e$\siii\ relation consistent with previous works. 

We used the $T_e$\oiii\--$T_e$\siii\ relation for estimating the temperature associated with the high-ionisation zone (traced by the O$^{2+}$ ion) in PHANGS--MUSE data, where we do not have access to \oiii$\lambda 4363$. In principle, we could have alternatively re-calibrated the $T_e$\oiii--$T_e$\nii\ relation to infer $T_e$\oiii\ , but \cite{rogers2021} reported that the  $T_e$\oiii--$T_e$\siii\ relation shows a tighter correlation than the $T_e$\oiii--$T_e$\nii\ relation.

\subsection{Ionic abundances determination} \label{section-abundance-determination}

\begin{table}
\caption{Temperature estimates and collisionally excited line fluxes used for measuring ionic abundances.}
    \centering
    \adjustbox{max width=0.8\textwidth}{
    \begin{tabular}{ccc}
        \toprule
        Ion  & Adopted $T_e$ &  Emission lines for\\
         &  &  ionic abundances\\
        \midrule
        N$^+$ & $T_e$\nii & \nii$\lambda 6584$\\
        S$^+$ & $T_e$\nii & \sii$\lambda \lambda 6717,31$\\
        O$^+$ & $T_e$\nii & \oii$\lambda \lambda 7320,30$\\
        S$^{2+}$ & $T_e$\siii & \siii$\lambda 9069$\\
        O$^{2+}$ & $T_e$\siii + Eq. \ref{eq-teOIII-teSIII-rel} & \oiii$\lambda \lambda 4959,5007$\\
        \bottomrule
    \end{tabular}}
    \tablefoot{$T_e$\nii\ and $T_e$\siii\ were directly measured from temperature-sensitive line ratios, while $T_e$\oiii\ was obtained using Eq. \ref{eq-teOIII-teSIII-rel}.}
    \label{table-summary-for-ionic-abund}
    \vspace{6pt} 
\end{table}

   \begin{figure}
   \centering
   \includegraphics[width=\linewidth]{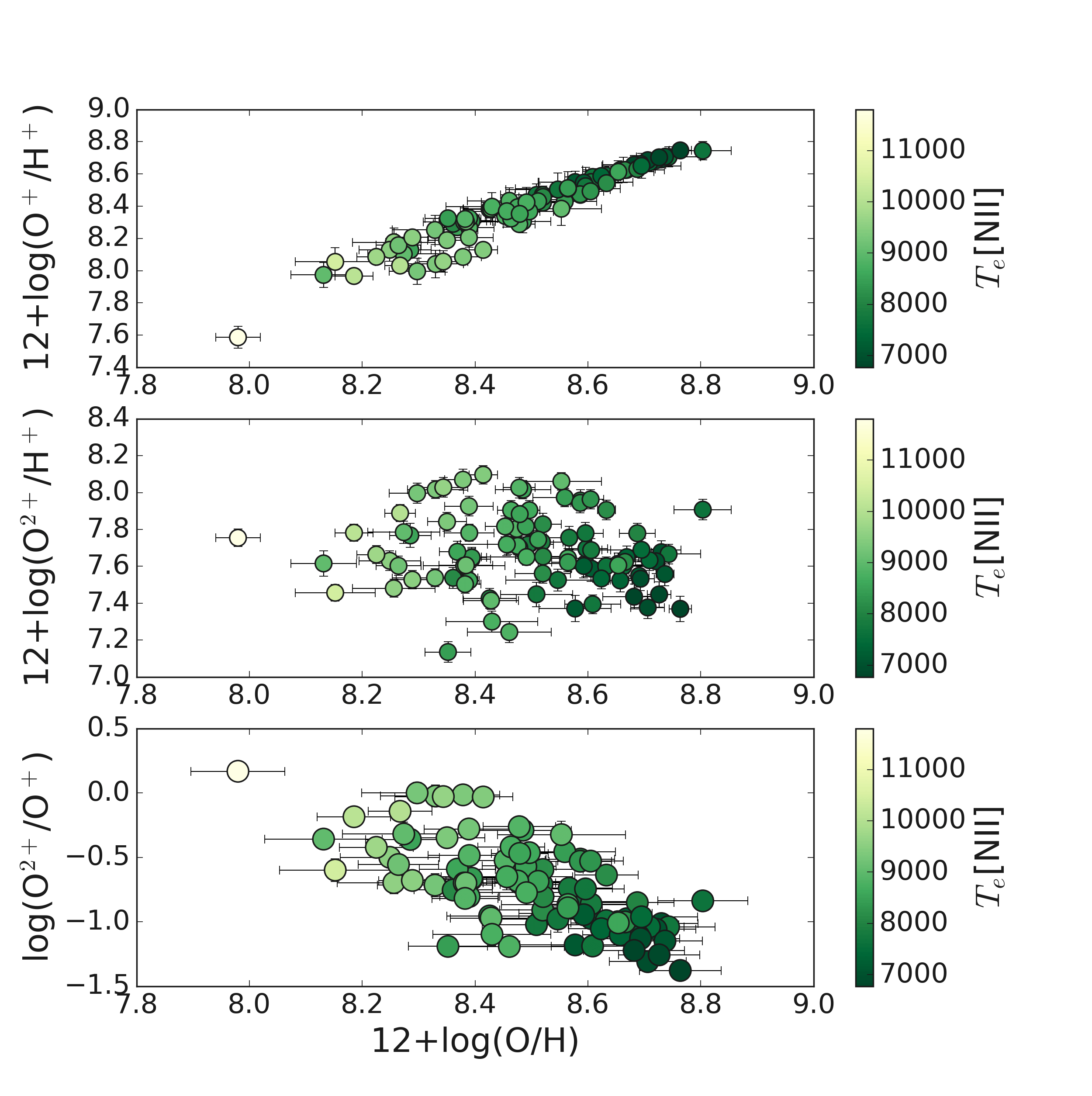}
      \caption{O$^+$ abundance (upper panel), O$^{2+}$ abundance (middle panel), and relative ionic abundance of the two species (bottom panel) as a function of the total oxygen abundance for the $95$ selected \hii\ regions from the PHANGS-MUSE nebular catalogue. 
              }
         \label{figure-plots-as-in-curti}
   \end{figure}

\begin{figure*}
   \centering
   \includegraphics[width=\linewidth]{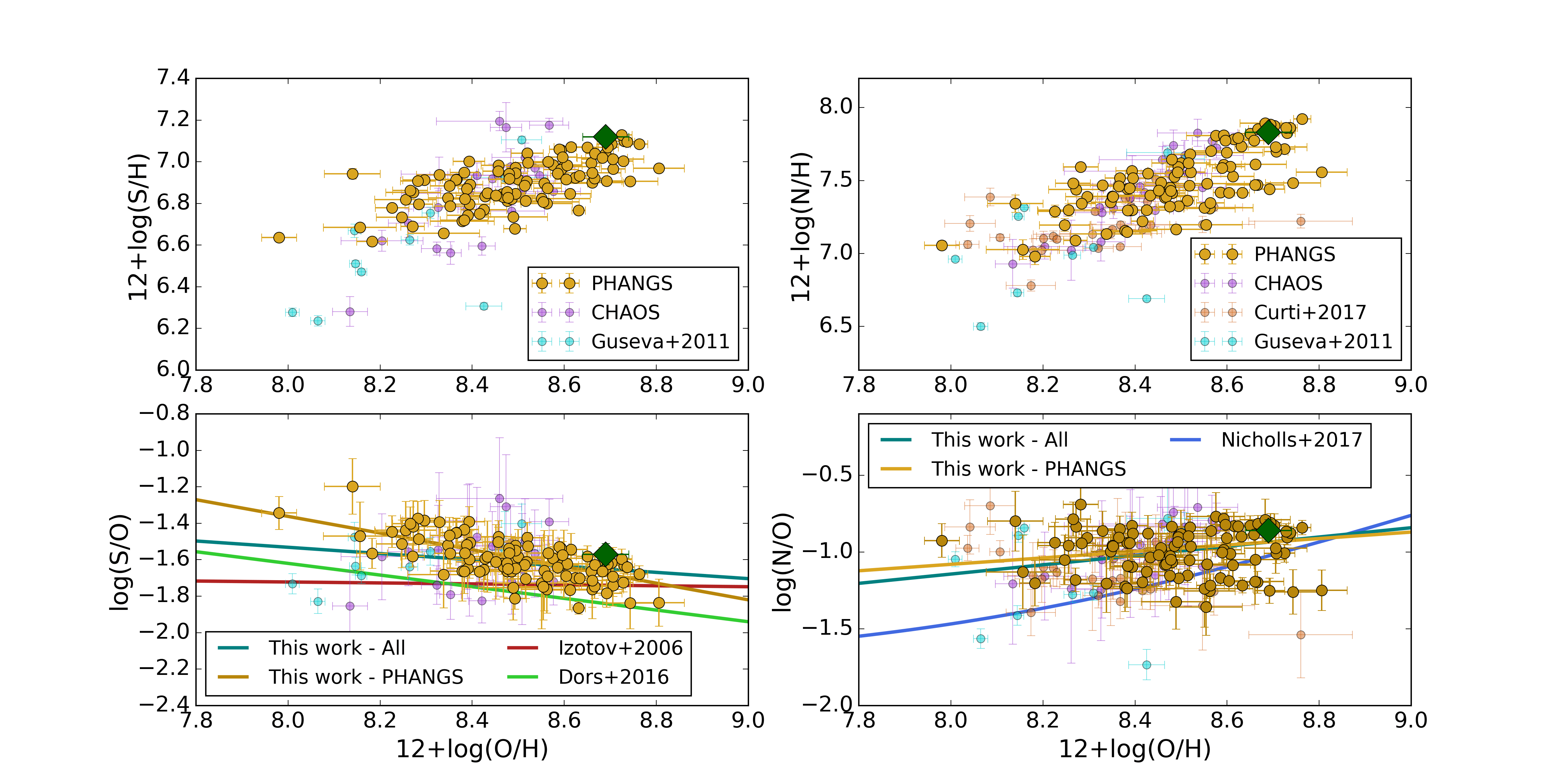}
      \caption{\textit{Upper panels}: Sulphur and nitrogen chemical abundances versus total system metallicity for the PHANGS--MUSE sample and the other analysed literature catalogues.
      \textit{Lower panels}: Relative abundances of sulphur and nitrogen with respect to oxygen, as a function of the total system metallicity. Solar values from \cite{asplund2009} are reported as teal points for comparison. We display linear trends calibrated either using only the PHANGS-MUSE data or the entire dataset, which also includes literature data. The other log(S/O) versus 12+log(O/H) trends are from \cite{izotov2006} and \cite{dors2016}, with the former calibrated over BCD galaxies and the latter over a compilation of literature emission-line intensities of \hii\ regions and star-forming galaxies and while considering sulphur ICF from \cite{thuan1995}. The log(N/O) versus 12+log(O/H) trend is taken from \cite{nicholls+2017} and it is calibrated over a compilation of literature stellar data.}
         \label{figure-sulfur-nitrogen-abundances}
   \end{figure*}
   
Once the electron temperatures and densities were measured, we could estimate ionic abundances through the ratio of a collisionally excited emission line to a Balmer recombination line, usually \hb\  \citep{garnett1992, perez-montero2017}. All the ionic abundances discussed in this section were evaluated with \textsc{Pyneb}, using the \textsc{getIonAbundance} function. 
In particular, we adopted the \textsc{Pyneb} default atomic data and the dust-corrected \hb\ as the reference hydrogen recombination line. 

Errors on ionic abundances (and on electron temperatures and densities) were computed with Monte Carlo simulations: for each of the $95$ selected \hii\ regions from the PHANGS--MUSE sample, we generated $500$ realisations of emission line fluxes of interest, according to a Gaussian distribution centred on the measured flux and with $\sigma$ equal to the measured error. For each realisation, we carried out dust correction, evaluated line ratios, and measured $T_e$, $n_e$, and ionic abundances. We took the median and the standard deviation of the resulting parameter distributions as the best value and corresponding error to associate with each parameter, respectively. 

When possible, ionic abundance measurements were carried out exploiting nebular lines rather than auroral lines, because they are brighter and detected at higher S/N. 
As is summarised in Table \ref{table-summary-for-ionic-abund}, this is the case for N$^+$, S$^+$, S$^{2+}$, and O$^{2+}$. However, the ionic abundance of O$^+$ was measured with the \oii$\lambda \lambda 7320,7330$ auroral lines, as the \oii$\lambda \lambda 3726,3729$ nebular lines are not covered by MUSE. The \oii\ auroral lines could be contaminated due to recombination \citep{rubin1986, perez-montero2017}, so to demonstrate the validity of this approach we compared the O$^+$ ionic abundance evaluated from auroral and nebular lines for those catalogues for which the two sets of lines are both available (CHAOS, \citetalias{curti2017} and \citetalias{guseva2011} catalogues). This comparison (see Fig. \ref{figure-O2_abund_comparison_auroral_nebular}) shows a consistency between the two measurements, validating our approach.
   
We computed the oxygen abundance as the sum of the two ionic species, O$^+$ and O$^{2+}$: 
\begin{equation}
    \frac{\text{O}}{\text{H}} = \frac{\text{O}^+}{\text{H}^+} + \frac{\text{O}^{2+}}{\text{H}^+},
\end{equation}
neglecting potential contributions from O$^{3+}$, which is typically minimal in \hii\ regions \citep{Draine}.

Figure \ref{figure-plots-as-in-curti} reports the O$^+$ (upper panel) and O$^{2+}$ (middle panel) abundances and the O$^{2+}$/O$^+$ fraction (lower panel) as a function of the total oxygen abundance, with PHANGS--MUSE data points colour-coded according to their $T_e$\nii\ value. From the first panel of Fig. \ref{figure-plots-as-in-curti}, we observe that \hii\ regions with lower temperatures are characterised by higher metallicities and higher O$^+$/H$^+$ values, with O$^{2+}$ representing only a small contribution of $20-30\%$ when $12+\log (\text{O/H}) \gtrsim 8.4$. Unlike the top panel, no evident trends are observable for O$^{2+}$ abundance (middle panel of Fig. \ref{figure-plots-as-in-curti}), although the O$^{2+}$/O$^+$ fraction appears to decrease when the total oxygen abundance increases (bottom panel of Fig.  \ref{figure-plots-as-in-curti}). Similar trends were also presented by \citetalias{curti2017}. 

While the two most abundant ionic species of oxygen exhibit bright emission lines in the optical, most other elements lack bright lines from all of predominant ionic states within the same wavelength range. Therefore, determining their total abundances requires an ionisation correction. In particular, under typical \hii\ region physical conditions nitrogen is found mostly in the form of N$^+$ and N$^{2+}$, and sulphur in the form of S$^+$, S$^{2+}$ and possibly S$^{3+}$, but the N$^{2+}$ and S$^{3+}$ ions do not present emission lines in the spectral range covered by MUSE. To calculate the total abundance of these elements, we introduced an ionisation correction factor (ICF): 
\begin{equation}
    X |_{tot} = \text{ICF} \big( X|_{ion} \big) \cdot X|_{ion}, 
\end{equation}
where for simplicity we have introduced the notation $ X = N(X)/N(H)$. 

Various sulphur and nitrogen ICF parametrisations exist in the literature, generally expressed as a function of O$^+$/O. In \cite{stasinska1978}, a first example of sulphur ICF was provided, which was subsequently updated in \cite{perez-montero2006}, \cite{kennicutt2003}, and \cite{dors2016}. \cite{peimbert1969} reported the first expression for the nitrogen ICF, subsequently adopted in \cite{thuan1995} and in the works from the CHAOS team. \cite{izotov2006} proposed new expressions for both the sulphur and the nitrogen ICFs as a function of O$^+$/O and further introduced a dependence on the system metallicity or, analogously, on its electron temperature. 


For the sulphur and nitrogen abundances, we adopted the following parametrisations from \cite{izotov2006}: 
\begin{align}
    \text{IC}(\text{S}^+ + \text{S}^{2+}) &= -0.825 v + 0.718 + 0.853/v, \quad \quad \text{Low Z}, \label{eq:icfS-izotov-low}\\
    &= -0.809 v + 0.712 + 0.852/v, \quad \quad \text{Inter. Z}, \label{eq:icfS-izotov-inter}\\
    &= -1.476 v + 1.752 + 0.688 / v,  \quad \quad \text{High Z}, \label{eq:icfS-izotov-high} \\
    \nonumber \\
    \text{ICF}(\text{N}^+) &= -0.825 v + 0.718 + 0.853/v, \quad \quad \text{Low Z}, \label{eq:icfN-izotov-low}\\
    &= -0.809v + 0.712 + 0.852/v ,\quad \quad \text{Inter. Z}, \label{eq:icfN-izotov-inter}\\
    &= -1.476 v +1.752 + 0.688/v, \quad \quad \text{High Z}, \label{eq:icfN-izotov-high}
\end{align}
where the low-metallicity regime is defined as $12 + \log (\text{O}/\text{H} ) \leq 7.6$ and the high-metallicity regime as $12 + \log (\text{O}/\text{H} ) \geq 8.2$, and  $v =$ O$^+$/O parameter. 
We found that the majority of \hii\ regions from the PHANGS--MUSE sample do not require corrections for sulphur ICF. In most cases, the ionisation correction is negligible compared to typical errors on ionic abundances, and applying it would only increase their uncertainty.
The sulphur ICF is significant only in 8 of 95 \hii\ regions. Even in these cases, the correction remains minimal, reaching a maximum value of 1.06 that corresponds to a relative variation in the sulphur abundance of only $\sim $6\%.  
In contrast, nitrogen ICF has a more significant effect on all the \hii\ regions of the sample, especially the ones at lower metallicity. 
ICF(N$^+$) is characterised by minimum, median and maximum values of 1.06, 1.33, and 2.45, respectively, which correspond to relative variations in nitrogen abundance of 5\%, 25\% and 60\%. The maximum ICF(N$^+$) value of 2.45 is found for the only PHANGS-MUSE \hii\ region with metallicity $<8.1$.

Trends between oxygen and sulphur and nitrogen abundances are reported in Fig. \ref{figure-sulfur-nitrogen-abundances} for both the PHANGS-MUSE sample and the other literature samples described in Sect. \ref{section-other-data}. The upper panels show that both sulphur and nitrogen abundances increase with oxygen abundance. The lower panels report instead the logarithmic S/O and N/O ratios, again as a function of the system metallicity, so that the relative abundance of sulphur and nitrogen with respect to oxygen can be quantified. 
We see a net decreasing trend of log(S/O) with log(O/H), while log(N/O) appears to be almost constant, despite a significant scatter. More quantitatively, we carried out a linear fit of log(S/O) and log(N/O) as a function of log(O/H) and found:
\begin{gather}
    \log (\text{S}/\text{O}) = (-0.46 \pm 0.06 ) (12 + \log (\text{O}/\text{H})) + (2.3 \pm 0.5), \\
    \log (\text{N}/\text{O}) = (0.21 \pm 0.10 ) (12 + \log (\text{O}/\text{H})) + (-2.8 \pm 0.9), 
\end{gather}
when considering only PHANGS-MUSE data, and:
\begin{gather}
    \log (\text{S}/\text{O}) = (-0.17 \pm 0.05 ) (12 + \log (\text{O}/\text{H})) + (-0.2 \pm 0.5), \\
    \log (\text{N}/\text{O}) = (0.30 \pm 0.07 ) (12 + \log (\text{O}/\text{H})) + (-3.6 \pm 0.6), 
\end{gather}
when considering the available literature catalogues too. 

A decreasing trend for log(S/O) versus O/H, similar to the one presented in Fig. \ref{figure-sulfur-nitrogen-abundances}, has previously been reported in the literature \citep{perez-montero2006, dors2016, diaz+zamora2022}. In \cite{perez-montero2006}, the decreasing trend in log(S/O) appears evident in particular for \hii\ galaxies and giant extragalactic \hii\ regions (their Fig. 5). Instead, in \cite{dors2016} and \cite{diaz+zamora2022} a sample of star-forming galaxies and \hii\ regions was analysed, and they both found a log(S/O) versus log(O/H) trend similar to ours, both in shape and normalisation, although \cite{dors2016} measurements show a slight offset towards lower log(S/O). 
However, other works in the literature \citep{izotov2006, maciel2017} found a roughly constant trend of S/O with metallicity.
Since both sulphur and oxygen are $\alpha$ elements primarily produced through type II supernovae, this flat trend is generally expected.
The variation observed here and in other works could indicate a metallicity dependence in the production of $\alpha$ elements by these massive stars, as suggested in \cite{guseva2011}.
This hypothesis finds further confirmation in the fact that such a trend is particularly evident for single \hii\ regions rather than whole galaxies, suggesting a strict connection with local stellar nucleosynthesis and chemical evolution processes. 

Moving now our focus onto nitrogen, we observe that, despite the large scatter, a similar range of values is found in \cite{gusev2012} or \cite{perez-montero+contini2009}, while other works, such as \cite{castellanos2002}, \cite{pilyugin2003} and \cite{pilyugin2010}, found an increasing trend of log(N/O) with log(O/H) at high metallicities and a flat behaviour in the low-metallicity regime, with the turning point roughly located at metallicities of $ \sim 8.4$. 
The independence of log(N/O) on metallicity in the low-metallicity regime suggests that nitrogen is a primary element, that is, its production does not depend on the presence of other heavy elements. However, this metallicity regime is not covered by our PHANGS--MUSE data.
The increasing trend found in some literature works at higher metallicity may be due to a secondary, metallicity-dependent production of nitrogen, as was first proposed in \cite{vila-costas+edmunds1993} and then adopted, for example, in \cite{nicholls+2017}. The reason for the large scatter observed in our data cannot be readily identified.

\subsection{Ionisation parameter measurements}
\label{section-ionisation-parameter-measurements}

The ionisation parameter is typically inferred by comparing two emission lines from the same atomic species originating from different ionisation states, with the most sensitive diagnostics coming from two states with the largest difference in ionisation potentials. The line ratios most commonly used to trace the ionisation parameter in the optical wavelength range are \siii$\lambda \lambda 9069, 9531$/\sii$\lambda \lambda 6717, 6731$ and \oiii$\lambda 5007$/\oii$\lambda \lambda 3726, 3729$. 
In the PHANGS--MUSE spectral range we estimate the flux of the \siii$\lambda9531$ from the observed \siii$\lambda 9069$ flux, assuming a fixed ratio of $2.469$. 
With this arrangement, the sulphur line ratio can be evaluated for PHANGS-MUSE, CHAOS and \citetalias{guseva2011} catalogues, while the oxygen line ratio is available for CHAOS, \citetalias{curti2017}, \citetalias{guseva2011} and \citetalias{nakajima2022} catalogues, as well as for high-redshift data from \cite{Sanders2023} and \cite{laseter2023}. 

There are various existing literature calibrations for determining the ionisation parameter, based on comparisons between diagnostic line ratios and photoionisation models. 
We test $\log (U)$ calibrations from \cite{morisset2016}, \cite{kewleydopita2002} and \cite{kewley2019-1}, using both oxygen and sulphur line ratios according to the line availability in the various catalogues.
We find significant discrepancies in $\log (U) $ estimates obtained with different literature calibrations for the same line ratio, mostly in terms of a normalisation offset. Additionally, we also notice inconsistencies between ionisation parameter estimates obtained with calibrations from the same publication, but based on different line ratios when both sulphur and oxygen ratios were measured.

\begin{figure}
    \centering
    \includegraphics[width = 0.8\linewidth]{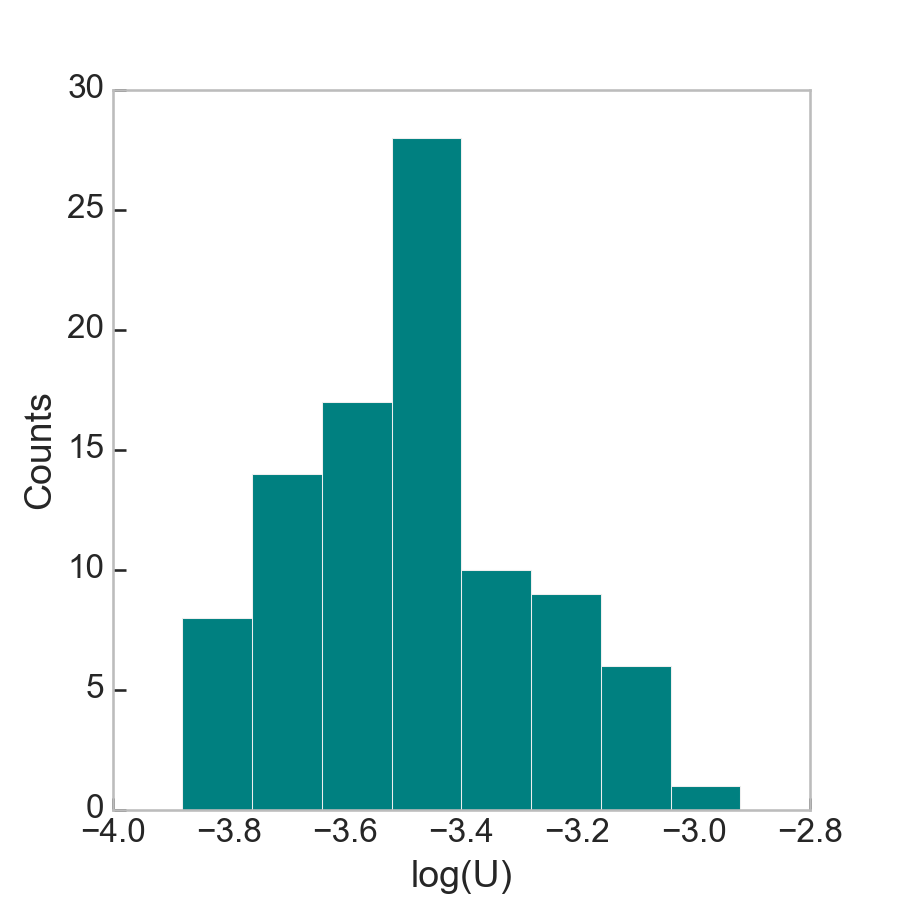}
    \caption{Measurements of ionisation parameter, $\log (U)$, for our sample of $95$ \hii\ regions.}
    \label{figure-ionisation-param-histogram}
\end{figure}

Among the calibrations considered, we adopted as fiducial the one from \cite{kewley2019-1}, because their prescription produces the smallest discrepancies in the $\log (U)$  estimates from the oxygen and sulphur lines. The measured $\log (U)$ values for the 95 regions in PHANGS--MUSE with auroral line detections are shown in Fig. \ref{figure-ionisation-param-histogram}. 
In general, the $\log (U)$ values for the selected PHANGS-MUSE \hii\ regions are slightly lower than the ones reported in literature (where typically $\log (U) $ spans from $-$3.6 to $-$2.6, see e.g. \citealp{poetrodjojo+2018} and \citealp{grasha+2012}), with a peak around $\log (U) \sim -3.5$.
This discrepancy may partially be due to the fact that PHANGS--MUSE \hii\ regions are characterised by higher metallicities with respect to the majority of analysed \hii\ regions in the literature, or it may be a consequence of using sulphur line ratio for inferring $\log (U)$, since the \siii/\sii\ line ratio tends to underestimate the ionisation parameter with respect to the oxygen line ratio \citep{kewleydopita2002}. 

\section{Results} 
\label{section-results}

\subsection{Empirical calibration of strong-line diagnostics}

\begin{figure*}[!ht]
  \centering
  \begin{subfigure}{0.48\linewidth}
    \centering
    \includegraphics[width=\linewidth]{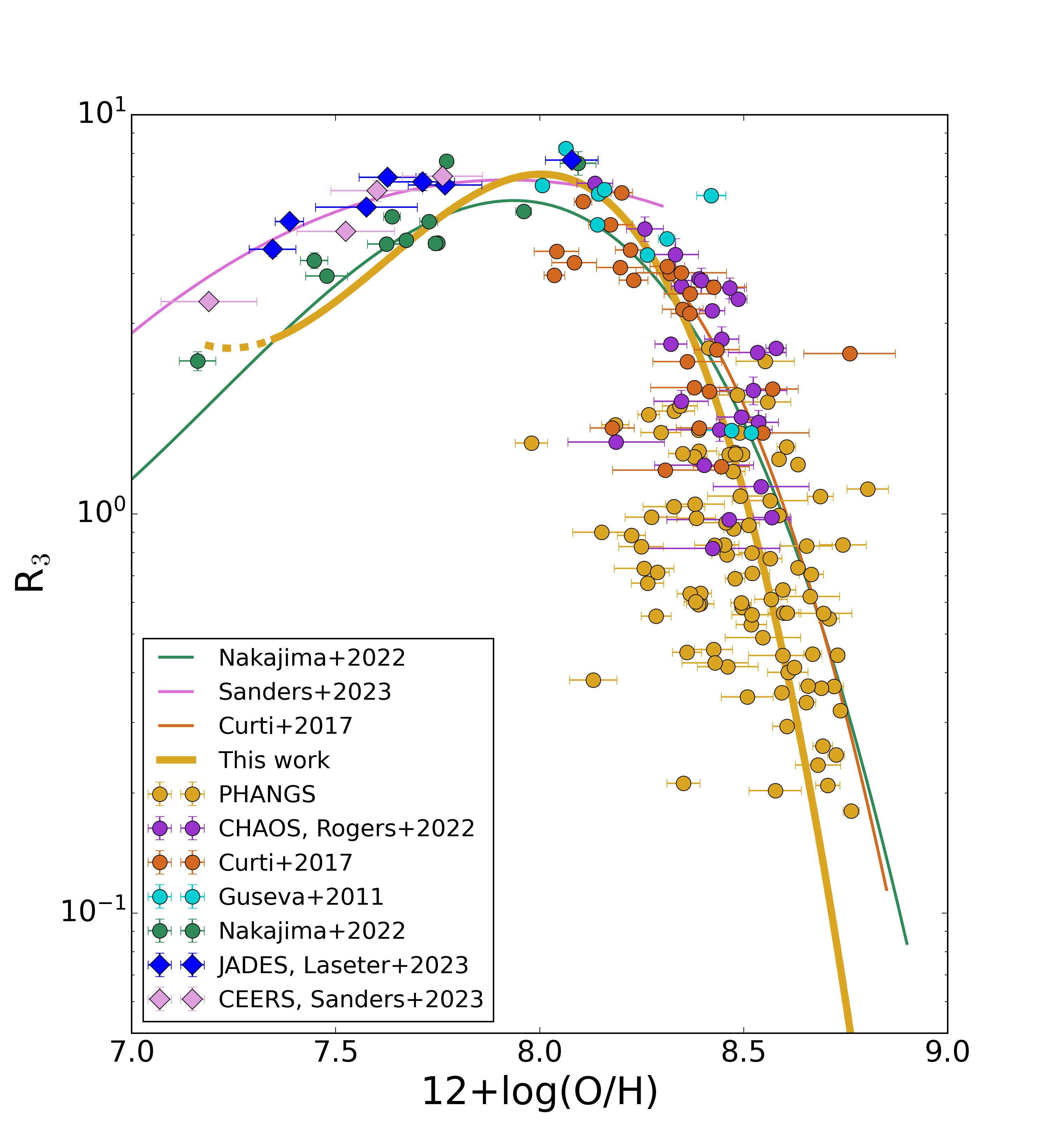}
    \caption{}
    \label{figure-diagnostics-R3}
  \end{subfigure}
  \begin{subfigure}{0.48\linewidth}
    \includegraphics[width=\linewidth]{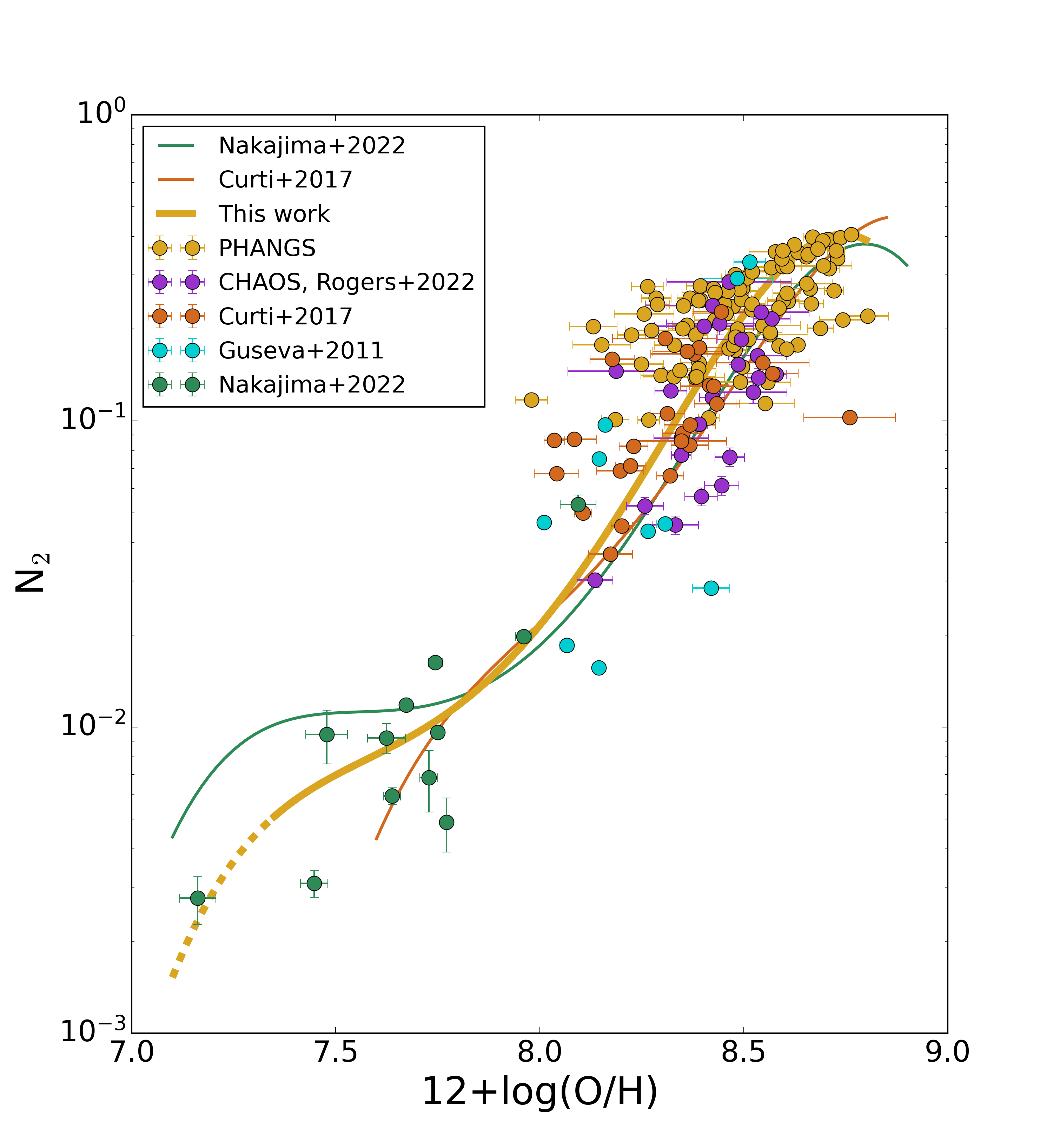}
    \caption{}
    \label{figure-diagnostics-N2}
  \end{subfigure}
  \medskip
  \begin{subfigure}{0.48\linewidth}
    \includegraphics[width=\linewidth]{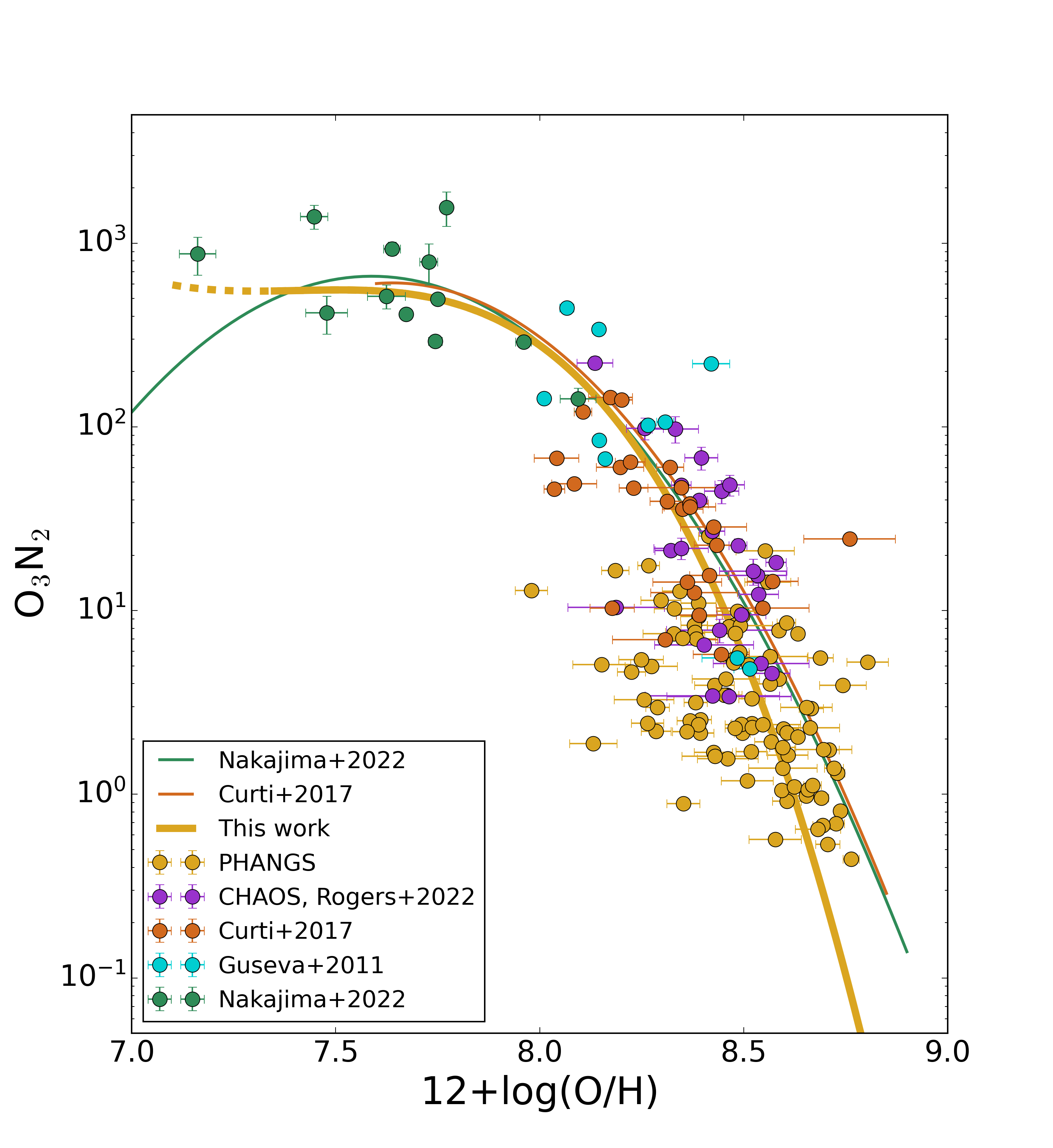}
    \caption{}
    \label{figure-diagnostics-O3N2}
  \end{subfigure}
  \begin{subfigure}{0.48\linewidth}
    \includegraphics[width=\linewidth]{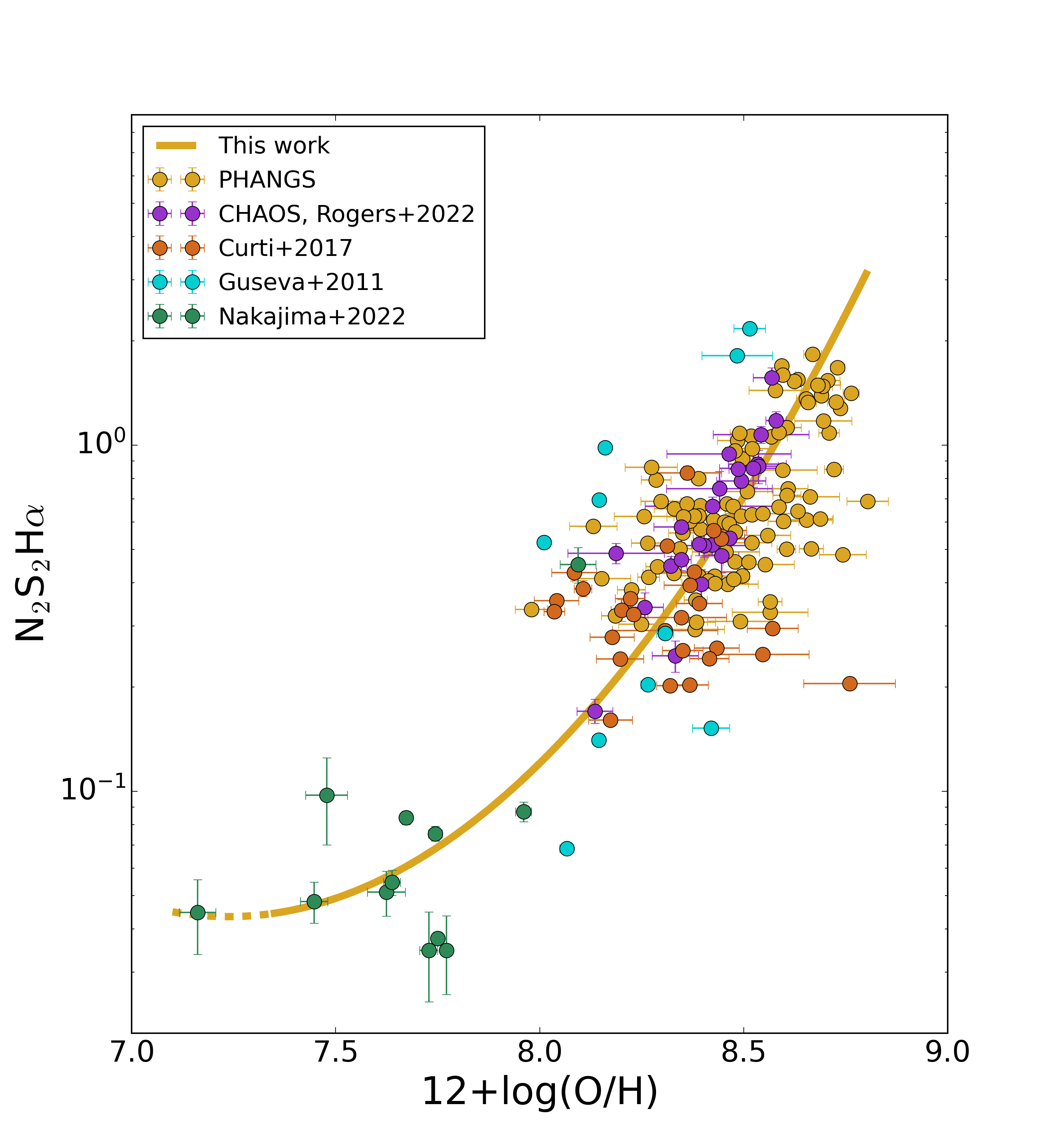}
    \caption{}
    \label{figure-diagnostics-N2S2Ha}
  \end{subfigure}
    
   \caption{Calibration of $R_3 = \oiii\lambda 5007 / \hb$ (\textit{Panel a}), $N_2 = \nii \lambda 6584 / \ha$ (\textit{Panel b}), $O_3N_2 = R_3 / N_2$ (\textit{Panel c}), and $N_2S_2H\alpha = N_2 / S_2\cdot N_2 ^ {0.264}$  (\textit{Panel d}) diagnostics carried out in this work (solid golden curves). Data points are colour-coded according to the original source catalogues, with golden points associated with PHANGS--MUSE data. The other curves report calibration from the literature  for comparison. The dashed extension at low-metallicity shows the extrapolation of our calibrations.}
  \label{figure-strong-line_diagnostics}
\end{figure*}

In this section we present the calibration of strong-line diagnostics using a total sample of $168$ direct metallicity measurements: $95$ from PHANGS--MUSE, $24$ from CHAOS, $27$ from \citetalias{curti2017}, $10$ from \citetalias{guseva2011} and $12$ from \citetalias{nakajima2022} catalogues. The 95 HII regions from PHANGS--MUSE sample provide numerous data points extending into the solar metallicity regime. The combination of these data allows for the calibration of diagnostics simultaneously valid over a wide range of metallicities ($7.4 \leq 12 + \log ( \text{O}/\text{H}) \leq 8.8$\footnote{However, the calibration in the $12 + \log ( \text{O}/\text{H}) \lesssim 8.0 $ regime should be used with caution, considering that it relies on only a few sources available at these low metallicities.}) and for different astrophysical sources, since our dataset is composed of both single \hii\ regions and whole galaxies. 

We calibrate $R_3 = \oiii\lambda 5007$/\hb, $N_2 = \nii \lambda 6584$/\ha, $O_3N_2 = R_3 / N_2$ and $N_2S_2 H \alpha = N_2 / S_2 \cdot N_2 ^ {0.264}$ (first proposed in \citealp{dopita2016} in its logarithmic form, with $N_2 = \nii \lambda 6584 / \ha$ and $S_2 = \sii\lambda \lambda 6717,6731 / \ha$) diagnostics as a function of metallicity, as is shown in Figs. \ref{figure-diagnostics-R3}, \ref{figure-diagnostics-N2}, \ref{figure-diagnostics-O3N2}, and \ref{figure-diagnostics-N2S2Ha}, respectively.  
Because of the absence of \oii$3726,3729$ nebular lines in PHANGS--MUSE spectra, we did not consider the popular $R_2$=[OII]$3726,3729/\hb$ diagnostic and all the others derived from it, such as $O_{32} = R_3 / R_2$ or $R_{23} = R_2 + R_3$.  

We fit each of the considered diagnostics using polynomials of the form:
\begin{equation}
    \log (R) = \sum _{n=0} ^N c_n x ^n, \label{eq-polynomial-fit}
\end{equation}
where we test  $N= 2, 3, 4$, $R$ is the diagnostic, and $x$ is the oxygen abundance normalised to the solar value from \cite{asplund2009}: $x = 12 + \log ( \text{O}/ \text{H}) - 8.69$.
We fit these relations using the \textsc{odr} python package, which performs orthogonal distance regression, taking into account the errors on both independent and dependent variables. 
We then select the polynomial that better fits our data by selecting the fit with the minimum value of $\chi ^2$; in addition, we also applied the Bayesian information criterion (BIC) to confirm the model selection and a binning procedure that aims to give the proper weight to the few available low-metallicity data points. 
When these metrics are similar, we selected preferentially lower order polynomials.
With the binning procedure, we divide the 12+log(O/H) range into 10 bins of equal width, and for each bin we evaluate the median and standard deviation of line ratios and oxygen abundances. 
These binned values are then exploited to check manually the consistency of the different polynomial fits with the data points over the whole metallicity range. In particular, they allow for a better visualisation of the general trends both in the high-metallicity regime, where there are numerous single data points but characterised by significant scatter, and in the low-metallicity regime, where there are few but fundamental data points. 
We do not repeat the polynomial fitting procedure on the binned values, but we use them for visual confirmation of the best polynomial order to use while carrying out the fit over the whole sample of single data points.

Our best-fitting  $c_n$ coefficients associated with the polynomial functional form from Eq. \ref{eq-polynomial-fit} are reported in Table \ref{table-diagnostic-coefficients} and the newly calibrated diagnostics are displayed as golden curves in Fig. \ref{figure-strong-line_diagnostics}. We also show calibrations from the literature for comparison. 
We consider the validity regime of our calibrations to cover the range 12+log(O/H)=[7.4-8.9], given the increasingly small number of datapoints at lower and higher metallicities.

\begin{table}
\caption{$c_n$ coefficients for our strong-line diagnostic calibrations.}
    \centering
    \adjustbox{max width=0.5\textwidth}{
    \begin{tabular}{c|ccccc|c|c}
        \toprule
        Diagnostic &  $c_0$ & $c_1$  &  $c_2$ & $c_3$ & $c_4$ & $\sigma$ & RMS\\
        \midrule
        $R_3$ &  $-0.84 $ & $-5.86$  &  $-6.27$ & $-1.95$ & / &0.22 & 0.44\\
        \midrule
        $N_2$ &  $-0.41$ & $0.57$  &  $-4.91$ & $-5.81$ & $-1.95$ & 0.14 & 0.25\\
        \midrule
        $O_3N_2$ &  $-0.51 $ & $-7.74 $  &  $-6.12$ & $-1.60$ & / &0.18 &0.66\\ 
        \midrule
        $N_2S_2 H \alpha$ & $0.24 $ & $2.21$ & $0.76$ & / & / &0.17 & 0.26\\
        
        \bottomrule
    \end{tabular}}
    \tablefoot{The $\sigma$ column represents the scatter in metallicity, expressed as 12+log(O/H), at fixed line ratio. The RMS column reports the root-mean-square of residuals of the fit; that is, the scatter in line ratios at fixed metallicity. }
    \label{table-diagnostic-coefficients}
    \vspace{6pt}
\end{table}

Specifically, $R_3$ is fitted using a $N=3$ polynomial to take into account the well-known double branch behaviour. In general, there is a good agreement between our $R_3$ calibration and the ones from \cite{curti2017} and \cite{nakajima2022}, although our result suggests a steeper decrease at $12 +\log ( \text{O}/ \text{H}) \gtrsim 8.5$. 
At these metallicities most of the data come from our new PHANGS--MUSE detections.
Additionally, the high-redshift data reported in this plot (blue and pink dots, taken respectively from \citealt{laseter2023} and \citealt{Sanders2023}) show slightly elevated $R_3$ with respect to local data in the same metallicity regime. The pink curve, which provides an excellent fit to these data, is one of the first high-redshift diagnostic calibrations, presented in \cite{Sanders2023} and valid over a redshift range spanning from Cosmic Noon to the Epoch of Reionisation and over the metallicity range $12 +\log ( \text{O}/ \text{H} ) = [7.0 - 8.3]$. 


We fit the $N_2 $ diagnostic using a $N = 4$ polynomial, in agreement with previous literature calibrations, while $O_3N_2$ is fitted using a $N=3$ polynomial, differing from both \cite{curti2017} and \cite{nakajima2022} where a $N=2$ polynomial is adopted. 
The difference in the choice of the polynomial order for $O_3N_2$ is only evident in the low-metallicity regime, which is still poorly populated. Lastly, the $N_2 S_2 H \alpha $ diagnostic is fitted using a $N=2 $ polynomial, and our calibration is consistent with the linear trend found in \cite{dopita2016} for its logarithmic form. 
Despite the fact that our sample encompasses only data points from \cite{nakajima2022} at 12+log(O/H)$\leq 8.0$, our diagnostic calibrations differ from those reported in the same work; this is due to the different high-metallicity samples adopted in the two analysis.  
In fact, \cite{nakajima2022} integrate their low-metallicity sample of EMPGs with data from \cite{curti2017}, while we also consider the CHAOS sample and especially the PHANGS-MUSE data points. 
This translates into significantly different final samples, with our dataset more biased towards the high-metallicity regime. 

Apart from these four diagnostics, we additionally calibrated $S_2$, $N_2 S_2 = N_2 / S_2$, $RS_{23} = R_3 + S_2$ and $O_3S_2 = R_3 / S_2$, but we did not include them in our analysis for the following reasons: 1) they are characterised by significantly larger scatter with respect to the four diagnostics reported in Fig. \ref{figure-strong-line_diagnostics}; 2) for those diagnostics where a calibration was carried out, we compared the indirect metallicity estimates obtained by combining only $R_3$, $N_2$, $O_3N_2$ and $N_2S_2 H \alpha$ diagnostics with the results obtained by further adding $O_3S_2$, $RS_{23}$ and $N_2S_2$ in all the possible combinations, and we found no significant difference in the final values. 
Thus we conclude that these diagnostics do not contribute to an improvement of indirect estimates of metallicities, but only add uncertainties due to their larger scatter. We therefore do not consider them further.

\begin{figure}
   \centering
   \includegraphics[width=0.9\linewidth]{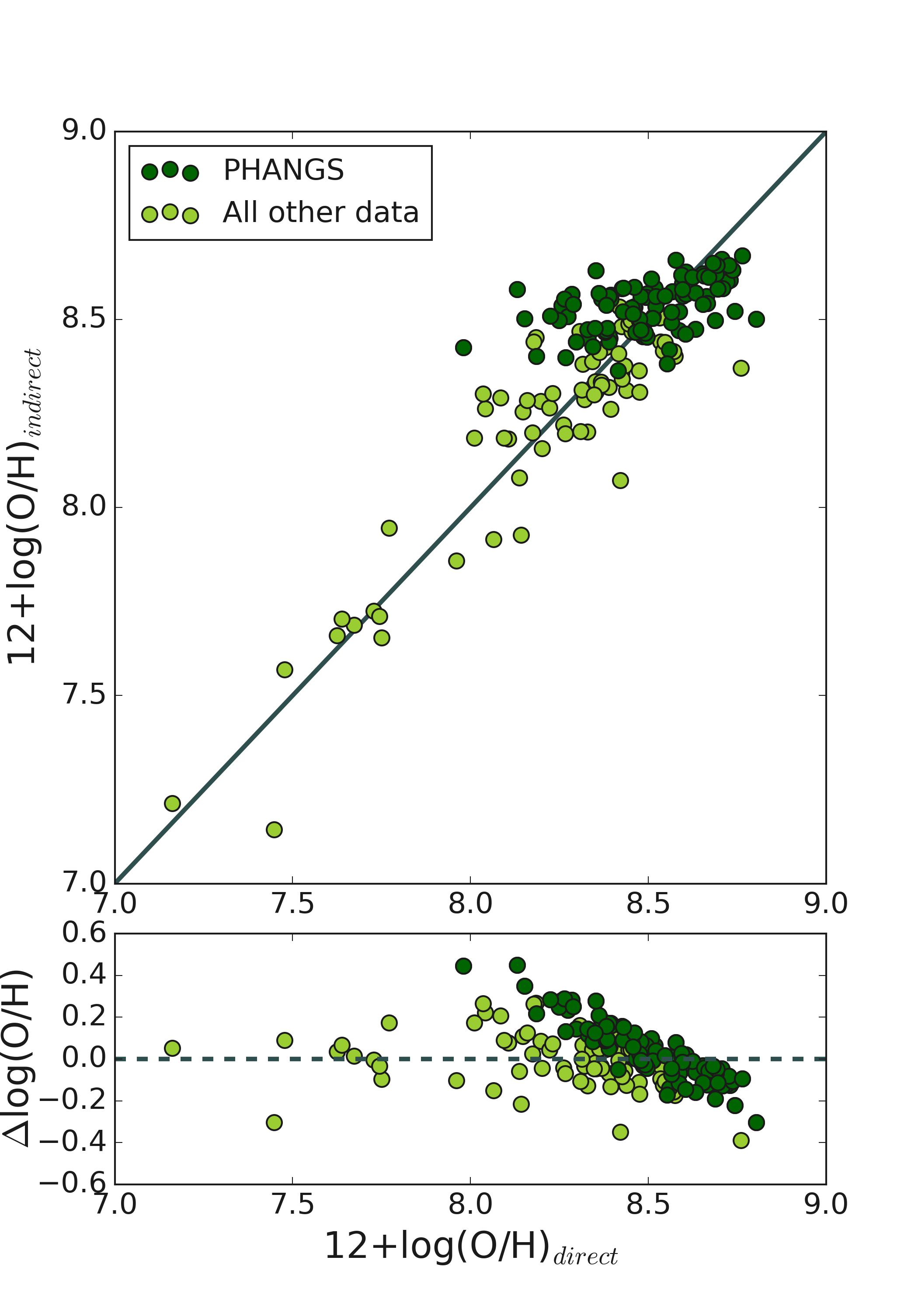}
      \caption{Comparison between direct metallicity estimates (\textit{x} axis) and indirect metallicity estimates (\textit{y} axis) obtained using the calibration based on $R_3$, $N_2$, $O_3N_2$ and $N_2 S_2 H \alpha$ diagnostics proposed in this work. Data points are divided into PHANGS--MUSE data (dark green points) and all the other data taken from the literature catalogues (light green points). The solid line is the 1:1 relation. 
    }
         \label{figure-indirect-direct-comparison}
   \end{figure}

To validate our analysis, we compared the direct metallicity measurements, as carried out in Sect. \ref{section-abundance-determination}, with the corresponding indirect estimates obtained by minimising the chi-square defined simultaneously by all the selected diagnostics ($R_3$, $N_2$, $O_3N_2$ and $N_2S_2 H \alpha$) as: 
\begin{equation}
    \chi ^2 = \sum _i \frac{\big( R_{{
    \rm obs},i} - R_{{\rm cal},i} \big)^2}{\sigma ^2 _{{\rm obs},i}}, \label{eq-indirect-measurements}
\end{equation}
where the sum is carried out over the different diagnostics, $R_{{\rm obs},i}$ are the observed line ratios, $\sigma_{{\rm obs},i}$ are the associated uncertainties, and $R_{{\rm cal},i}$ are the values predicted by each calibration for a given metallicity. 
In principle, one could also consider the intrinsic dispersion of each strong-line diagnostic calibration (the RMS column in Table \ref{table-diagnostic-coefficients}), in order to attribute a major weight to the diagnostics with lower intrinsic scatter. However, because such scatter is comparable for all the diagnostics and always dominant by one or two orders of magnitude with respect to the observed line ratio uncertainties, we find that considering the intrinsic scatter does not change the final $\chi ^2$ minimisation procedure.

This comparison is reported in Fig. \ref{figure-indirect-direct-comparison}. In general, there is a good agreement between direct and indirect metallicity estimates over the whole metallicity range, although at high metallicity ($ 12 + \log (\text{O/H} ) \gtrsim 8.0$) the scatter around the 1:1 relation becomes significant (RMS of residuals: 0.14 dex). If we consider only PHANGS--MUSE data (dark green points), there exists a trend of residuals with metallicity, with indirect measurements overestimating the corresponding $T_e$-based metallicity at $12+\log ( \text{O/H}) \lesssim 8.5$ and underestimating them at higher metallicities. These data points at $12+ \log ( \text{O/H} ) \lesssim 8.5$ that significantly deviate from the 1:1 relation are the same that significantly deviate from the best fit relations in Fig. \ref{figure-strong-line_diagnostics}. 
As such deviations are already evident in the calibration plots, we conclude they are not due to potential inaccuracies in our calibrations but to limitations in the direct measurements themselves, the primary ones being the simplification in the assumed three-zone theoretical model and the hypothesis of an homogeneous ISM. 
To a lesser extent, some effects may arise from selection biases since working with auroral lines limits the study to the brightest \hii\ regions. 
Nonetheless, we verify that increasing the ANR threshold to 7 or 10, instead of 5, still yields consistent results without reducing the scatter in the data points.  
We conclude that, because strong-line ratios are influenced by many factors beyond the source metallicity — such as the physical and geometrical structures of ionised nebulae —, their calibration remains complex and highly dependent on the specific sources. 
Addressing this requires abandoning the current simplifications and conducting a comprehensive analysis of these regions, extending beyond the narrow focus on the single chemical aspect.

Inspired by the recent work from \cite{easeman2024}, we also compared our direct metallicity measurements with the indirect ones obtained using only the $N_2 S_2 H\alpha$ diagnostic, which \cite{easeman2024} consider the most reliable strong-line diagnostic when working with MUSE data. However, we did not find significant differences in indirect estimates by using only the $N_2 S_2 H \alpha$ diagnostic or by combining it with $R_3$, $N_2$ and $O_3N_2$. We therefore prefer to use in our fiducial calibration the combination of all four line ratios we have calibrated. This could also help minimise spurious effects arising from the use of a single strong line diagnostic due to secondary dependencies on sulphur and nitrogen abundances \citep{dopita2016, MaiolinoMannucci2019}. 


\begin{figure*}[!ht]
  \centering
  \begin{subfigure}{\linewidth}
  \hspace{-0.6cm}
    \includegraphics[width=\linewidth]{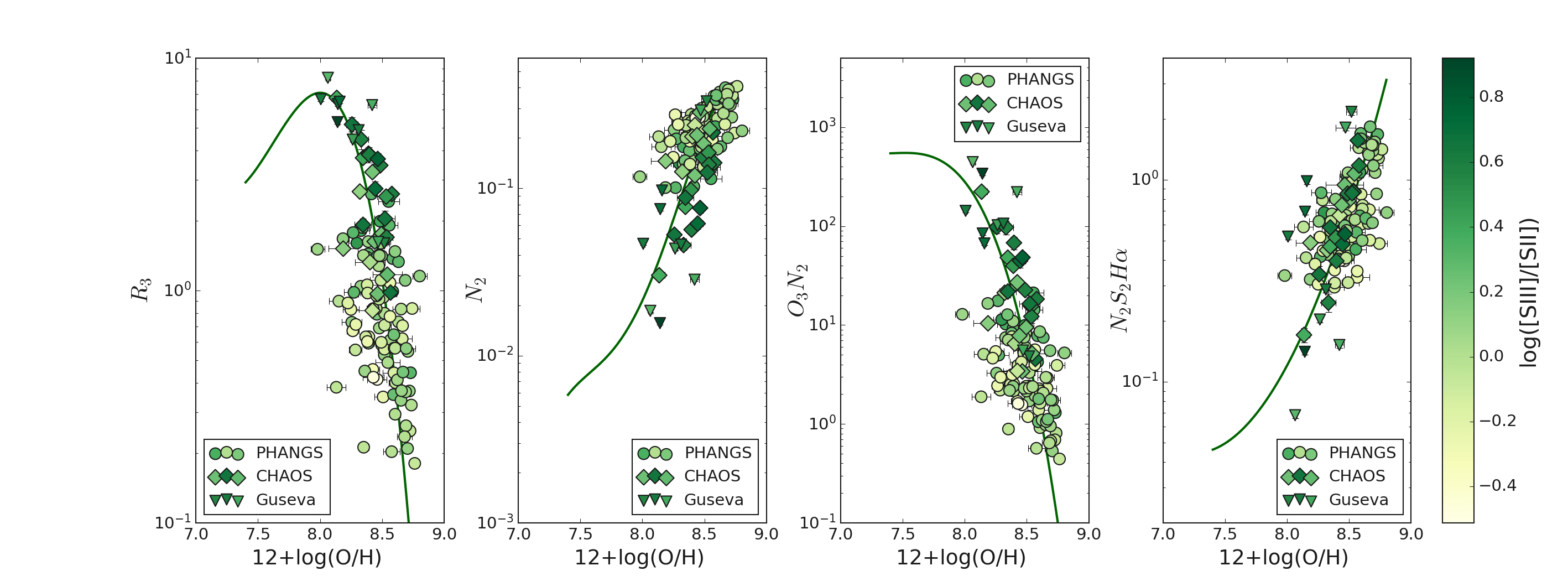}
    \label{figure-photoion_dependence_S}
  \end{subfigure}
 
  \begin{subfigure}{\linewidth}
  \hspace{-0.6cm}
    \includegraphics[width=\linewidth]{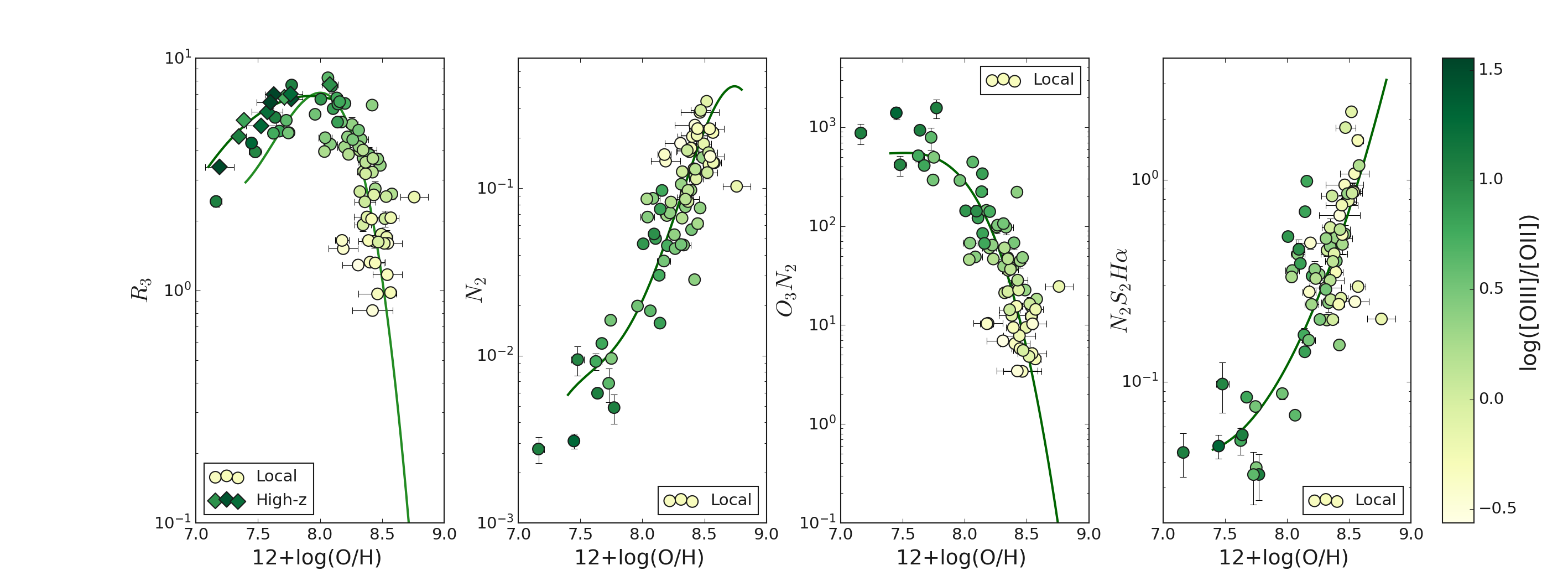}
    \label{figure-photoion_dependence_O}
  \end{subfigure}
  
   \caption{Dependence of strong-line diagnostics on the ionization parameter. \textit{Upper panels}: Strong-line diagnostics colour-coded according to the \siii$\lambda \lambda 9069, 9531$/\sii$\lambda \lambda 6717, 6731$ line ratio. \textit{Lower panels}: Strong-line diagnostics colour-coded according to the \oiii$\lambda 5007$/\oii$\lambda \lambda 3726, 3729$ line ratio. We have labelled as `local' the data from CHAOS, Curti’s, Guseva’s and Nakajima’s catalogues and as `high-z' data from \cite{Sanders2023} and \cite{laseter2023}. The plotted curves are our diagnostic calibrations plus the high-redshift calibration from \cite{Sanders2023} for $R_3$.}
  \label{figure-strong-line_diagnostics_color-coded-U}

   \hspace{-1.2cm}
    \includegraphics[width=\linewidth]{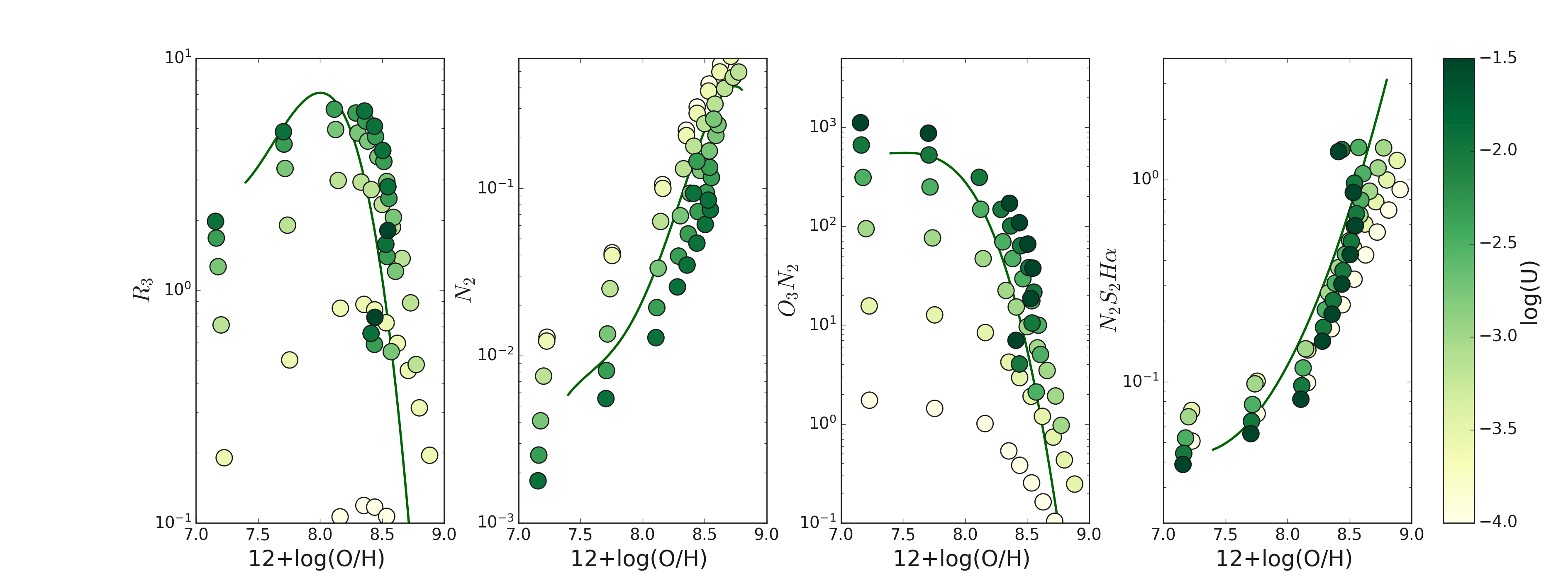}
    \caption{Dependence of strong-line ratios on $\log (U)$ resulting from the photoionisation model analysis carried out in this work. To realise this figure, in particular, we have analysed a $0.5$ Myr SSP input spectrum. Data points are colour-coded according to their $\log (U)$ values.}
    \label{figure-photoionisation-models}
\end{figure*}




\subsection{Dependence of strong-line diagnostics on ionisation parameter}

In this section we test whether the scatter observed in the strong-line calibration plots of Fig. \ref{figure-strong-line_diagnostics} can be attributed to the degeneracy between metallicity and ionisation parameter, as these two parameters are known to be strongly anti-correlated \citep{morisset2016, ji+2022}.
We study the dependence of strong-line diagnostics on $\log (U)$ by both evaluating the ionisation parameter for our dataset in terms of sulphur and oxygen line ratios, as was discussed in the previous section, and by analysing photoionisation models. 

In Fig. \ref{figure-strong-line_diagnostics_color-coded-U} we report the same diagnostics as in Fig. \ref{figure-strong-line_diagnostics}, but with data points colour-coded according to the sulphur and oxygen line ratios used as a proxy for the ionisation parameter. 
We can identify some clear trends in these plots, particularly evident for the \oiii/\oii\ line ratio, with data points at lower metallicities characterised by higher \oiii/\oii\ and \siii/\sii\ line ratios, which correspond to higher ionisation parameter values. 
On the other hand, at higher metallicities the situation becomes more complex and the scatter is significant. 
The same line ratio value on the \textit{y} axis can be associated with different metallicity values because of the degeneracy of metallicity with $\log (U)$, implying that the scatter observed in these plots encodes the further dependence of strong-line ratios on other physical properties apart from the metallicity of the source. 
To further investigate this additional dependence, we turn to photoionisation models.

We consider Cloudy photoionisation models\footnote{available at \url{https://github.com/francbelf/python\_izi/tree/master/grids}} described in \cite{belfiore2022}. They used input spectra generated with the Flexible Stellar Population Synthesis (FSPS) code \citep{conroy+2009}, using simple stellar population (SSP) models \citep{byler+2019}, with ages between $0.5$ Myr and $20$ Myr, spanning a metallicity range of [Z/H]=[$-0.6$, $ 0.4$], and an ionisation parameter range of $\log (U)$ = [$-5$, $-1$] (local \hii\ regions showing typical $\log(U)$ values of [$-3.6$,$-2.6$], see e.g. \citealt{grasha+2012, poetrodjojo+2018}). The corresponding input gas-phase metallicities vary in the range 12+log(O/H) = [$6.5$, $9.0$].
As we are interested in studying star-forming regions, we focus on young SSP spectra with age of  $0.5$ Myr, although very similar results are obtained by considering ages up to $\sim 5$ Myr.
The median age of PHANGS-MUSE stellar clusters associated with \hii\ regions is $\sim 4$ Myr \citep{scheuermann2023}.
We applied the same data analysis procedure to the models as to the observational data. In particular, we estimated $T_e$ and ionic abundances for each model following the same procedure as in Sect. \ref{section:methods}. In Fig. \ref{figure-photoionisation-models}, we show the resulting line ratios as a function of our estimate of the strong-line metallicity applying the calibration developed in this work to the models.

\begin{figure*}[!t]
    \centering
    \includegraphics[width=\linewidth]{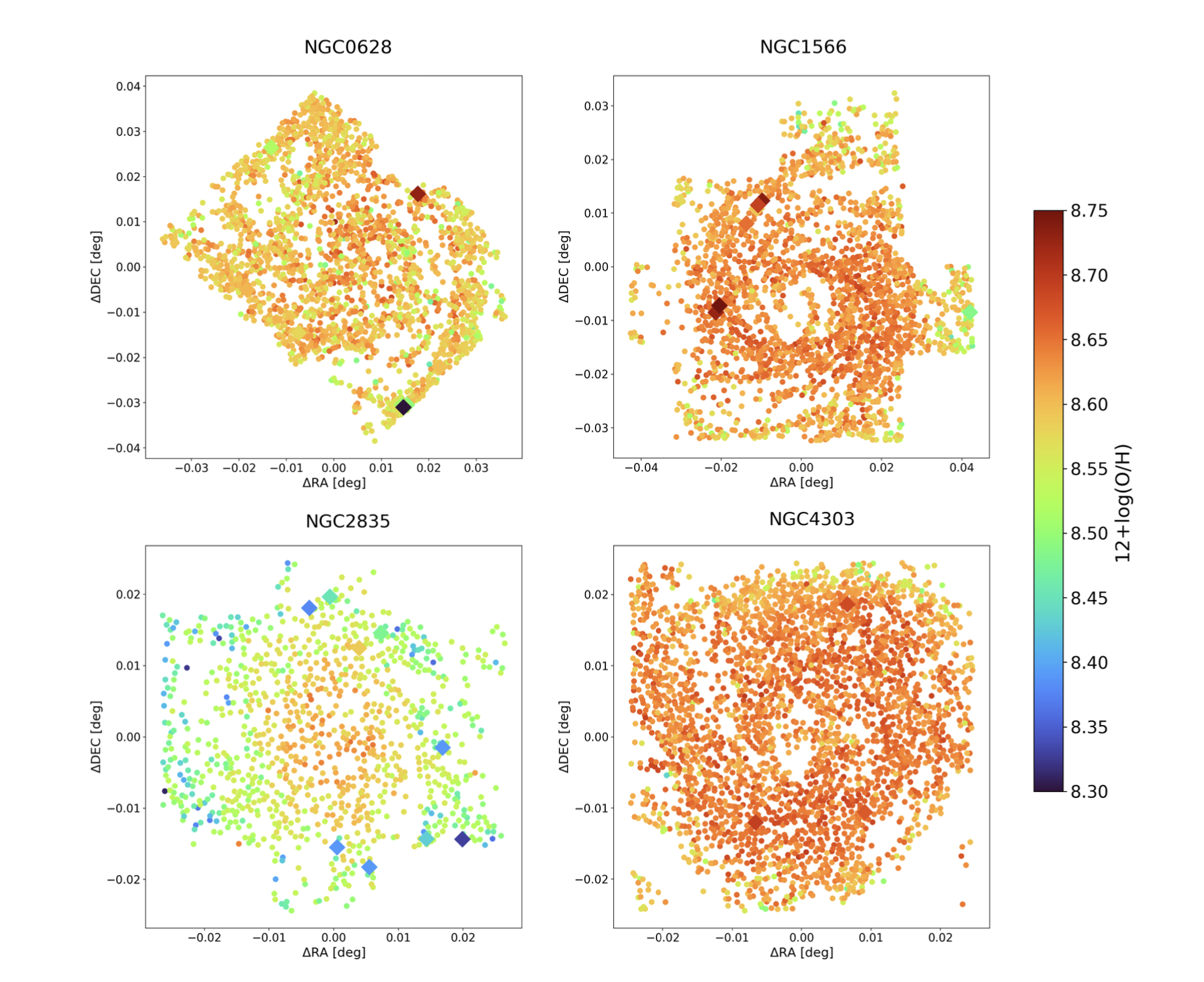}
    \caption{2D metallicity distribution of \hii\ regions in NGC0628, NGC1566, NGC2835 and NGC4303. \hii\ regions in each galaxy are colour-coded according to their metallicity. Indirect estimates are shown with dots, direct measurements with diamonds.}
  \label{figure-2d_met_grad}
\end{figure*}

We observe that differences in ionisation parameter can account for the scatter present in strong-line diagnostics, especially $R_3$ and $O_3N_2$, which show the largest dependence on $\log(U)$ \citep{MaiolinoMannucci2019}. 
We observe that the metallicity estimates reported on the \textit{x} axis are the output metallicities, that is, the ones obtained by applying the direct method to the input line ratios. 
We compared these metallicity values with the input ones and found a general consistency that breaks only at high metallicities (12+log(O/H) $\gtrsim 8.5$), with this effect more evident for higher ionisation parameters. 
This discrepancy, according to which photoionisation models systematically overestimate metallicity estimates with respect to empirical methods, is well known in literature (\citealp{marconi2024} and references therein). 

\section{Effect on metallicity gradients} \label{section-metallicity-gradients}

In this section we consider the effect of our revised calibration on the measurements of metallicity gradients. Specifically, we compare with the metallicity gradients measured for the PHANGS--MUSE sample by \cite{groves2023}.
For each galaxy in the PHANGS--MUSE sample, we select all the nebulae from the catalogue of \cite{groves2023} with S/N $>5$ on the lines that define the $R_3$, $N_2$, $O_3N_2$ and $N_2S_2H \alpha$ diagnostics, namely: \ha, \hb, \oiii$\lambda 5007$ \nii$\lambda 6584$, \sii$\lambda \lambda 6717, 6731$, \siii$\lambda 9069$. We select \hii\ regions by applying the same BPT diagram selection mentioned in Sect. \ref{section-sample-selection}. These cuts result in the selection of $23704$ \hii\ regions out of $31497$ initial entries.
We then use Eq. \ref{eq-indirect-measurements} to indirectly estimate metallicity according to our calibration. 

In Fig. \ref{figure-2d_met_grad} we report the 2D distribution of \hii\ regions metallicities for four of the 19 galaxies of the PHANGS--MUSE sample, to illustrate the variety of distributions. Each \hii\ region is colour-coded according to its strong-line metallicity estimate, apart from those with a direct measurement available, which are shown as large diamonds. 

\begin{table*}
  \centering
   \caption{Radial metallicity gradients for the 19 galaxies in the PHANGS--MUSE sample. }
  \label{table-metallicity-gradients}
  \adjustbox{max width=1.0\textwidth}{
 
  \begin{tabular}{cccccccc}
    \toprule
    Galaxy & Intercept & Slope [dex/($r/ \rm R_{eff})$] & Intercept (cut in $r$) & Slope  [dex/($r/ \rm R_{eff})$] (cut in $r$) & $\sigma $(O/H)\\
    \midrule
    IC5332 & 8.596 $\pm$ 0.003 & -0.080 $\pm$ 0.004 & 8.595 $\pm$ 0.007 & -0.077 $\pm$ 0.009 & 0.035\\
    NGC0628 & 8.617 $\pm$ 0.002 & -0.024 $\pm$ 0.002 & 8.618 $\pm$ 0.002 & -0.024 $\pm$ 0.002 & 0.031\\
    NGC1087 & 8.615 $\pm$ 0.002 & -0.048 $\pm$ 0.002 & 8.610 $\pm$ 0.003 & -0.045 $\pm$ 0.002 & 0.022\\
    NGC1300 & 8.655 $\pm$ 0.003 & -0.039 $\pm$ 0.002 & 8.663 $\pm$ 0.003 & -0.046 $\pm$ 0.003 & 0.031\\
    NGC1365 & 8.652 $\pm$ 0.003 & -0.072 $\pm$ 0.004 & 8.714 $\pm$ 0.006 & -0.145 $\pm$ 0.007 & 0.039\\
    NGC1385 & 8.602 $\pm$ 0.002 & -0.030 $\pm$ 0.001 & 8.591 $\pm$ 0.002 & -0.021 $\pm$ 0.002 & 0.022\\
    NGC1433 & 8.639 $\pm$ 0.003 & -0.008 $\pm$ 0.001 & 8.637 $\pm$ 0.003 & -0.008 $\pm$ 0.002 & 0.033\\
    NGC1512 & 8.638 $\pm$ 0.005 & -0.006 $\pm$ 0.003 & 8.639 $\pm$ 0.006 & -0.006 $\pm$ 0.004 & 0.030\\
    NGC1566 & 8.662 $\pm$ 0.001 & -0.023 $\pm$ 0.001 & 8.666 $\pm$ 0.002 & -0.025 $\pm$ 0.001 & 0.028\\
    NGC1672 & 8.634 $\pm$ 0.002 & -0.010 $\pm$ 0.001 & 8.630 $\pm$ 0.002 & -0.008 $\pm$ 0.001 & 0.029\\
    NGC2835 & 8.626 $\pm$ 0.003 & -0.074 $\pm$ 0.002 & 8.622 $\pm$ 0.003 & -0.071 $\pm$ 0.003 & 0.037\\
    NGC3351 & 8.625 $\pm$ 0.005 & 0.009 $\pm$ 0.004 & 8.611 $\pm$ 0.006 & 0.018 $\pm$ 0.004 & 0.049\\
    NGC3627 & 8.636 $\pm$ 0.002 & 0.001 $\pm$ 0.001 & 8.633 $\pm$ 0.002 & 0.003 $\pm$ 0.001 & 0.025\\
    NGC4254 & 8.654 $\pm$ 0.001 & -0.0190 $\pm$ 0.0005 & 8.660 $\pm$ 0.001 & -0.021 $\pm$ 0.001 & 0.025\\
    NGC4303 & 8.661 $\pm$ 0.001 & -0.018 $\pm$ 0.001 & 8.670 $\pm$ 0.002 & -0.023 $\pm$ 0.001 & 0.027\\
    NGC4321 & 8.629 $\pm$ 0.002 & -0.001 $\pm$ 0.002 & 8.628 $\pm$ 0.002 & -0.001 $\pm$ 0.002 & 0.031\\
    NGC4535 & 8.623 $\pm$ 0.003 & -0.004 $\pm$ 0.003 & 8.637 $\pm$ 0.004 & -0.018 $\pm$ 0.004 & 0.036\\
    NGC5068 & 8.573 $\pm$ 0.002 & -0.039 $\pm$ 0.001 & 8.575 $\pm$ 0.002 & -0.041 $\pm$ 0.002 & 0.032\\
    NGC7496 & 8.649 $\pm$ 0.002 & -0.053 $\pm$ 0.002 & 8.664 $\pm$ 0.004 & -0.064 $\pm$ 0.003 & 0.034\\
\bottomrule
\end{tabular}}
\tablefoot{The cut in $r$ is done at $0.5 \rm \ R_{eff}$ (more details in the text). The $\sigma $(O/H) is the standard deviation of the $\Delta$(O/H) distribution, with $\Delta$(O/H) being the difference between the indirect metallicity estimate obtained by using our strong-line diagnostic calibrations and the metallicity estimate expected from the metallicity gradients at the given distance.}
\end{table*}

\begin{figure*}
    \centering
    \includegraphics[width=\linewidth]{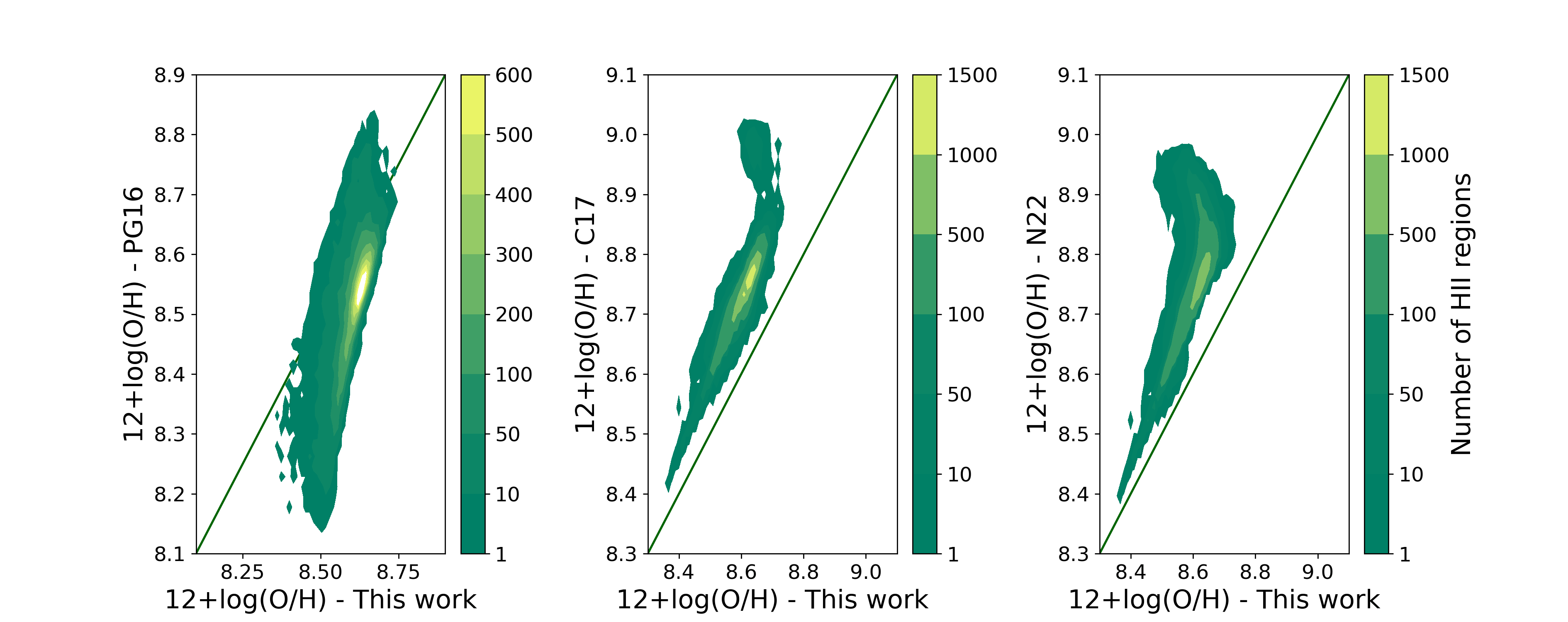}
    \caption{Comparison between indirect metallicity estimates based on our calibration and on \citetalias{PG16}, \citetalias{curti2017} and \citetalias{nakajima2022} calibrations (respectively: \textit{left}, \textit{central}, and \textit{right} panels) for all \hii\ regions in the PHANGS--MUSE sample. For each figure, the colourbar describes the contour lines of the metallicity distribution. The solid line indicates the 1:1 relation.}
    \label{figure-comparison-indirect-met-estimates}
\end{figure*}

We fit linear slopes to the metallicity gradient in units of effective radius ($\rm R_{eff}$). The best fits are reported in Table \ref{table-metallicity-gradients} for each galaxy.
We fit two different radial gradients for each galaxy: one that considers all \hii\ regions and one that excludes those within $\rm r<0.5 \ R_{eff}$, as was first proposed in \cite{sanchez-menguiano2018} and then adopted in \cite{groves2023}. 
For each galaxy, we also evaluate the standard deviation of the metallicity residuals with respect to the best-fit linear gradient, $\sigma$(O/H), which we report in the last column of Table \ref{table-metallicity-gradients}. We find that $\sigma$(O/H) varies from 0.022 to 0.039 dex, with a mean value across the sample of $\sigma$(O/H) of $0.031$ dex.

\subsection{Comparison with literature calibrations}

We compare the metallicity gradients obtained by using our strong-line diagnostics for indirect metallicity estimates with the same gradients based on the literature calibrations from \citetalias{curti2017}, \citetalias{nakajima2022}, and \cite{PG16}, abbreviated as \citetalias{PG16} in the following.
In Fig. \ref{figure-comparison-indirect-met-estimates} we show a comparison of indirect metallicity estimates based on our, \citetalias{curti2017}, \citetalias{nakajima2022} and \citetalias{PG16} calibrations for the whole PHANGS--MUSE \hii\ region sample. 
In general, the \citetalias{PG16} calibration provides lower metallicities with respect to those estimated in this work, or from the calibrations of \citetalias{curti2017} and \citetalias{nakajima2022} (left panel of Fig. \ref{figure-comparison-indirect-met-estimates}), with this discrepancy more evident at lower metallicities (12+log(O/H) $\lesssim8.3$). Conversely, the \citetalias{curti2017} and \citetalias{nakajima2022} calibrations provide slightly higher metallicity values with respect to our calibration (central and right panels of Fig. \ref{figure-comparison-indirect-met-estimates}). 
Despite the clear presence of an offset between the two calibrations, they present a 1:1 relation that breaks only at high metallicities (12+log(O/H) $\gtrsim 8.7$).

To understand the origin of the discrepancy between the \citetalias{PG16} and our calibration, we tested the dependence of residuals on a number of secondary parameters: ionisation parameter (from the sulfur line ratio, estimated as in Sect. \ref{section-ionisation-parameter-measurements}), $\chi ^2$ (as defined in Eq. \ref{eq-indirect-measurements}), and \ha\ flux, velocity dispersion, and equivalent width. 
We found no dependence on \ha\ flux, velocity dispersion, and equivalent width or $\chi ^2$, and only a mild inverse correlation with ionisation parameter (Pearson correlation coefficient: $0.3)$. 

\begin{figure*}[!ht]
    \centering
    \includegraphics[trim={4.0cm 4.2cm 3cm 3cm}, width=0.7\linewidth]{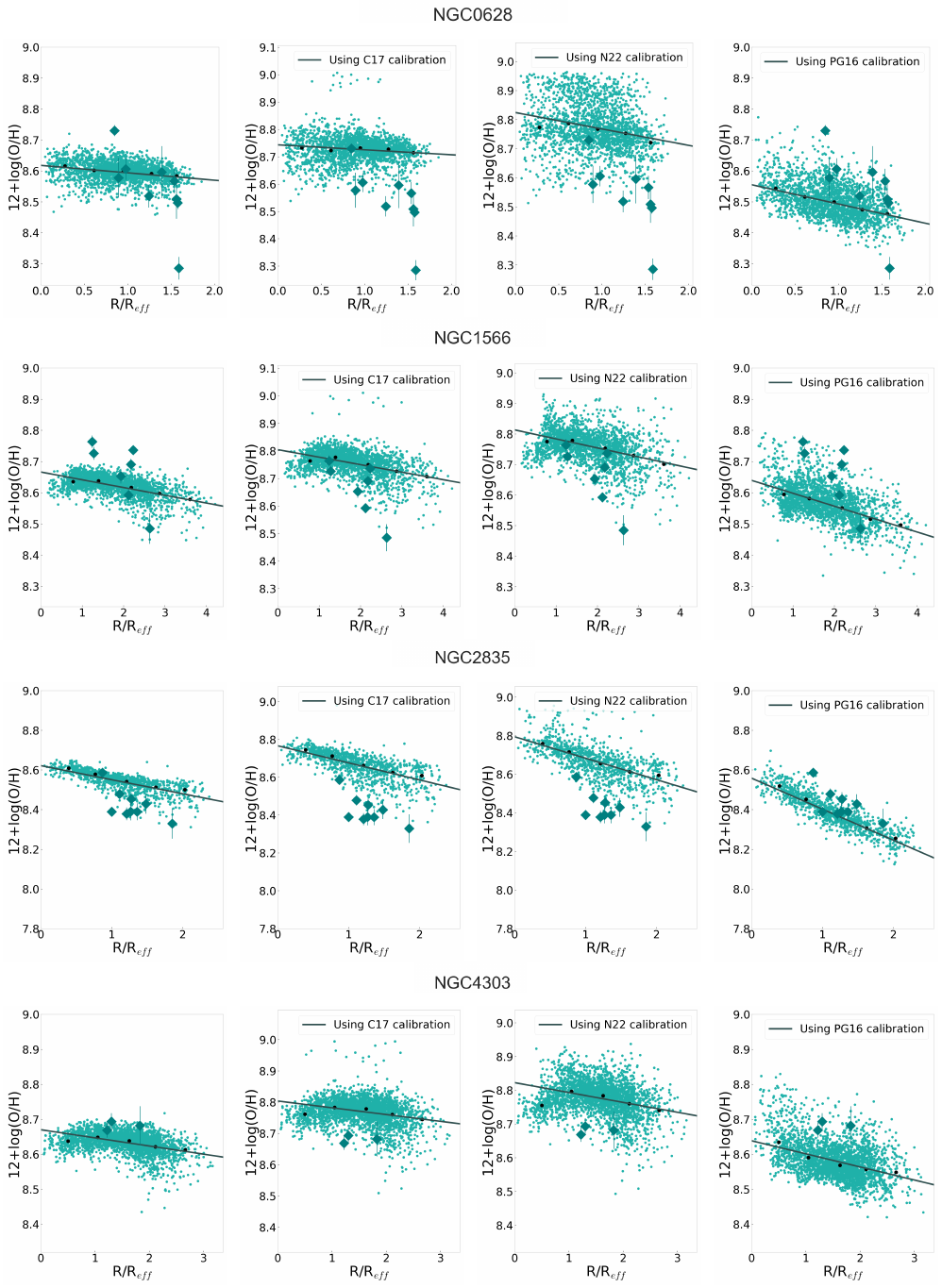}
    \caption{Radial metallicity gradients for galaxies NGC0628, NGC1566, NGC2835, and NGC4303. Metallicity estimates are obtained with our, \citetalias{curti2017}, \citetalias{nakajima2022} and \citetalias{PG16} calibrations (from left to right) and reported as small dots in each panel. Big diamonds are the direct measurements, not used to calibrate gradients but reported for visualisation purposes. The solid lines represent the linear best fit performed by excluding data at $r < 0.5 \rm r_{eff}$, and small black dots are the indirect metallicity median values evaluated in $0.36 r_{eff}$, $0.81 r_{eff}$, $0.46 r_{eff}$, and $0.62 r_{eff}$ wide bins, respectively. }
    \label{figure-met-grad-cal}
\end{figure*}

We compare the metallicity gradients derived from different calibrations in Fig. \ref{figure-met-grad-cal} for four galaxies of the PHANGS--MUSE sample.
The \citetalias{nakajima2022} calibration shows gradients characterised by significant, probably unphysical, scatter. The scatter is slightly reduced with \citetalias{curti2017} calibrations, and the lowest scatter is obtained when using the calibrations from this work or \citetalias{PG16} calibrations. In a few cases (one to three \hii\ regions per galaxy) the \citetalias{PG16} calibration provides unphysically low metallicity estimates (typically 12+log(O/H) $\lesssim 7.5$), which are not shown in Fig. \ref{figure-met-grad-cal}.

In Fig. \ref{figure-slope-comparison} we provide a further comparison between the gradient slopes obtained when using the different calibrations for metallicity estimates. 
We first verify that the slopes obtained when using \citetalias{PG16} calibration are consistent with the ones provided in \cite{groves2023}, where the same calibration is adopted. 
Hence, in the upper panel of Fig. \ref{figure-slope-comparison} we directly compare the slopes presented in \cite{groves2023} with the ones based on the strong-line diagnostic calibrations presented in this paper. 
We observe that \citetalias{PG16} calibration provides systematically steeper gradient slopes with respect to ours, with this effect being stronger as the slopes become more negative.  
On the other hand, there is a substantially better agreement between our results and the ones obtained with \citetalias{nakajima2022} and especially \citetalias{curti2017} calibrations (in the lower and central panels of Fig. \ref{figure-slope-comparison}, respectively). 
From Fig. \ref{figure-comparison-indirect-met-estimates}, we observe that our diagnostics tend to overestimate PHANGS--MUSE \hii\ region metallicities at 12+log(O/H)$\lesssim 8.4$, suggesting a shrinking in the metallicity coverage relative to the direct method. 
This effect can lead to a slight underestimate in gradient slopes, and may be partially responsible for the discrepancies observed in Figs. \ref{figure-met-grad-cal} and \ref{figure-slope-comparison} when comparing our results with the literature, especially \citetalias{PG16}. 
Nevertheless, our results from Table \ref{table-metallicity-gradients} are consistent with the general trends of metallicity gradients observed in local star-forming galaxies, as reported in other studies.
In \cite{belfiore2017}, for example, typical metallicity gradients for the SDSS IV Mapping Nearby Galaxies at Apache Point Observatory (MaNGA; \citealp{bundy2015}) survey span from 0.0 to -0.1 dex/($r/ \rm R_{eff}$) for $z\sim 0$ galaxies with masses $\log (M/M_\odot ) = 9.0-11.5$, encompassing the mass range covered by the PHANGS--MUSE galaxies \citep{leroy2021a}. 
The CHAOS collaboration finds metallicity slopes around -0.1 dex/($r/ \rm R_{eff}$) \citep{rogers2021, rogers2022}. 
Similar results are also reported in \cite{sanchez2014} for the Calar Alto Legacy Integral Field Area (CALIFA) Survey \citep{sanchez2012}, where gradient slopes are found to vary from -0.2 to 0.0 dex/($r/ \rm R_{eff}$) with a well-defined characteristic value of -0.1 dex/($r/ \rm R_{eff}$) and standard deviation of 0.09 dex/($r/ \rm R_{eff}$). 


\section{Conclusions}
\label{section-conclusions}

In this work, we provide a chemical analysis of \hii\ regions from the PHANGS-MUSE nebular catalogue compiled in \cite{groves2023}. 
We fitted all the spectra from the catalogue following an innovative procedure based on single-region spectral fitting, with the intent of better constraining the stellar continuum. We carried out a selection procedure that aimed to select without any prior bias the brightest sources of the catalogue, then we exploited the measured line fluxes to estimate the electron temperatures and densities, and hence the ionic abundances. 
We complemented the PHANGS-MUSE emission line catalogue with other emission line compilations from the literature, which have been carefully re-analysed in order to obtain a homogeneous and comprehensive sample of direct metallicity estimates covering a wide range in metallicity ($7.4 \leq 12+ \log (\text{O/H}) \leq 8.8$). 
We then exploited the total dataset to empirically re-calibrate some of the most widely used strong-line diagnostics for the determination of the oxygen abundance, and we investigated their dependence on the ionisation parameter. 
Lastly, we used our newly calibrated diagnostics to infer the metallicity of all \hii\ regions within the PHANGS-MUSE sample, and for each galaxy of the sample we evaluated the radial metallicity gradients.  
We summarise our main results as follows.

\begin{figure}
    \centering
    \includegraphics[width=0.9\linewidth]{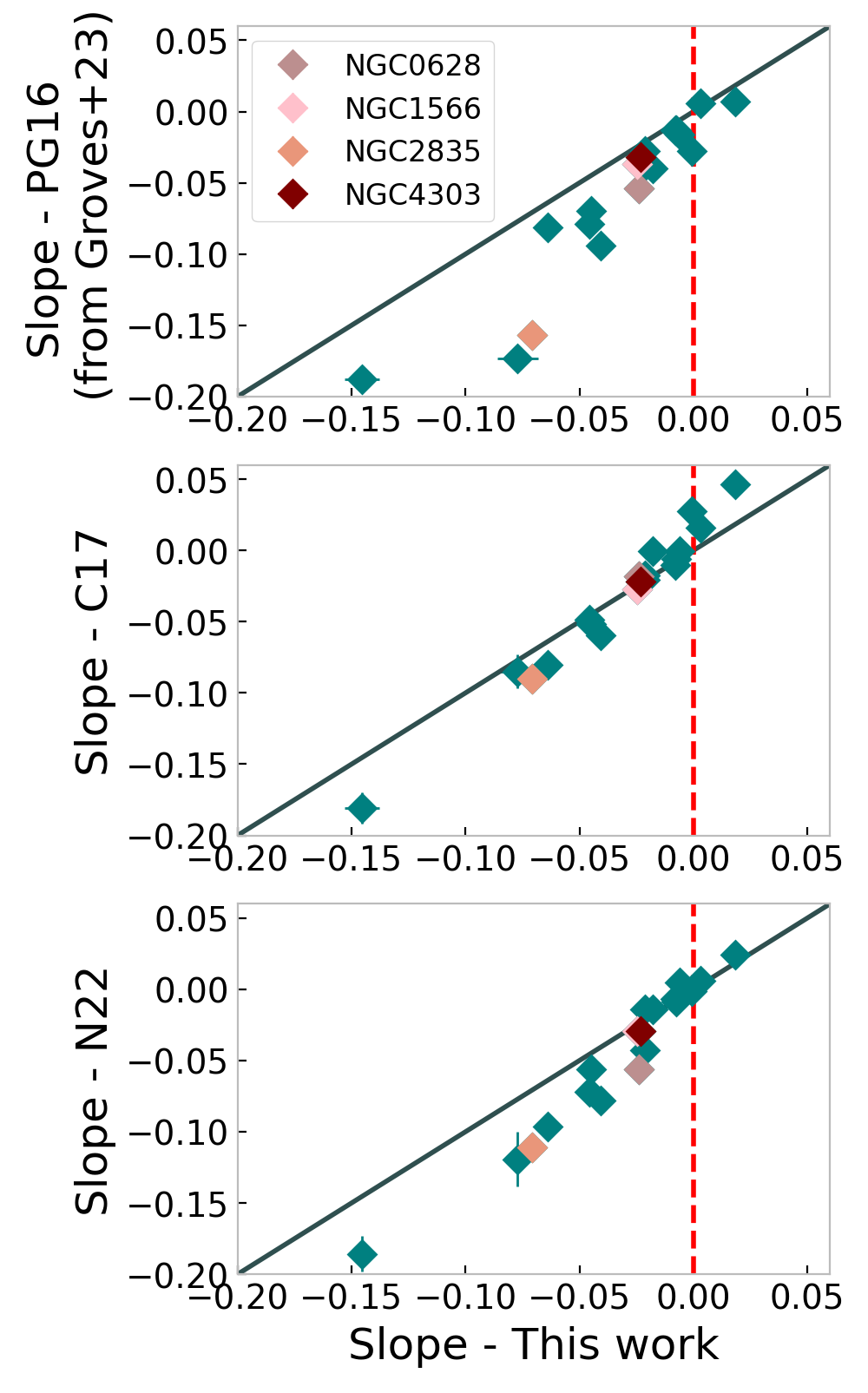}
    \caption{Comparison between the slopes of metallicity radial gradients fitted on different metallicity estimates based on our PG16 (\textit{upper panel}), C17 (\textit{middle panel}), and N22 (\textit{lower panel}) calibrations. The solid line is the 1:1 relation. For visual clarity, we have highlighted with different colours the four galaxies from Fig. \ref{figure-2d_met_grad}. }
    \label{figure-slope-comparison}
\end{figure}

\begin{itemize}
    \item PHANGS-MUSE galaxies are characterised by significantly high gas-phase metallicities, with 12+log(O/H) $\geq 8.0$. Oxygen is found mostly in its singly ionised state (Fig. \ref{figure-plots-as-in-curti}). 
    Sulphur and nitrogen abundances are typically sub-solar (Fig. \ref{figure-sulfur-nitrogen-abundances}). The S/O ratio drops significantly as the source metallicity increases, while the N/O ratio does not show any evident trend with metallicity. Both are consistent with previous literature results, as was discussed in Sect. \ref{section-abundance-determination}. 
    \item We used PHANGS-MUSE data to re-calibrate the $T_e$[SIII]-$T_e$[NII] relation, and we find consistent results with the ones from the CHAOS collaboration (Fig. \ref{figure-TSIII-TNII}). 
    \item We present new calibrations of strong-line diagnostics covering a wide range in metallicity and calibrated on both \hii\ regions (e.g. from PHANGS-MUSE, CHAOS, or \citetalias{guseva2011} catalogues) and single or stacked galaxies (e.g. from \citetalias{curti2017} and \citetalias{nakajima2022} catalogues). The new diagnostics provide reasonable metallicity estimates over the whole analysed metallicity range (Fig. \ref{figure-comparison-indirect-met-estimates}) and they are consistent with previous literature results, especially the ones from \cite{curti2017}, although they were obtained with different methods. 
    This confirms the validity of both.
    \item The strong-line diagnostics from Fig. \ref{figure-strong-line_diagnostics} show significant scatter, especially evident at higher metallicities where there are more data points available. This scatter then translates into some residuals deviations from the 1:1 relation in Fig. \ref{figure-indirect-direct-comparison}. 
    The choice of imposing S/N > 5 on all the emission lines of interest, auroral lines included, guarantees that this scatter is not attributed to the inclusion in our dataset of spurious data such as noise spikes, which are the most common source of errors when dealing with faint lines. 
    Hence, we have investigated the dependence of strong-line ratios on the ionisation parameter using both an empirical and theoretical approach, with the former based on the calculation of some specific line ratios as a proxy of the ionisation parameter, $U$, and the latter based on the use of photoionisation models. 
    From Figs. \ref{figure-strong-line_diagnostics_color-coded-U} and \ref{figure-photoionisation-models}, we conclude that the scatter observed in strong-line diagnostics is most likely attributed to their additional dependence on $U$. 
    \item We have applied our newly calibrated strong-line diagnostics to all the \hii\ regions within the PHANGS-MUSE nebular catalogue, and we have compared our indirect metallicity measurements with the same estimates based on different literature calibrations (Fig. \ref{figure-comparison-indirect-met-estimates}). We do not find a precise correspondence between the different calibrations. The \citetalias{curti2017} and \citetalias{nakajima2022} calibrations show a clear offset towards higher metallicities with respect to our ones, and when comparing with \citetalias{PG16} calibration we completely lose the 1:1 expected relation. These discrepancies may be due to the different approaches and calibration sets used for indirectly estimating 12+log(O/H).
    \item We carried out a linear fit to estimate the metallicity radial gradients within each galaxy of the PHANGS-MUSE sample (Table \ref{table-metallicity-gradients}). Hence, we compared our findings with the same gradients fitted on different indirect metallicity estimates by exploiting \citetalias{PG16}, \citetalias{curti2017} and \citetalias{nakajima2022} strong-line diagnostic calibrations. 
    We observe that our and \citetalias{PG16} calibrations are the ones that guarantee the best agreement between direct and indirect estimates, and the least scatter in the indirect estimates. With respect to our calibration, \citetalias{PG16} provides steeper gradients, with this discrepancy more evident at higher gradient slope absolute values (Fig. \ref{figure-slope-comparison}). The same effect is still present, although less evident, when comparing with \citetalias{curti2017} and \citetalias{nakajima2022} calibrations. 
   
\end{itemize}

\begin{acknowledgements}
This work has been carried out as part of the PHANGS
collaboration. It is  based on observations collected
at the European Southern Observatory under ESO programmes 094.C-0623 (PI: Kreckel), 095.C-0473, 098.C0484 (PI: Blanc), 1100.B-0651 (PHANGS-MUSE; PI: Schinnerer), as well as 094.B-0321 (MAGNUM; PI: Marconi), 099.B-0242, 0100.B-0116, 098.B-0551 (MAD; PI: Carollo) and 097.B-0640 (TIMER; PI: Gadotti).
FB acknowledges support from the INAF Fundamental Astrophysics program 2022. Part of this work was supported by the German \emph{Deut\-sche For\-schungs\-ge\-mein\-schaft, DFG\/} project number Ts~17/2--1. G.A.B. acknowledges the support from the ANID Basal project FB210003. The publication has been produced with co-funding from the European Union - Next Generation EU.
RSK acknowledges financial support from the European Research Council via the ERC Synergy Grant ``ECOGAL'' (project ID 855130),  from the German Excellence Strategy via the Heidelberg Cluster of Excellence (EXC 2181 - 390900948) ``STRUCTURES'', and from the German Ministry for Economic Affairs and Climate Action in project ``MAINN'' (funding ID 50OO2206). 
KK and JEMD gratefully acknowledge funding from the Deutsche Forschungsgemeinschaft (DFG, German Research Foundation) in the form of an Emmy Noether Research Group (grant number KR4598/2-1, PI Kreckel) and the European Research Council’s starting grant ERC StG-101077573 (“ISM-METALS"). 

\end{acknowledgements}

\bibliographystyle{aa}
\bibliography{bib}

\newpage
\appendix

\section{Comparison between spectral fits performed within small wavelength regions and the full spectral range}
\label{appendixA}

\begin{figure}[hbt]
    \centering
    \begin{subfigure}{\linewidth}
        \includegraphics[width=0.85\linewidth]{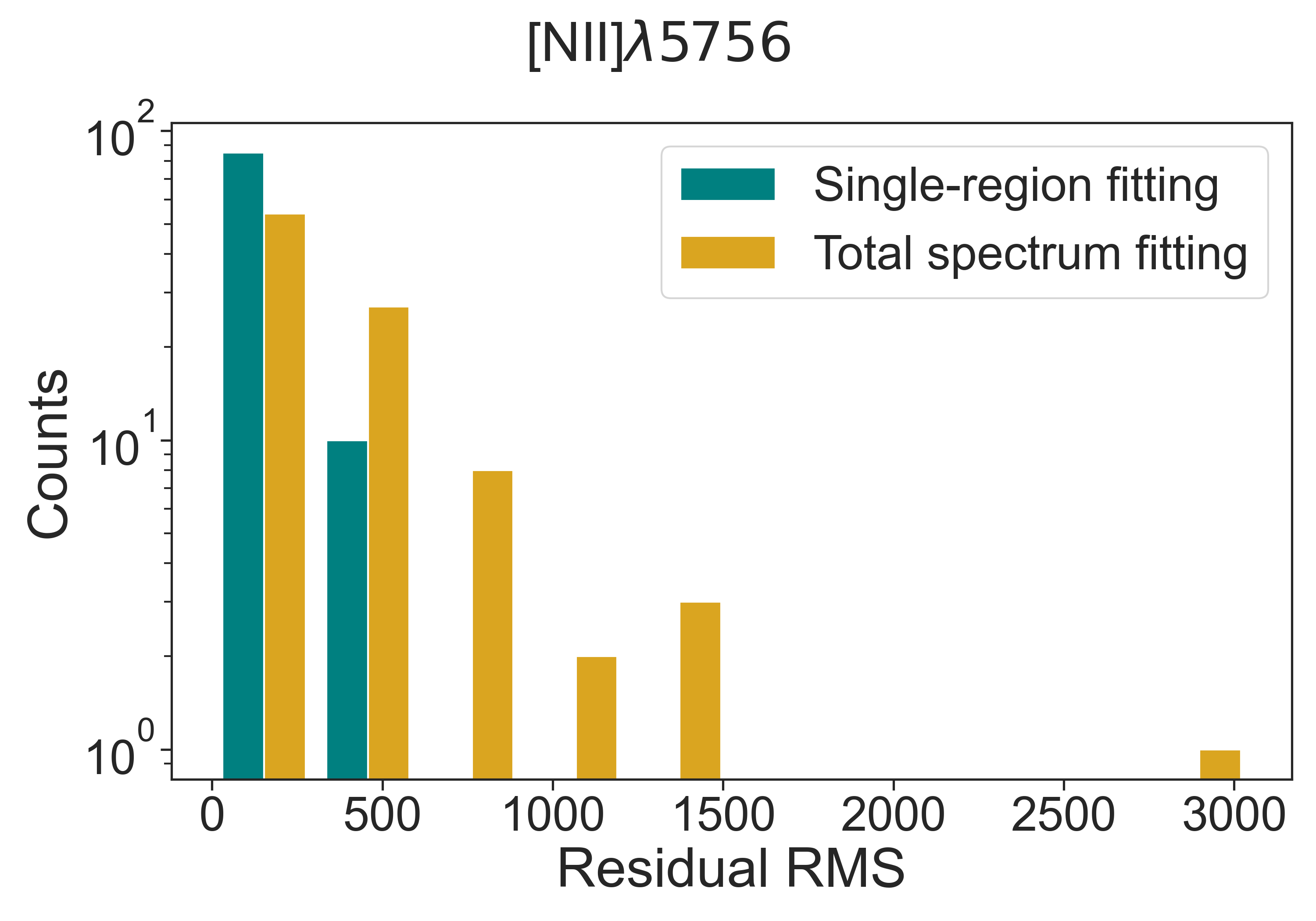}
    \label{fig-NII-residual-rms-comparison}
    \end{subfigure}
    \begin{subfigure}{\linewidth}
        \includegraphics[width=0.85\linewidth]{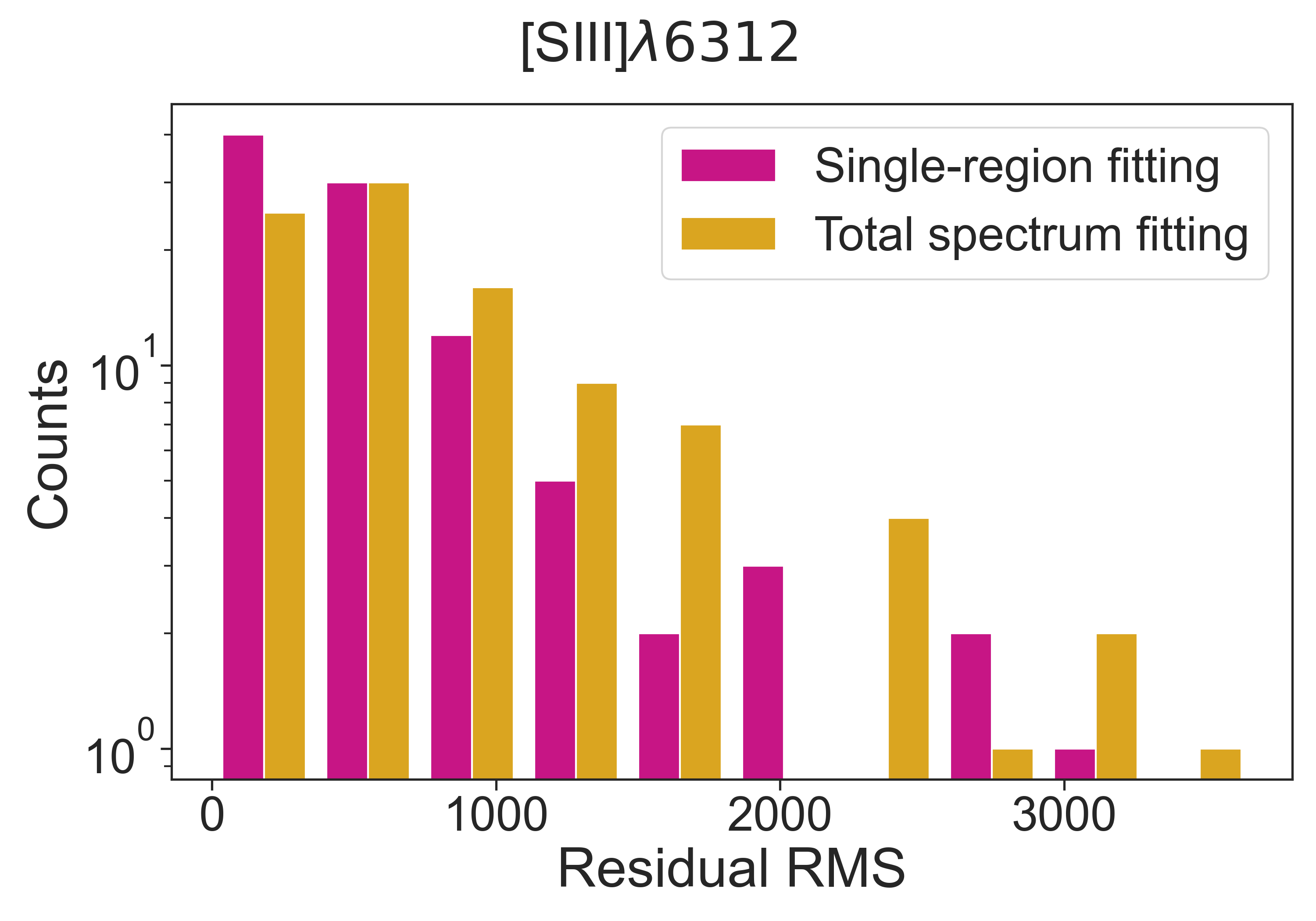}
    \label{fig-SIII-residual-rms-comparison}
    \end{subfigure}
    \begin{subfigure}{\linewidth}
        \includegraphics[width=0.85\linewidth]{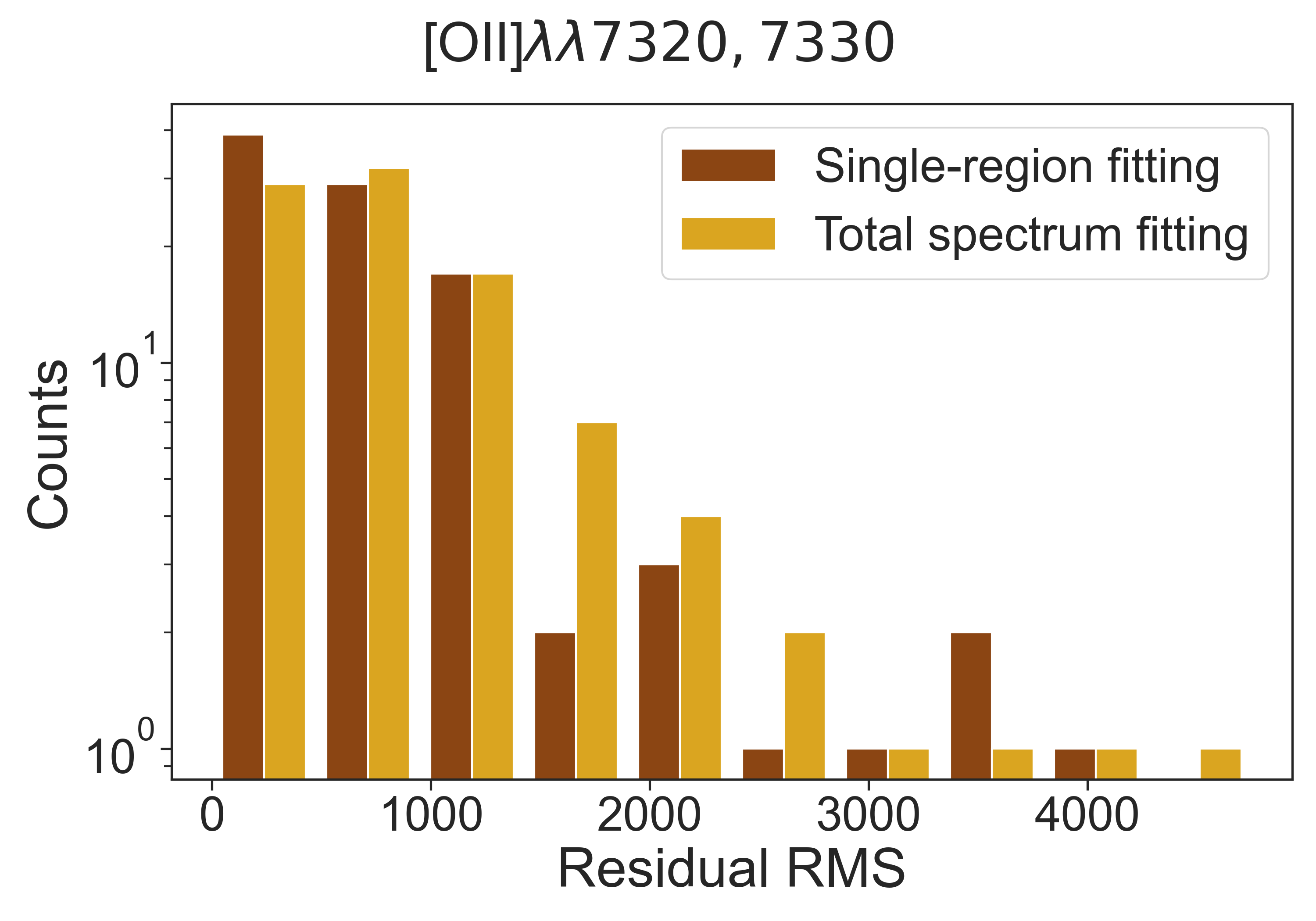}
    \label{fig-OII-residual-rms-comparison}
    \end{subfigure}
    \caption{Histograms describing the distribution of residual RMS around the auroral lines analysed in this work (from left to right: [NII]$\lambda 5756$, [SIII]$\lambda 6312$ and [OII]$\lambda \lambda 7320,7330$) for the 95 selected \hii\ regions from the PHANGS-MUSE nebular catalogue. 
    The two fitting procedures (single-region and total spectrum) are highlighted with different colours in each panel.}
    \label{figure-rms-comparison-hist}
\end{figure}

\begin{figure}[ht]
    \centering
    \includegraphics[width=\linewidth]{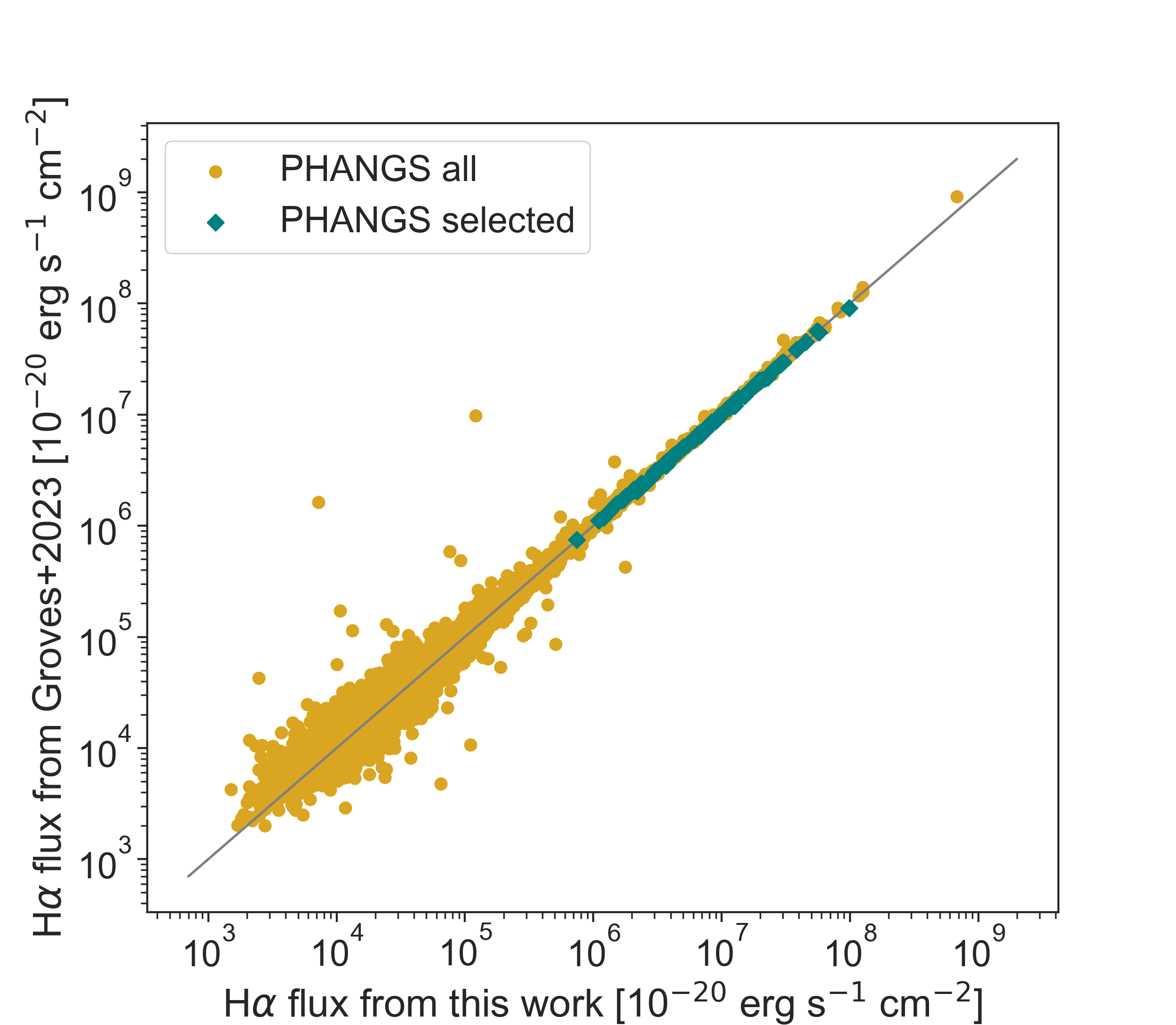}
    \caption{Comparison between dust-corrected \ha\ fluxes from \cite{groves2023} and from our single-region fitting procedure for the 95 selected PHANGS-MUSE \hii\ regions (teal points), as well as for the whole PHANGS-MUSE \hii\ region sample with S/N(\ha) > 5 (golden points). 
    The two fitting procedures produce comparable \ha\ fluxes, especially for the most brilliant regions 
    (\ha\ flux $\gtrsim 5 \cdot 10^{-15}$ 
    erg s$^{-1}$ cm$^{-2} \AA^{-1}$) where their relative difference is always $< 10 $\%. }
    \label{fig:ha-comparison-this-work-vs-groves}
\end{figure}

In this appendix we motivate our choice of deriving auroral line fluxes by performing spectral fits in small wavelength regions around the line of interest instead of fitting the entire wavelength range simultaneously. For this comparison, we fit the spectra from the PHANGS--MUSE nebular catalogue using the entire wavelength range and compare the residuals with the fits obtained in Sec. \ref{section-spectral-fitting}. We focus on on the residuals around the \nii$\lambda 5756$, \siii$\lambda 6312$ and \oii$\lambda \lambda 7320,7330$ auroral lines. 
We computed the RMS of residuals over the whole wavelength ranges from Table \ref{table-spectral-regions}. We only excluded the spectral region at $\lambda \gtrsim 5775 \AA$ for galaxies observed with ground-layer adaptive optics, as in those objects part of the wavelength range is masked because of the sodium emission of the lasers. 

In Fig. \ref{figure-rms-comparison-hist} we report the results of RMS computation for our 95 selected \hii\ regions from PHANGS--MUSE nebular catalogue (Sect. \ref{section-sample-selection}).
We find systematically higher residual RMS around auroral lines when fitting the entire spectrum. The largest differences in RMS appear around the \nii$\lambda 5756 $ auroral line, meaning that for this line the two fitting procedures lead to significantly different results, concerning in particular the fit of stellar continuum, as the line fluxes show similar values in both the procedures (with relative differences $<5\%$ in almost all cases). 

\begin{figure*}
    \centering
    \begin{subfigure}{0.495\linewidth}
        \includegraphics[width=\linewidth]{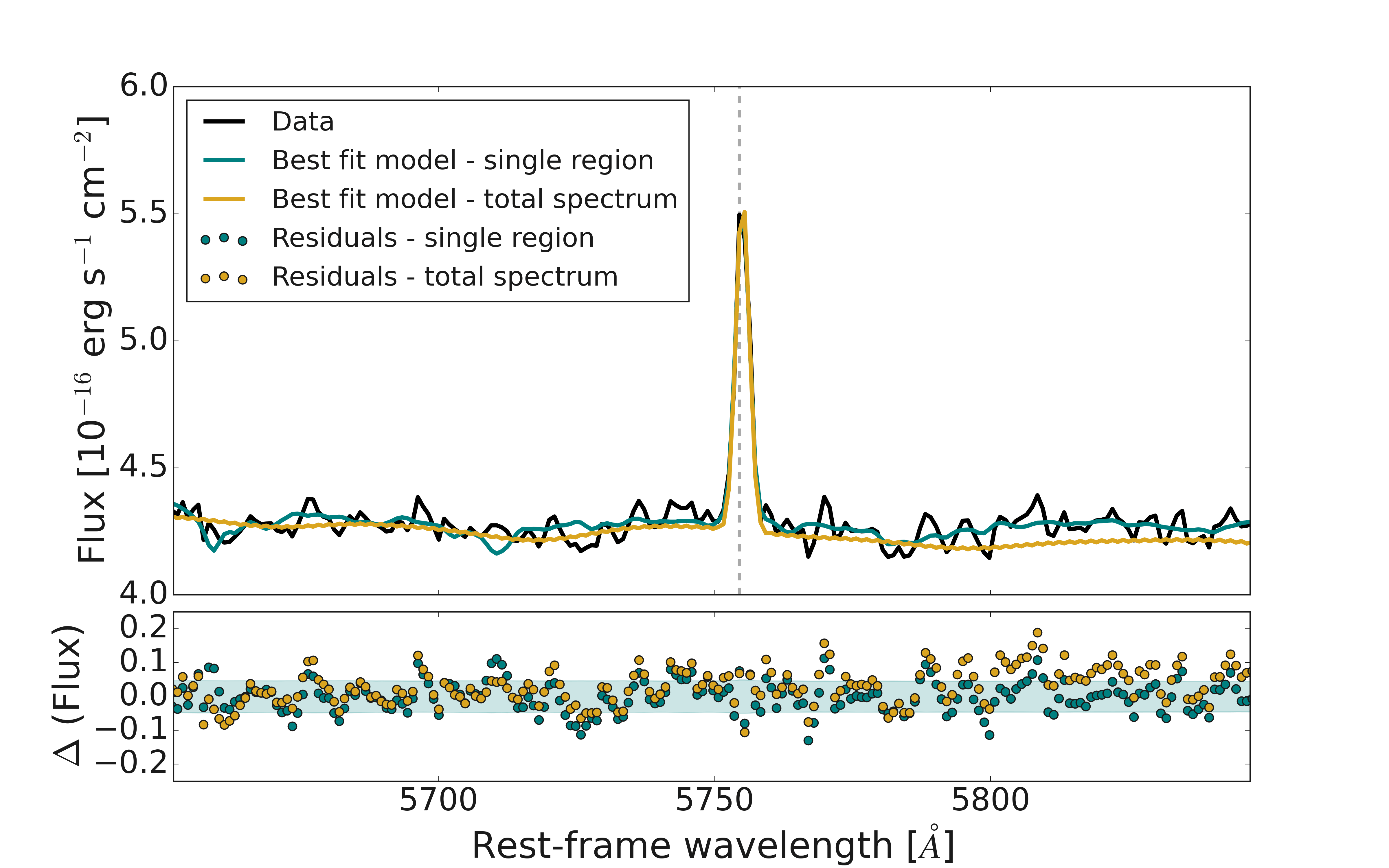}
    \end{subfigure}
    \begin{subfigure}{0.495\linewidth}
        \includegraphics[width=\linewidth]{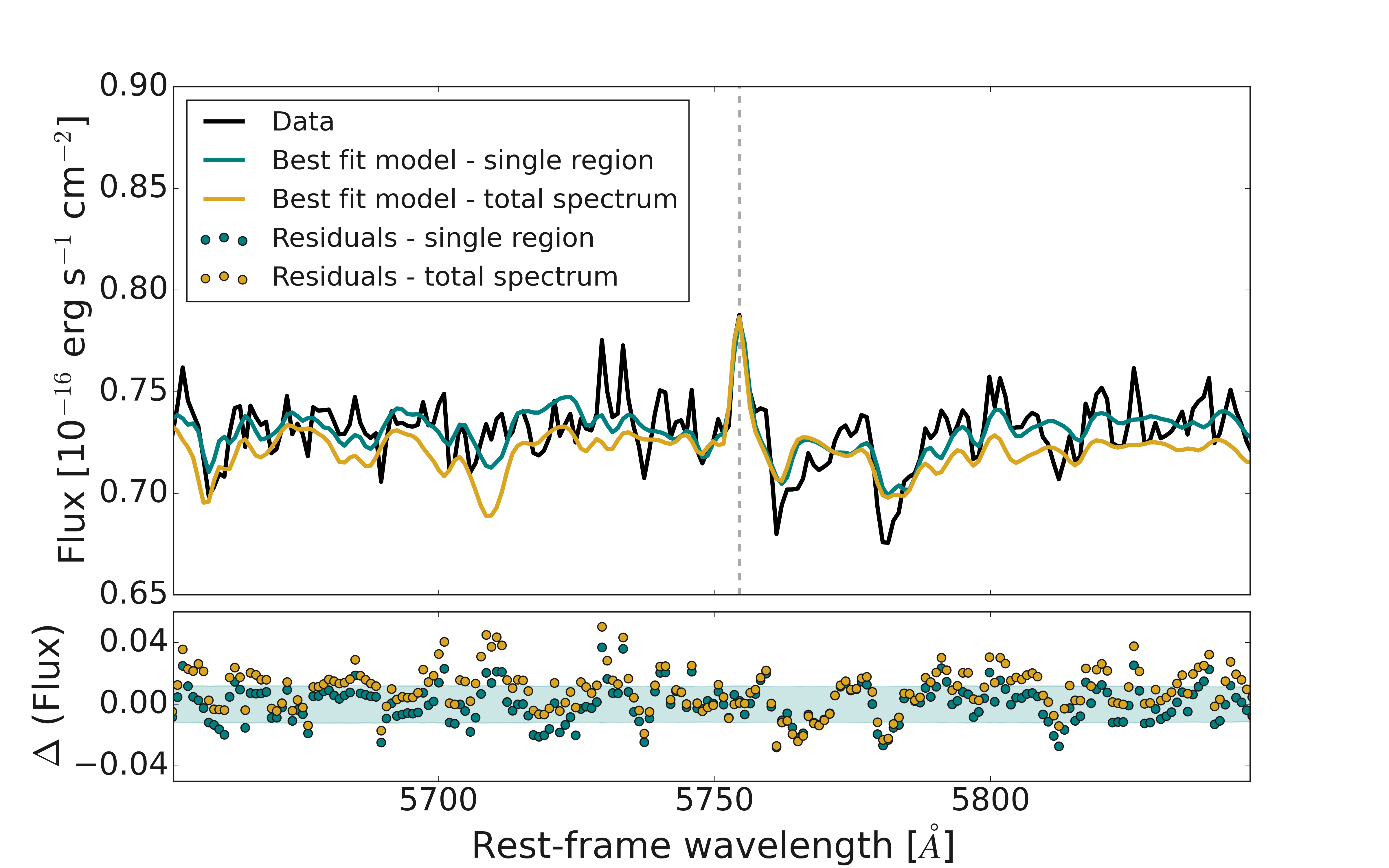}
    \end{subfigure}
    \begin{subfigure}{0.495\linewidth}
        \includegraphics[width=\linewidth]{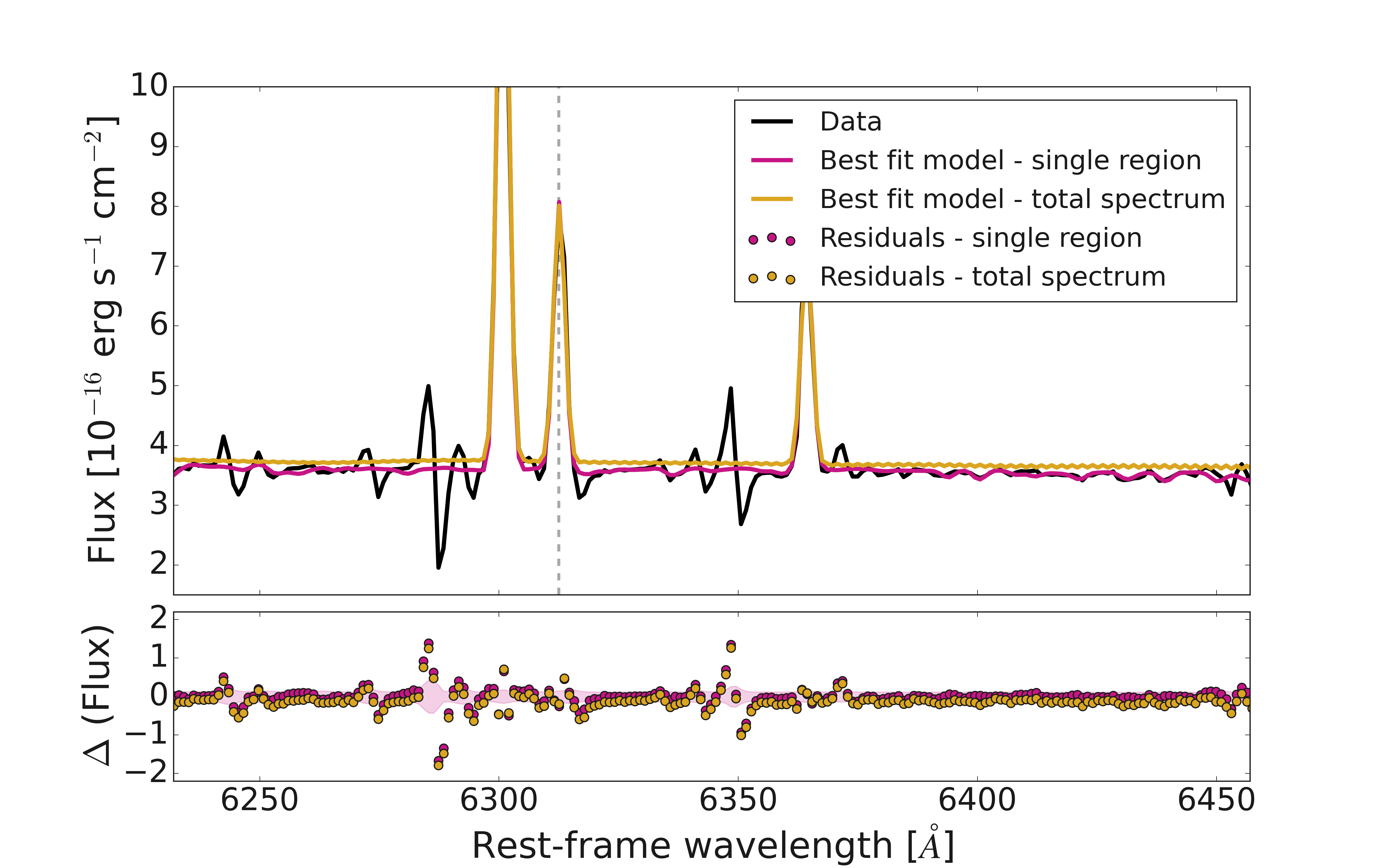}
    \end{subfigure}
    \begin{subfigure}{0.495\linewidth}
        \includegraphics[width=\linewidth]{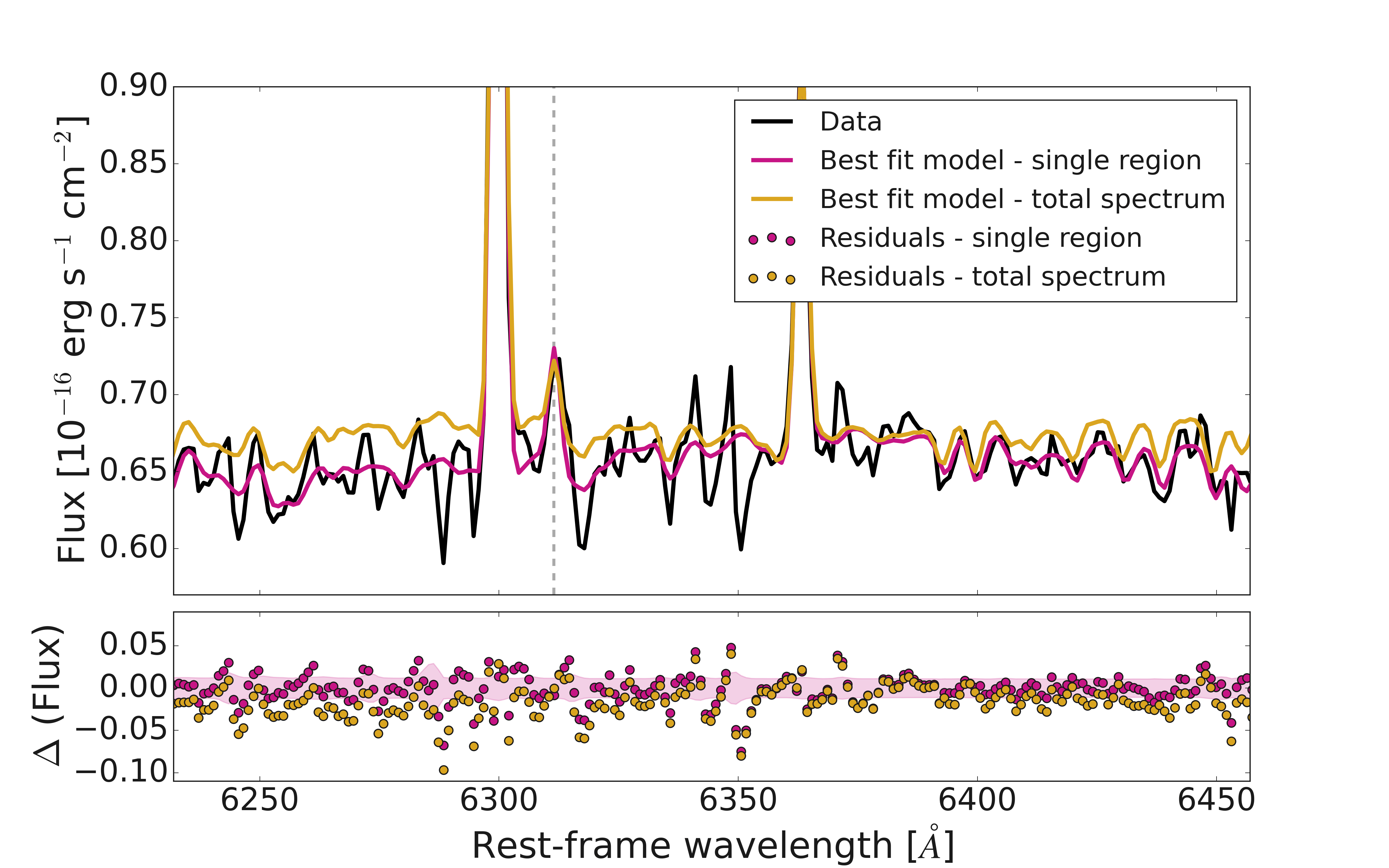}
    \end{subfigure}
    \begin{subfigure}{0.495\linewidth}
        \includegraphics[width=\linewidth]{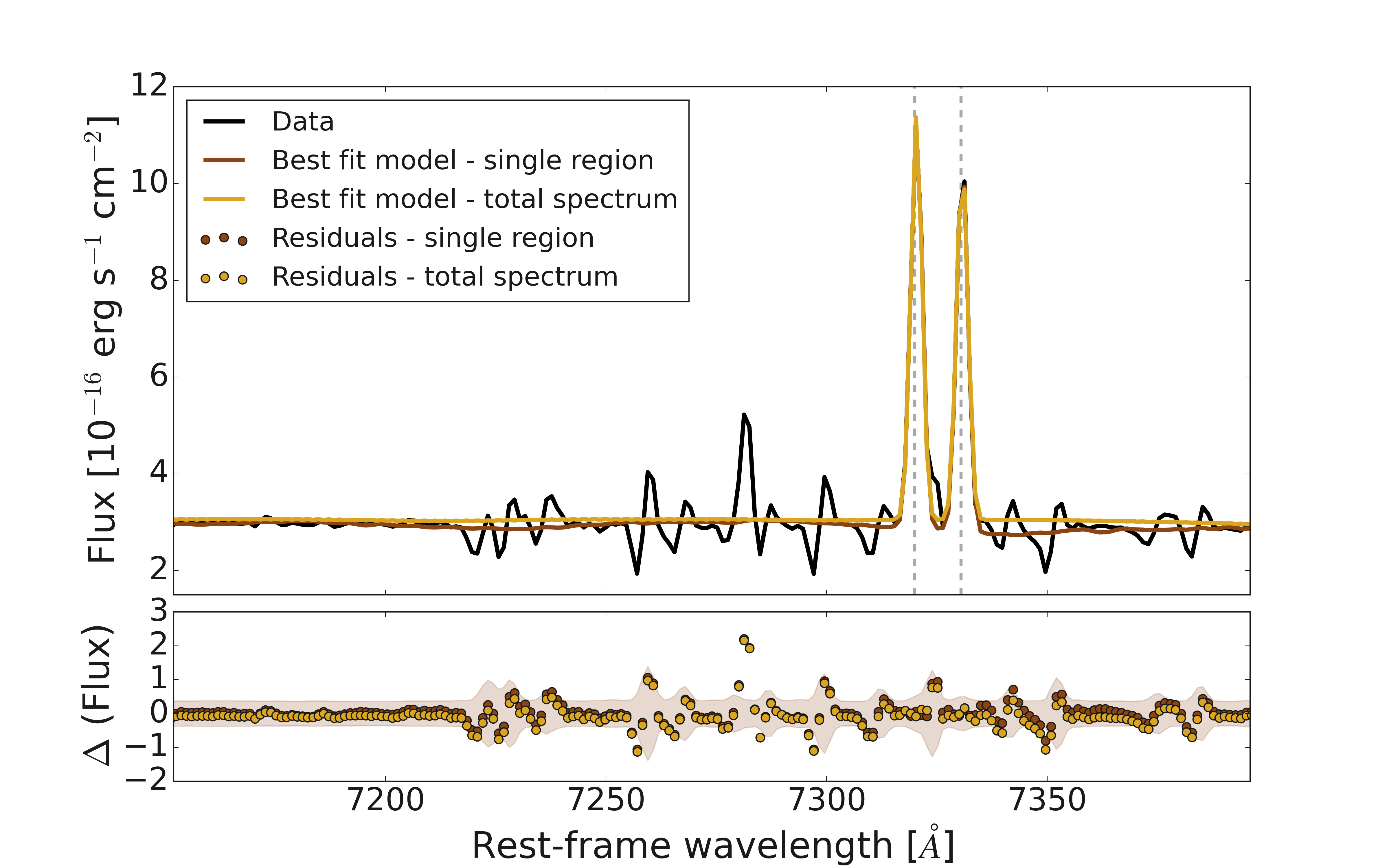}
    \end{subfigure}
    \begin{subfigure}{0.495\linewidth}
        \includegraphics[width=\linewidth]{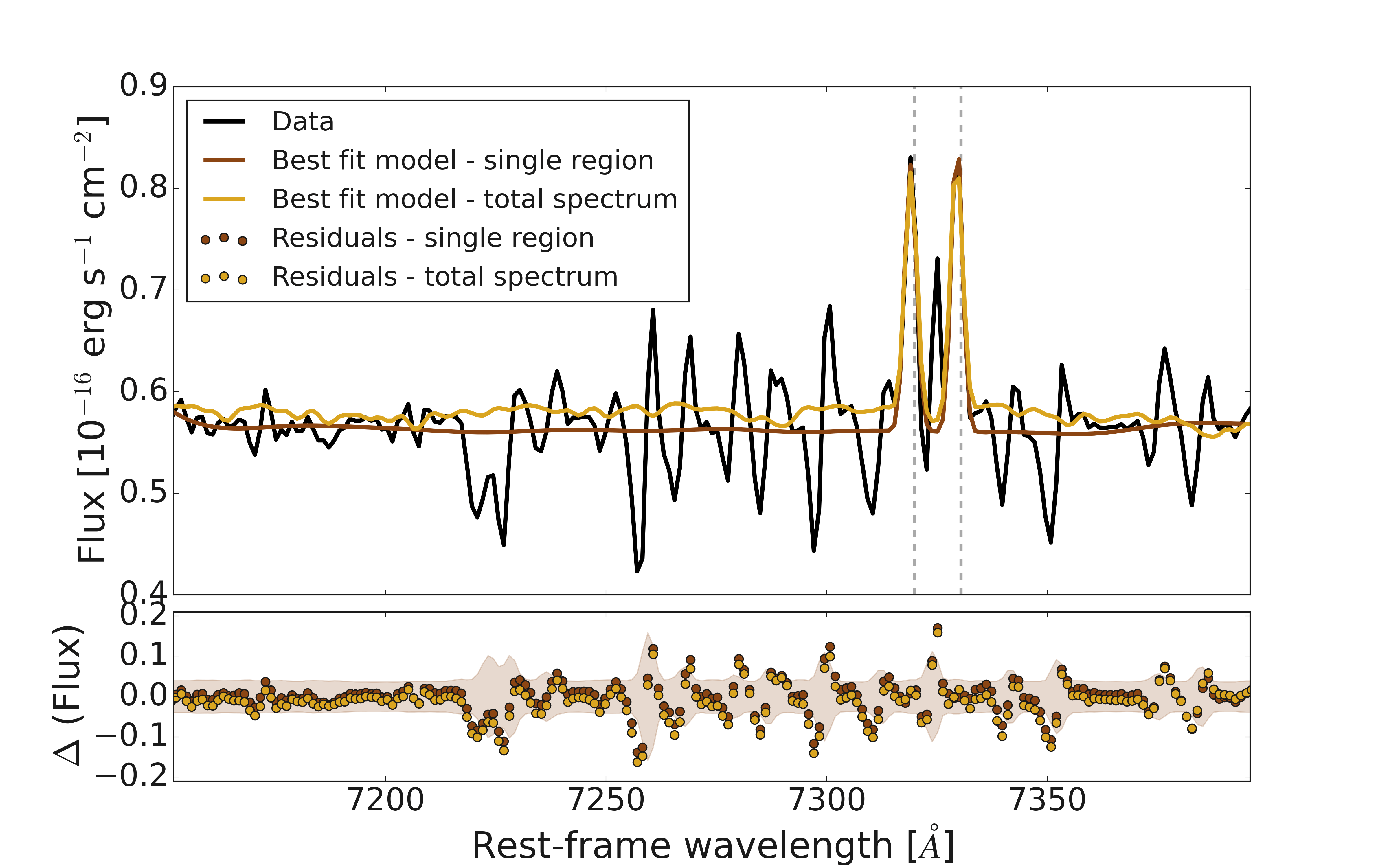}
    \end{subfigure}
    \caption{Comparison between total spectrum and our single-region spectral fitting procedures around the [NII]$\lambda 5756$ (\textit{upper panels}), [SIII]$\lambda 6312$ (\textit{middle panels}) and [OII]$\lambda \lambda 7320,7330$ (\textit{lower panels}) auroral lines for regions 268 (\textit{left column}) and 817 (\textit{right column}) in NGC5068, the same from Fig. \ref{figure-fitting-results-auroral-lines}. 
    When using single-region spectral fitting, the RMS of residuals between data and best-fit around the [NII]$\lambda 5756$, [SIII]$\lambda 6312$, and [OII]$\lambda \lambda 7320,7330$ lines is respectively of 464, 2781, and 3692 for region 268 and 122, 174, and 420 for region 817. When using total spectrum fitting procedure (golden fits), the RMS of residuals reaches values of 611, 3073 and 3798 for region 268 and 169, 253 and 460 for region 817. The corrected spectral error is represented as a shaded region in the residuals panels.  
    }
    \label{figure-ngc5068-comparison-single-total-fit}
\end{figure*}

\begin{figure*}
    \centering
    \begin{subfigure}{0.495\linewidth}
        \includegraphics[width=\linewidth]{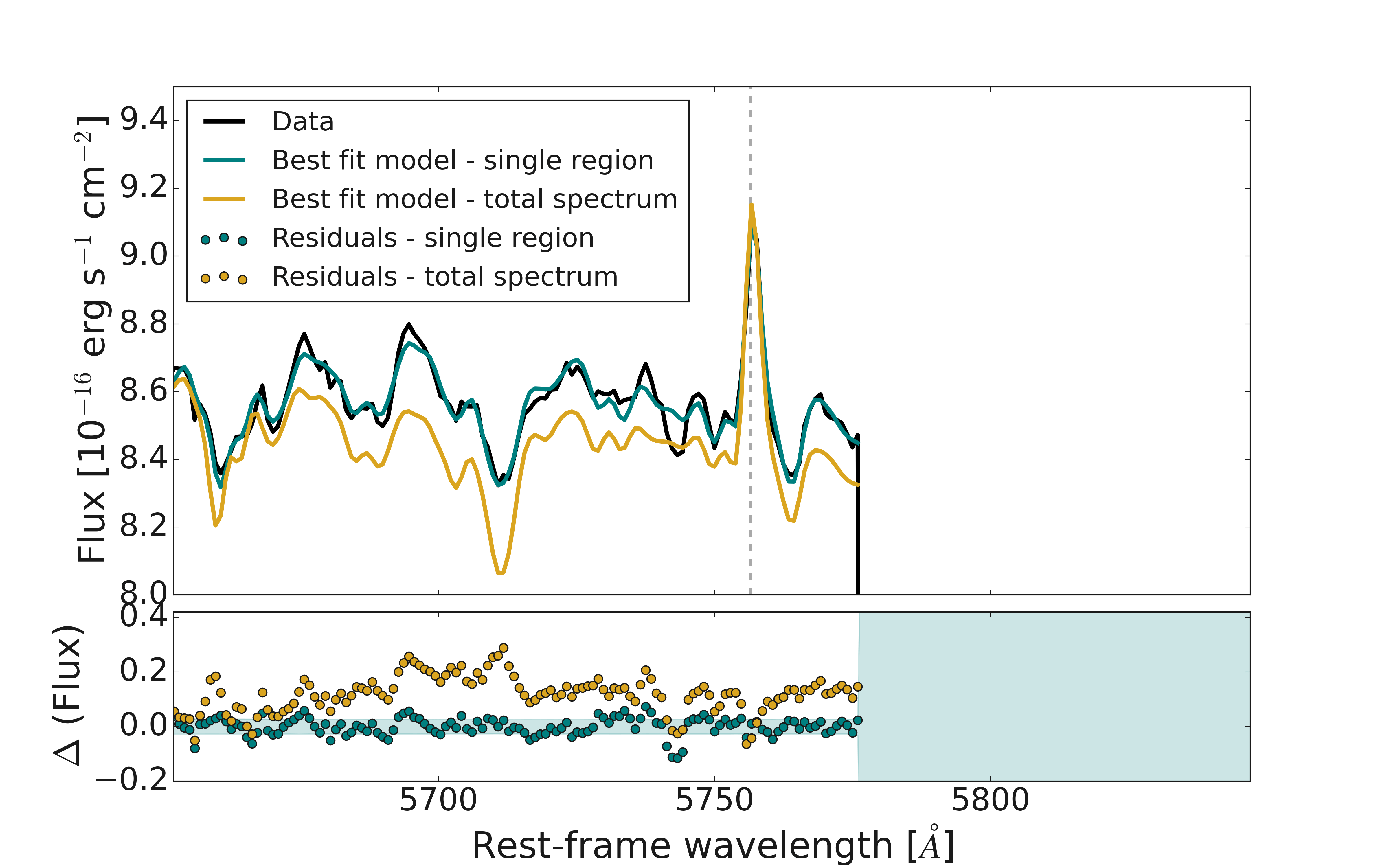}
    \end{subfigure}
    \begin{subfigure}{0.495\linewidth}
        \includegraphics[width=\linewidth]{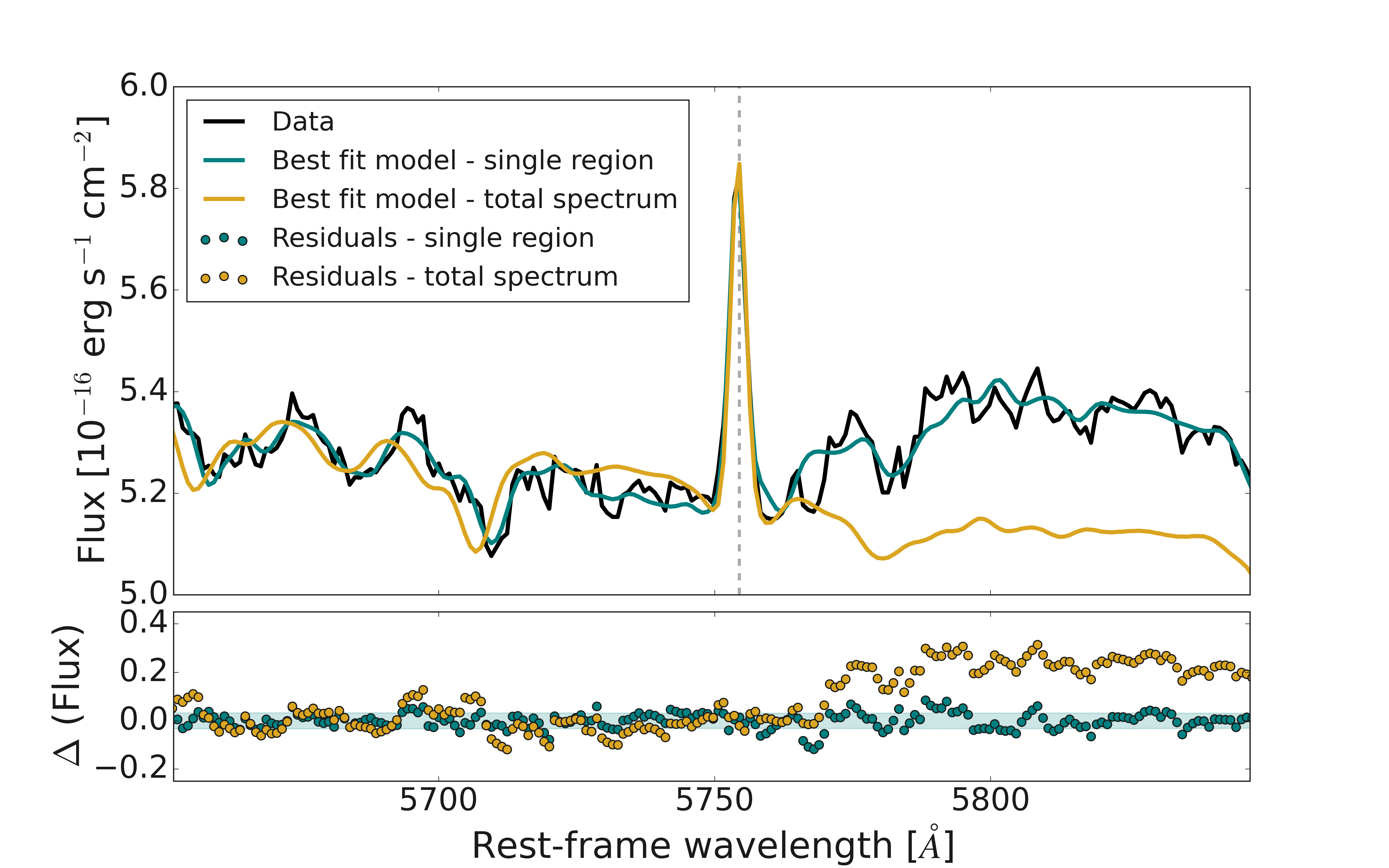}
    \end{subfigure}
    \begin{subfigure}{0.495\linewidth}
        \includegraphics[width=\linewidth]{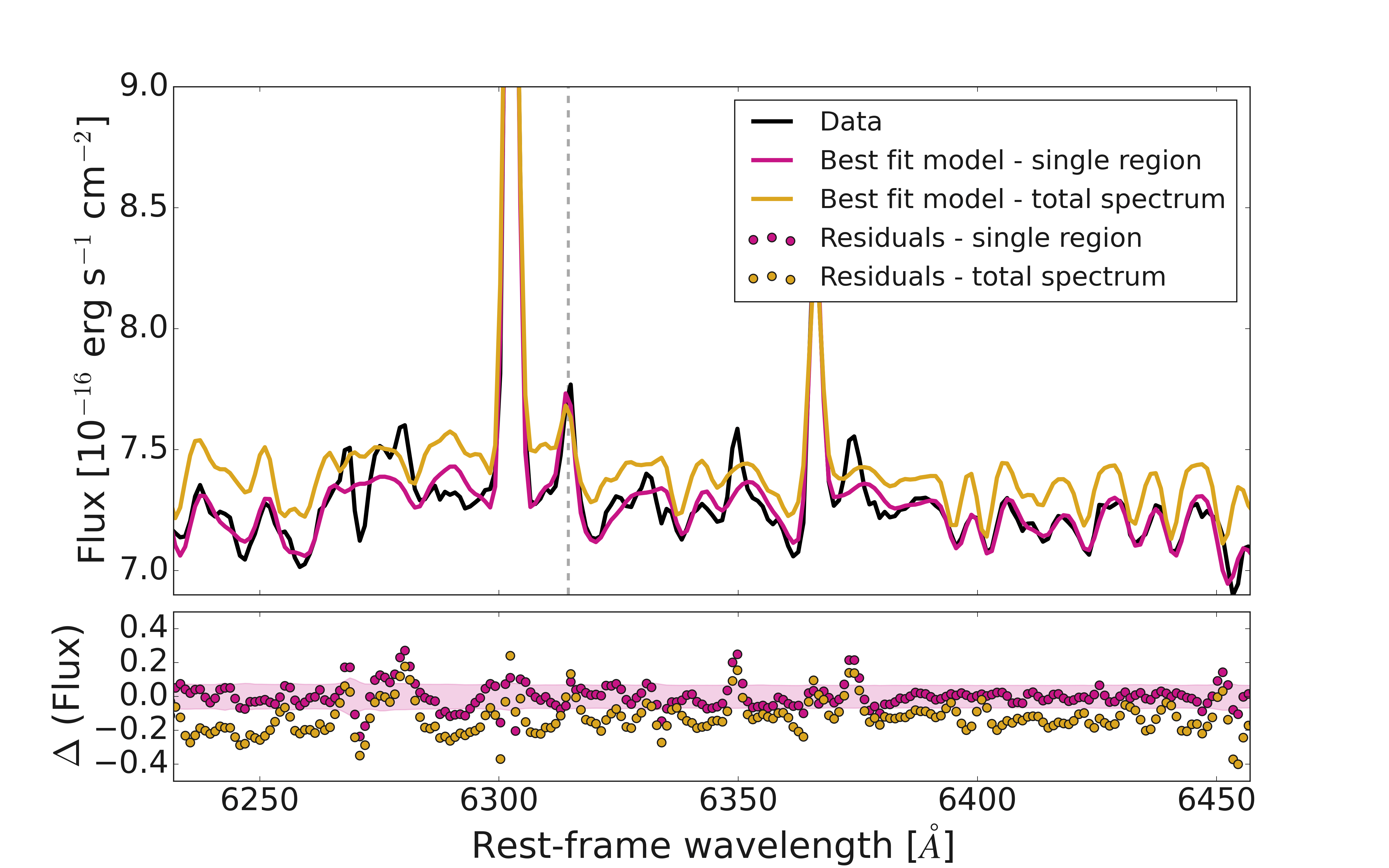}
    \end{subfigure}
    \begin{subfigure}{0.495\linewidth}
        \includegraphics[width=\linewidth]{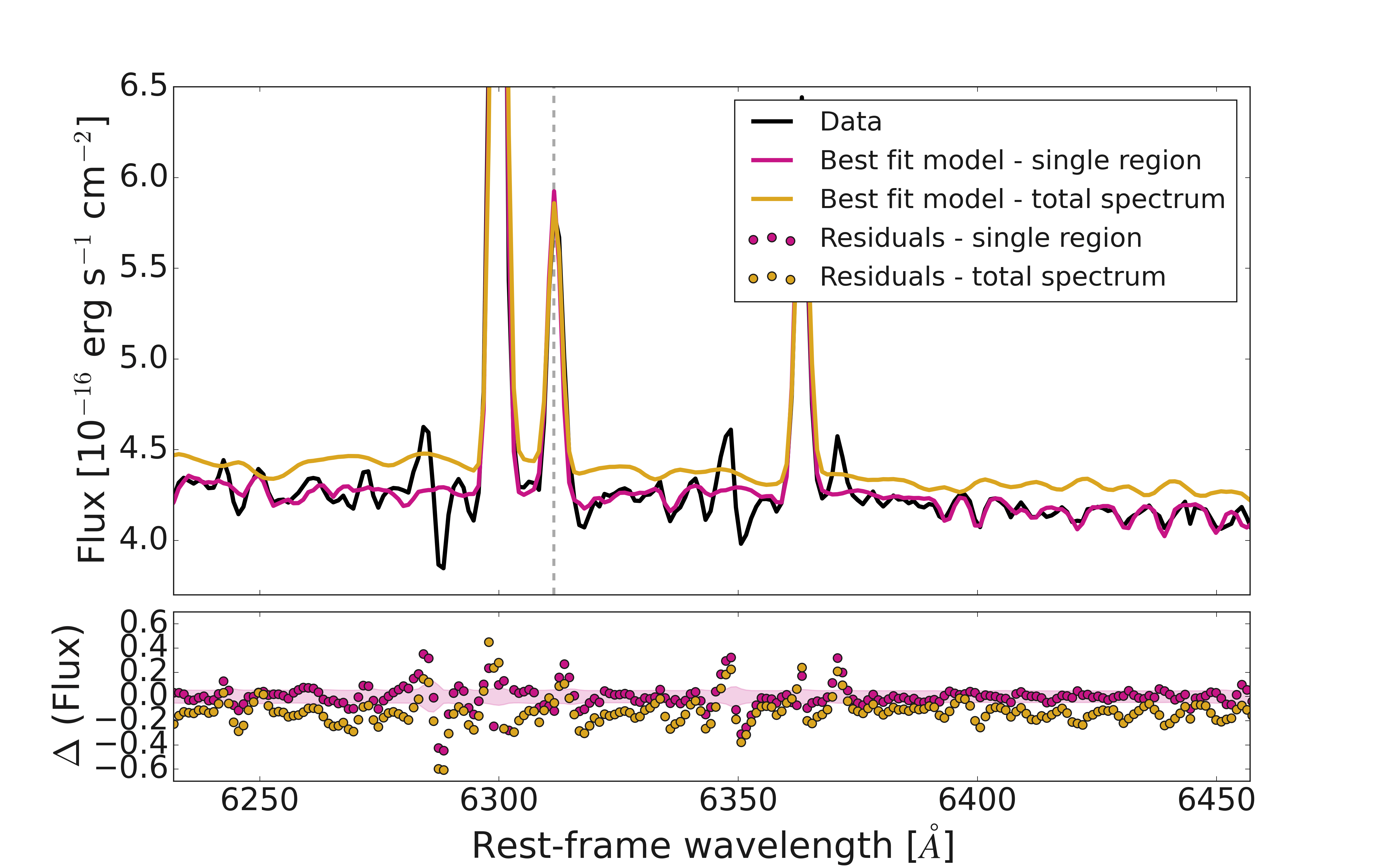}
    \end{subfigure}
    \begin{subfigure}{0.495\linewidth}
        \includegraphics[width=\linewidth]{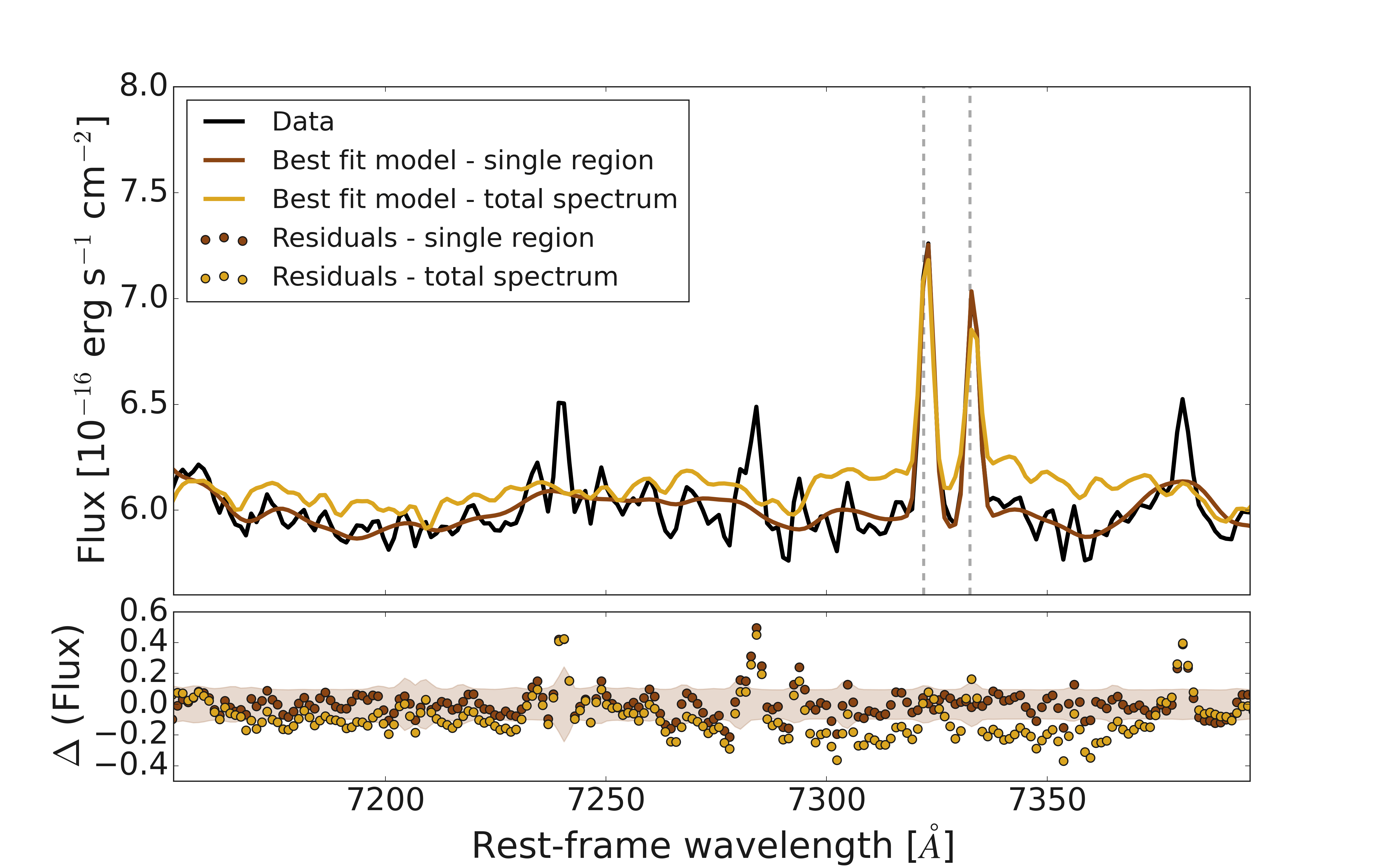}
    \end{subfigure}
    \begin{subfigure}{0.495\linewidth}
        \includegraphics[width=\linewidth]{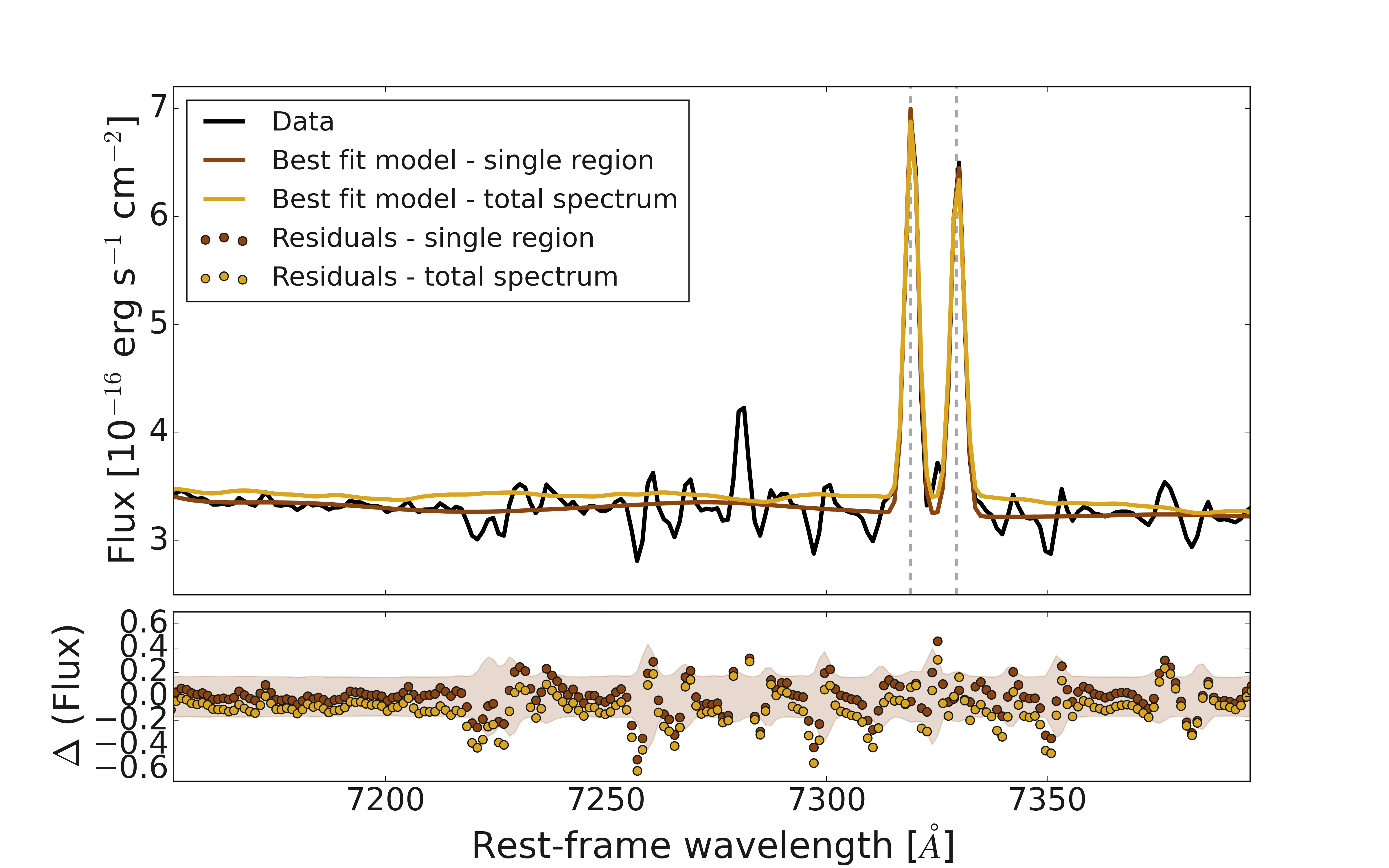}
    \end{subfigure}
    \caption{Two representative examples of \hii\ region spectral fitting around auroral lines where our single-region procedure leads to significantly better results than the standard total spectrum fitting. \textit{Left column:} region 498 in NGC1566. The RMS of residuals around the [NII]$\lambda 5756$, [SIII]$\lambda 6312$ and [OII]$\lambda \lambda 7320,7330$ auroral lines (from to bottom) is respectively of 283, 719 and 998 when considering single-region fitting, but reaches values of 1462, 1597 and 1611 when considering total spectrum fitting. For this and the other galaxies observed in AO-WFM, we exclude from RMS evaluation around  [NII]$\lambda 5756$ the wavelength region with $\lambda _{obs} \gtrsim 5775$\AA, where the flux is masked because of sodium absorption.  \textit{Right column:} region 513 in galaxy NGC5068. In this case, the RMS of residuals around the [NII]$\lambda 5756$, [SIII]$\lambda 6312$ and [OII]$\lambda \lambda 7320,7330$ auroral lines is respectively of 329, 241, and 1581 when considering single-region fitting, but reaches values of 1489, 1799, and 1933 when considering total spectrum fitting. The corrected spectral error is represented as a shaded region in the residuals panels.}
    \label{fig-some-examples-comparison-total-single-spectrum}
\end{figure*}

Some examples comparing the best continuum fits from the two procedures are reported in Figs. \ref{figure-ngc5068-comparison-single-total-fit} and \ref{fig-some-examples-comparison-total-single-spectrum}. 
In particular, the two \hii\ regions presented in Fig. \ref{figure-ngc5068-comparison-single-total-fit} are the same as in Fig. \ref{figure-fitting-results-auroral-lines}, while the two \hii\ regions in Fig. \ref{fig-some-examples-comparison-total-single-spectrum} have been selected to illustrate some of the cases where the two different fitting procedures lead to differences of a factor of $\gtrsim 4$ in residual RMS estimates. 
The presence of an offset between the two fits reflects the fact that the total spectrum fit, as it spans a wavelength range of $> 4000 \AA$, cannot exactly reproduce all the smallest details in restricted wavelength intervals of $\sim 200 \AA$ as the ones analysed in this work.
Interestingly, the bump at $\lambda \sim 8000 \AA$ evident in the top right panel of Fig. \ref{fig-some-examples-comparison-total-single-spectrum} can be attributed to the presence of Wolf-Rayet (WR) stars within the galaxy and in particular to their broad \civ$\lambda 5808$ emission (\citealp{lopez-sanchez+2010}, \citealp{monreal-ibero+2017}). 

\begin{table*}[ht]
    \centering
    \begin{tabular}{c|ccccc}
        \toprule
          &  PHANGS-MUSE & CHAOS  &  \citetalias{curti2017} & \citetalias{guseva2011} & \citetalias{nakajima2022}\\
        \midrule
        \textbf{O$^+$ abundance}  & \multirow{2}{*}{X} & \multirow{2}{*}{\checkmark} & \multirow{2}{*}{\checkmark} & \multirow{2}{*}{\checkmark} & \multirow{2}{*}{\checkmark *} \\ 
       {[OII]$\lambda \lambda 3726,3729$ + $T_e\nii$} & & & & & \\
        \midrule 
        \textbf{O$^+$ abundance}  & \multirow{2}{*}{\checkmark} & \multirow{2}{*}{\checkmark} & \multirow{2}{*}{\checkmark} & \multirow{2}{*}{\checkmark} &  \multirow{2}{*}{X}\\ 
       {[OII]$\lambda \lambda 7320,7330$ + $T_e\nii$} & & & & & \\
        \midrule 
        \textbf{O$^{2+}$ abundance}  & \multirow{2}{*}{\checkmark *} & \multirow{2}{*}{\checkmark} & \multirow{2}{*}{\checkmark} & \multirow{2}{*}{\checkmark} &  \multirow{2}{*}{\checkmark}\\ 
       {[OIII]$4959,5007$ + $T_e\oiii$} & & & & & \\
        \bottomrule
    \end{tabular}
    \caption{O$^+$ and O$^{2+}$ chemical abundance measurements for all the catalogues analysed in this work. The asterisk refers to indirect $T_e$ estimates using $T_e$-$T_e$ relations. In particular, in PHANGS-MUSE analysis $T_e\oiii$ is estimated from Eq. \ref{eq-te-te-rel}, while in Nakajima's analysis $T_e\nii$ is estimated from $T_e\oiii$ using the $T_e$-$T_e$ relation from \cite{rogers2021}.}
    \label{table-chemical-abundances-literature-catalogues}
\end{table*}

The \oii$\lambda \lambda 7320, 7330$ auroral lines are the ones that are less influenced by the different adopted fitting procedure. They lie in a region strongly affected by sky lines that lead to high residual RMS in both cases, and the stellar continuum under these lines is rather featureless. Nevertheless, our single-region fitting procedure provides slightly better results, as highlighted in the lower panel of Fig. \ref{figure-rms-comparison-hist}.

These results demonstrate that single-region fitting provides better stellar continua for the selected spectral
regions around auroral lines with respect to total spectrum fitting. 
There are, however, some issues that could arise when applying this procedure to spectral regions where there is a degeneracy between emission and absorption, for example, around Balmer lines where the stellar absorption becomes significant. 
To verify that our procedure does not bias the Balmer line fluxes because of incorrect stellar absorption modeling, we compare our measurements of dust-corrected \ha\ fluxes for both the 95 selected and the whole sample of PHANGS--MUSE \hii\ regions with the same measurements provided by \cite{groves2023}, where total spectrum fitting is instead carried out. 
The results of this comparison are reported in Fig. \ref{fig:ha-comparison-this-work-vs-groves}. The two fitting procedures always produce comparable \ha\ fluxes, and the correspondence becomes more precise at higher \ha\ fluxes, that is, for brighter \hii\ regions. 
Our 95 selected \hii\ regions all lie within this regime, as they have \ha\ fluxes $\gtrsim 7 \cdot 10^{-15}$ erg s$^{-1}$ cm$^{-2} \AA^{-1}$: their relative difference between \ha\ fluxes is always $< 10$\% and in most cases $<5$\%, which implies variations $< 10$\% in strong-line ratios, negligible with respect to the scatter of the strong-line relations. 
This suggests that the two procedures provide comparable results, and hence no biases are introduced in Balmer line fitting. 

Lastly, we tried to improve the total spectrum fit by increasing the order of the multiplicative polynomial to 12 (from our fiducial value of 8), and we found a marginal improvement in the final fits. 
However, because our single-region fitting procedure still provides better results and with a lower multiplicative polynomial order, we conclude that this is the best approach to fit the stellar continuum for analysis of this type, where it is not necessary to fit the entire spectrum of a source but it is sufficient to fit with high precision only restricted wavelength regions around the spectral features of interest.

\section{Temperature and abundance determination for the literature datasets}
\label{appendixB}

In this Appendix, we summarise the main differences in the analysis of the literature catalogues with respect to the PHANGS--MUSE data analysis presented in Sects. \ref{section-electron-temperatures} and \ref{section-abundance-determination}.
We attempted as far as possible to maintain a homogeneous procedure for determining abundances and temperature, but adaptations were needed due to the differences in line availability.

\paragraph{\textbf{Low-redshift data.} }
In Curti’s \citepalias{curti2017} catalogue the main difference from PHANGS-MUSE data is the lack of $\siii$ lines (Table \ref{table-literature-catalogues-lines}), so that $T_e\siii$ (necessary for estimating sulphur abundance) had to be estimated by exploiting the $T_e\oiii$-$T_e\siii$ relation from \cite{rogers2021}, since $T_e\oiii$ can be directly measured. We decided to employ this relation, rather than Eq. \ref{eq-te-te-rel}, for consistency with Nakajima’s catalogue analysis.

In Nakajima's \citepalias{nakajima2022} catalogue the reported line fluxes allow the measurement of only one electron temperature, $T_e$\oiii. We therefore infer $T_e$\nii\ and $T_e$\siii\ using the $T_e$\oiii-$T_e$\nii\ and $T_e$\oiii-$T_e$\siii\ relations from \cite{rogers2021}. The \oii$\lambda$37327,29 nebular lines are used to infer the O$^+$ abundance.

The CHAOS and Guseva’s \citepalias{guseva2011} catalogues include all the emission lines present in PHANGS-MUSE. For this two catalogues we therefore carried out the same analysis as for PHANGS--MUSE, even though the availability of additional lines would have allowed a more direct estimate of some quantities (e.g. $T_e$\oiii\ using \oiii$\lambda4363$)

\begin{figure}
   \centering
   \includegraphics[width=0.9\linewidth]{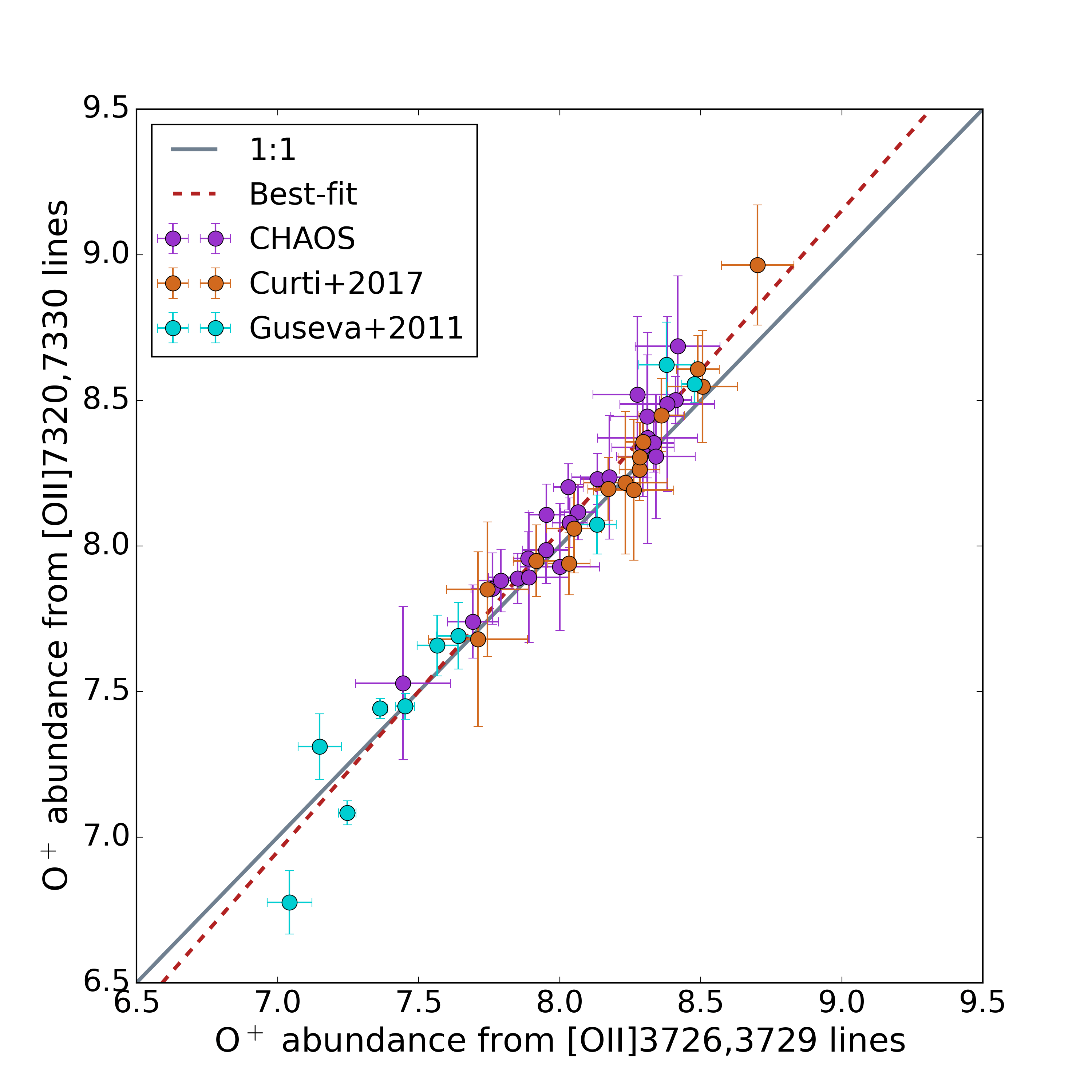}
      \caption{Comparison between O$^+$ abundances measured using auroral (\textit{y} axis) and nebular (\textit{x} axis) lines for CHAOS, \citetalias{curti2017} and \citetalias{nakajima2022} catalogues. 
              }
         \label{figure-O2_abund_comparison_auroral_nebular}
   \end{figure}

The main difference in the derivation of oxygen ionic abundance estimates are summarised in Table \ref{table-chemical-abundances-literature-catalogues}. In three catalogues (CHAOS, \citetalias{curti2017} and \citetalias{guseva2011}) we can measure O$^+$ chemical abundances using both the \oii\ auroral and nebular lines. We therefore verified that O$^+$ chemical abundance measurements do not depend on the \oii\ lines being used, that is, that we find consistent values by exploiting both nebular and auroral lines. This result is shown in Fig. \ref{figure-O2_abund_comparison_auroral_nebular}. 
Some deviations from the 1:1 relation appear at both high and low metallicities, as it appears from the linear best-fit to the data; however, in these metallicity regimes the limited number of data points makes it challenging to definitively conclude that the two O$^+$ abundance measurements are inconsistent. 
If present, such inconsistencies may be explained in terms of possible \oii\ auroral lines recombination contamination, as observed in \cite{perez-montero2017}, and/or different density sensitivity for the auroral and nebular \oii\ lines, as proposed in \cite{mendez-delgado+2023_density}. 
In fact, these authors find a median offset of $\sim 0.1$ dex between the two O$^+$ abundance measurements when using \sii$\lambda 6731 / \lambda 6716$ and \oii$\lambda 3726 / \lambda 3729$ as density estimators, but this offset disappears when adopting different density-sensitive diagnostics, such as \sii$\lambda \lambda 4069,4076\lambda  \lambda 6716,6731$ or \oii$\lambda \lambda 7320,7330\lambda  \lambda 3726,3729$ (see their Fig. 12).
They suggest that this occurs because the \sii$\lambda 6731 / \lambda 6716$ and \oii$\lambda 3726 / \lambda 3729$ line ratios do not account for the presence of high-density clumps within the nebulae, introducing a bias of $\sim 300$ cm$^{-2}$ towards lower densities.
This bias particularly affects low-density regions like the ones we are analysing, which typically have $n_e < 200$ cm$^{-2}$.
We do not further explore this hypothesis, as neither the \sii$\lambda \lambda 4069,4076$ nor the \oii$\lambda \lambda 3726, 3729$ lines are available for the PHANGS--MUSE \hii\ regions.

In our analysis, we find a median offset between auroral and nebular O$^+$ abundance estimates of $\sim 0.05$ dex, which is slightly lower than what is found in \cite{mendez-delgado+2023_density} and is also lower than the associated errors on both abundance measurements (median error on O$^+$ auroral estimates: $\sim 0.12$ dex; median error on O$^+$ nebular estimates: $\sim 0.08$ dex). 
Hence, we can conclude that the two estimates are comparable, and therefore we exploit the nebular lines, when available, to minimise the final errors on ionic abundances. 
Regarding instead the PHANGS--MUSE O$^+$ abundance estimates, which necessarily rely on O$^+$ auroral lines, we find that an eventual offset correction on O$^+$ abundance would produce final oxygen abundances still consistent with the old estimates within the errors (median offset on $12 + \log (\text{O}/\text{H})$ of 0.01 dex versus a median error of 0.03 dex). For this reason, we decide not to apply any offset correction. 

\begin{figure}
   \centering
   \includegraphics[width=\linewidth]{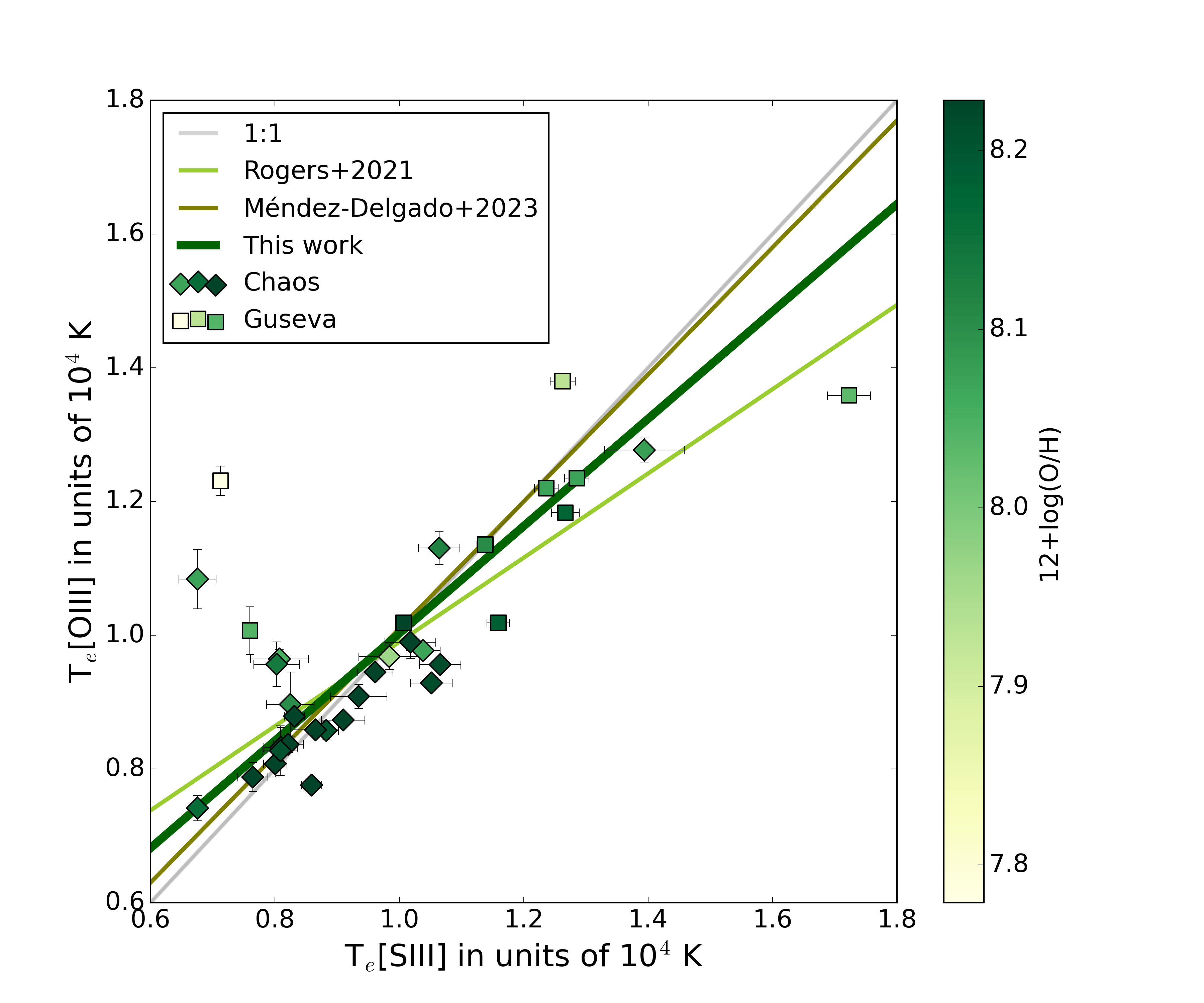}
      \caption{$T_e\oiii$-$T_e\siii$ relation calibrated over CHAOS and \citetalias{guseva2011} selected data, which are reported with different symbols and colour-coded according to their metallicity. The solid dark green line is the MCMC best-fit relation. For comparison, we also report the $T_e\oiii$-$T_e\siii$ relations from \cite{rogers2021} and \cite{mendez-delgado+2023_density}, and we see that there is a good agreement, especially at lower temperatures that is the physical region covered by the CHAOS sample. The grey line is the 1:1 line. 
              }
         \label{figure-TOIII-TSIII}
   \end{figure}

Another significant difference in data analysis is the adopted temperature estimate for the high ionisation zone. 
As the \oiii$\lambda 4363$ auroral line is not available in the PHANGS--MUSE sample, we used the $T_e$\oiii-$T_e$\siii\ relation to estimate $T_e$\oiii  (Sect. \ref{section-electron-temperatures}). We re-calibrated this relation using data from the CHAOS and \citetalias{guseva2011} catalogues, for which both $T_e$\oiii\ and $T_e$\siii\ can be directly measured. We follow a procedure analogue to the one adopted to fit the $T_e\siii$-$T_e\nii$ relation for PHANGS--MUSE data and we obtain (see Fig. \ref{figure-TOIII-TSIII}): 
\begin{equation}
    T_e \oiii = (0.80 \pm 0.02 )~ T_e \siii +(0.20 \pm 0.02),
    \label{eq-teOIII-teSIII-rel}
\end{equation}
with temperatures expressed in units of $10^4$ K. 
The intrinsic dispersion is $\sigma _{int} = 1270 \pm 170$ K, consistent with what was previously found in \cite{rogers2021}.

We test the use of this $T_e$\oiii\ on the determination of the O$^{2+}$ abundance. We find a few data points in both CHAOS and \citetalias{guseva2011} catalogues where the measured $T_e$\oiii\ is significantly higher than that predicted from the $T_e$\siii -$T_e$\oiii\ relation. This trend echoes the findings of \cite{vaught+2023}, who find several regions in PHANGS with abnormally high $T_e$\oiii. 
In \cite{mendez-delgado+2023_density} similar outliers are not present, and in fact they find a $T_e$\siii -$T_e$\oiii\ relation characterised by a smaller intrinsic dispersion of 830 K.

\paragraph{\textbf{High-redshift data.} }

Due to the smaller number of emission lines detected and/or reported in high-redshift publications, a significantly different approach was followed to derive temperature and ionic abundances. 
In particular, dust correction is carried out by exploiting the \hb/$\hg$ ($\lambda_{\hg} = 4341 \AA$) ratio fixed at 2.13 rather than \ha/\hb, as \ha\ falls out of NIRSpec wavelength coverage at $z$ > 6.6.
In both the \cite{laseter2023} and \cite{Sanders2023} datasets the \oiii\ nebular and auroral lines are reported, allowing a direct measurement of $T_e\oiii$. \oii\ nebular lines are also reported, while \oii\ auroral lines are limited to a few detections and some upper limits presented in \cite{Sanders2023}, as for numerous galaxies in the sample such emission lines fall outside of NIRSpec wavelength coverage. We thus decided to neglect the \oii\ auroral lines in our  analysis. 
Moreover, the \sii$\lambda \lambda 6717, 6731$ lines are not reported, so we fixed $n_e = 300$ cm$^{-3}$, which is a representative value for $z=2-3$ galaxies \citep{sanders+2016}.
As the \nii$\lambda 5756$ and \siii$\lambda 6312$ auroral lines are not available, we could not estimate either $T_e\nii$ or $T_e\siii$, and hence we had to rely on the direct measurement of $T_e\oiii$ and the indirect estimate of $T_e\oii$ (rather than  $T_e\nii$, as at high redshift we are dealing only with oxygen lines) from the $T_e\oiii$-$T_e\oii$ relation, as thoroughly described in the reference papers. However, we point out that the use of $T_e\oii$ rather than $T_e\nii$, while referring to the same low-ionisation zone, could introduce some discrepancies \citep{rogers2021, vaught+2023}. 

\end{document}